\DocumentMetadata{}
\documentclass[sigconf]{acmart}
\AtBeginDocument{%
  \providecommand\BibTeX{{%
    \normalfont B\kern-0.5em{\scshape i\kern-0.25em b}\kern-0.8em\TeX}}}

\newcommand{\userquote}[1]{\textit{``#1''}}

\usepackage{booktabs}  
\usepackage{amsmath}  
\usepackage{tabularx}
\usepackage{multirow}
\usepackage{arydshln}
\newcolumntype{R}{>{\raggedleft\arraybackslash}X} 

\usepackage{graphicx}
\usepackage{subcaption}
\usepackage{multirow} 
\usepackage{soul} 
\setstcolor{gray}

\usepackage{cleveref}
\usepackage{listings}
\usepackage{pifont}

\usepackage{xcolor}
\definecolor{grey}{RGB}{128,128,128}
\definecolor{neonfuchsia}{rgb}{1.0, 0.25, 0.39}
\definecolor{darkgreen}{rgb}{0.0, 0.5, 0.0} 
\definecolor{darkcyan}{HTML}{00717f}
\newcommand{\yiren}[1]{{\small\textcolor{neonfuchsia}{\bf [*** Yi-Ren: #1]}}}
\newcommand{\yun}[1]{{\small\textcolor{purple}{\bf [*** Yun: #1]}}}

\usepackage{soul}
\newcommand{\disrevised}[2][]{%
  \if\relax\detokenize{#1}\relax
    #2%
  \else
    #2%
  \fi
}

\newcommand{\disrevisedcamclean}[2][]{%
  \if\relax\detokenize{#1}\relax
    #2%
  \else
    #2%
  \fi
}

\newcommand{\systemName}{\textsc{PersonaFlow}}

\setcopyright{acmcopyright}
\copyrightyear{2024}
\acmYear{2024}
\acmDOI{XXXXXXX.XXXXXXX}

\acmConference[DIS '25]{}{}{}
\acmBooktitle{DIS 25} 
\acmPrice{15.00}
\acmISBN{978-1-4503-XXXX-X/18/06}

\begin{document}

\title{\systemName: Designing LLM-Simulated Expert Perspectives for Enhanced Research Ideation}

\author{Yiren Liu}
\email{yirenl2@illinois.edu}
\affiliation{%
  \institution{University of Illinois Urbana-Champaign}
  \country{USA}
}

\author{Pranav Sharma}
\email{pranav24@illinois.edu}
\affiliation{%
  \institution{University of Illinois Urbana-Champaign}
  \country{USA}
}

\author{Mehul Jitendra Oswal}
\email{mehuljo2@illinois.edu}
\affiliation{%
  \institution{University of Illinois Urbana-Champaign}
  \country{USA}
}

\author{Haijun Xia}
\email{haijunxia@ucsd.edu}
\affiliation{%
  \institution{University of California San Diego}
  \country{USA}
}

\author{Yun Huang}
\email{yunhuang@illinois.edu}
\affiliation{%
  \institution{University of Illinois Urbana-Champaign}
  \country{USA}
}



\begin{abstract}

Generating interdisciplinary research ideas requires diverse domain expertise, but access to timely feedback is often limited by the availability of experts. In this paper, we introduce \textit{PersonaFlow}, a novel system designed to provide multiple perspectives by using LLMs to simulate domain-specific experts. Our user studies showed that the new design 1) increased the perceived relevance and creativity of ideated research directions, and 2) promoted users’ critical thinking activities (e.g., \textit{interpretation}, \textit{analysis}, \textit{evaluation}, \textit{inference}, and \textit{self-regulation}), without increasing their perceived cognitive load. Moreover, users’ ability to customize expert profiles significantly improved their sense of agency, which can potentially mitigate their over-reliance on AI. This work contributes to the design of intelligent systems that augment creativity and collaboration, and provides design implications of using customizable AI-simulated personas in domains within and beyond research ideation.

\end{abstract}






\begin{teaserfigure}
    \centering
    \includegraphics[width=0.8\linewidth]{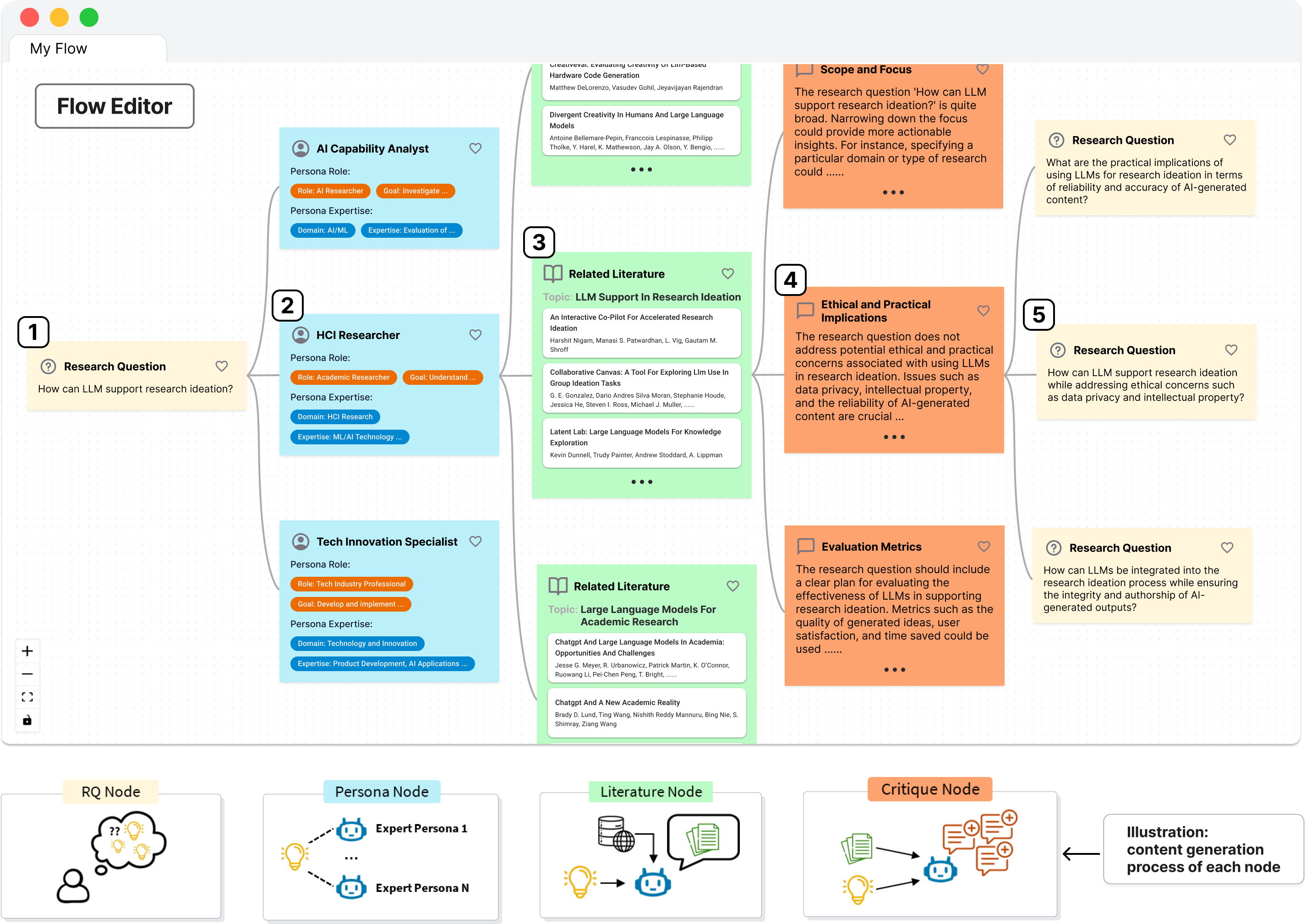}
    \vspace{-.2cm}
    \caption{
    An overview of \systemName\ system's interface design. The system follows a graph-based design that uses four types of nodes to resemble the process of gathering feedback on research ideas from human domain experts using AI-simulated expert perspectives. The design of the system allows users to create initial \textit{Research Question (RQ) Nodes} {\large \textcircled{\small 1}} where users can indicate their topics of interest for exploration. \textit{Persona Nodes} {\large \textcircled{\small 2}} represent AI-simulated expert perspectives that can suggest related literature retrieved from online publication database (\textit{Literature Nodes} {\large \textcircled{\small 3}}), and subsequently provide feedback and critiques (\textit{Critique Nodes} {\large \textcircled{\small 4}}) to users' initial research idea. Based on the critiques and identified literature, the system can further help revise users' initial idea into a revised RQ (\textit{RQ node} {\large \textcircled{\small 5}}). Users can perform this process iteratively and combine inputs from multiple expert personas until they discover satisfactory RQs of their interest.
    }
    \Description{
    An overview of the PersonaFlow system’s interface design. Two panel diagram: Upper panel demonstrates a graph-based design with four types of nodes to mimic the process of gathering feedback on research ideas from AI-simulated expert personas. The design starts with users creating Research Question (RQ) Nodes, where users input their initial research topics. Persona Nodes represent AI personas that suggest relevant literature pulled from online databases, demonstrated as Literature Nodes. These Literature Nodes are connected to Critique Nodes which provide critiques and feedback. These critiques help users iteratively refine their initial RQs into revised RQ Nodes (5). Through this process, users can explore and improve their research questions until they are satisfied with the results. The lower panel of the diagram illustrates the different node categories and corresponding illustrations to signal the node function
    }
    \label{fig:overview-diagram}
\end{teaserfigure}

\maketitle

\section{Introduction}

The development of novel research ideas often requires discussion and feedback collection with peer researchers~\cite{al2020supporting,saeed2021integrating,man2018understanding}, especially for researchers conducting research on interdisciplinary topics. Researchers who are unfamiliar with a research domain often require guidance from domain experts to quickly gain an overview of the field.
However, this requires timely and iterative input from experts, and can often be challenging due to the constraints of \disrevisedcamclean[accessibility and availability of experts]{experts' availability} willing to provide feedback~\cite{chugh2022supervisory}.
Meanwhile, feedback from peers with different backgrounds can be hard to interpret without domain knowledge, thus leading to communication gaps \cite{august2024know} between researchers from various disciplines when discussing research ideas. 
Moreover, researchers sometimes perceive expert critique as personal judgment, potentially leading to increased caution or defensiveness when seeking feedback~\cite{wang2011tell}.

Prior research has shown that AI can be used to deliver non-judgemental immediate feedback and interactions~\cite{lee2020hear,gratch2014s,ta2020user}. 
The concept of persona has been historically used in scenarios like design ideation~\cite{chen2011can,so2017does,salminen2020effect} to facilitate need-finding.
The use of AI-based personas for feedback gathering from an audience perspective has been widely discussed in existing works under various ideation contexts such as design~\cite{park2022social,li2024user}, journalism~\cite{petridis2023anglekindling} and group-based decision-making~\cite{chiang2024enhancing} as beneficial for ideation by simulating human thinking.
\disrevisedcamclean[Providing diverse perspectives using AI-simulated personas would analogically benefit interdisciplinary research ideation.]{Similarly, the use of AI-simulated personas may also benefit interdisciplinary research ideation when used for exploration of diverse perspectives or addressing blind spots due to the limitation of human expertise.}
However, scientific ideation~\cite{foster2004nonlinear} differs from design ideation, which often requires extensive domain knowledge and grounding in literature. 
It remains under-studied whether and how AI personas can be used as expert personas and provide knowledge-intensive feedback and opinions. 
In the context of using LLM for scientific ideation, prior studies have explored using LLMs to assist researchers in developing research ideas in domains such as HCI~\cite{liu2024ai}, ML/NLP~\cite{wang2023scimon} and Economics~\cite{korinek2023language}.
While recent research has shown that LLMs have the capability to simulate personas with different expertise and effectively boost their performance in both knowledge- and reasoning-demanding tasks \cite{wang2023unleashing,chan2023chateval,li2023metaagents}, there appears to be limited UX research exploring the use of LLMs to simulate the diverse perspectives of multidisciplinary experts with different backgrounds for facilitating research ideation, and how the different perspectives impact researchers' ideation processes.

To this end, our study proposed a novel system, \systemName, that aims to provide novel insights into how AI can help facilitate the peer feedback process through simulating diverse domain experts.
As shown in \Cref{fig:overview-diagram}, \systemName\ is an LLM-based research ideation system that facilitates human-AI co-creation between human researchers and LLM-simulated experts. 
Informed by a formative interview study, the novel design utilized a graph-based design that enables users to iteratively co-create and scaffold research ideas and questions. The system also adopts a mind map-styled design to visually organize information and ideation outcomes. 
To evaluate \systemName, we conducted a two-phase study. First, we ran a task-based study with 21 participants, 
evaluating users' perceptions of and interactions with these experts, and examining how these changed temporally as they engaged with more expert perspectives temporally during the study.
Second, we conducted an optional open-ended exploration session where 10 participants were invited to freely explore the system without tasks or constraints, to uncover their in-situ use cases of \systemName\ in supporting the ideation process.

The contribution of this paper to the HCI community and the literature on human-AI co-creation is three-fold:

\begin{itemize}
    \item First, we designed, developed, and evaluated a novel LLM-based system, called \systemName, which aims to facilitate human-AI collaboration in research ideation through graph-based interactions. The system supports interdisciplinary research ideation through four key steps, each associated with a specific node type: 1) framing research questions (\textit{RQ Node}), 2) seeking help from domain experts (\textit{Persona Node}), 3) gathering literature recommendations (\textit{Literature Node}), and 4) obtaining feedback (\textit{Critique Node}) from domain experts. 
    

    \item Second, our user evaluation studies provided new empirical evidence, demonstrating its effectiveness in supporting research ideation. Specifically, users perceived multiple perspectives offered by the simulated experts as more relevant and insightful. The design also enabled users to edit experts' profiles, enhancing their sense of agency.

    \item Third, the case studies illustrated the system's versatility in supporting users' \textit{critical thinking} activities, e.g., \textit{interpretation}, \textit{analysis}, \textit{evaluation}, \textit{inference}, and \textit{self-regulation}. For example, we observed that participants engaged in \textit{inference} thought processes after receiving new relevant papers recommended by the (\textit{Literature Node}); and they exhibited \textit{self-regulation} by rephrasing their initial RQ after reviewing the (\textit{Critique Nodes}) from different AI-simulated experts.
    
    \item Fourth, we proposed design implications for mitigating potential confirmation and social biases that were revealed during participants' use of the system. We further discussed the broader applications of such design platforms for scaffolding ideas from multiple perspectives in other domains, e.g., service design and creative writing. 
    
\end{itemize}

\section{Related Work}
In this section, to show how our work contributes to the existing literature, we provide a detailed review of relevant existing works from three perspectives.
First, we present how LLMs have been used to support research ideation. 
Then, we show the feasibility of LLM-based persona simulation to support both task-oriented and creative applications. 
Finally, we review prior literature on cognitive theories that inform the dynamics of collaborative ideation.

\subsection{LLM for Research Ideation and Scaffolding}
Researchers have started exploring methods to use LLMs to augment research ideation and brainstorming processes for scientific writing \cite{baek2024researchagent,wang2023scimon,shen2023convxai,korinek2023language}. For instance, \citet{nigam2024acceleron} employed an LLM-Agentic workflow that utilizes Retrieval-Augmented Generation (RAG) to identify similar papers based on an initial research topic provided by the user, utilizing colleague and mentor profile-based agents to perform motivation validation and method synthesis iteratively to develop a research proposal. Similarly, \citet{gu2024generation} used a knowledge graph to identify sub-graphs relevant to a researcher's interest and uses these sub-graphs alongside a model to suggest scientific research ideas based on initial research topics given by researchers. \citet{guo2024exploring} explored the usage of AI assistants on user ideation, specifically, analyzing the impact of AI to introduce bias in user ideation.
Studies \cite{memmert2023towards, gonzalez2024collaborative, shaer2024ai} have also discussed the use of LLMs for augmenting ideation and brainstorming processes for creative and group ideation. Other works, such as \citet{sandholm2024randomness}'s work employed a non-linear thinking methodology that utilizes a problem-solution dataset and fine-tuned LLMs to infer new problem statements from the original user problem statements. \citet{lozano2023clinfo}'s work used RAG and fine-tuned LLMs to build a question-answering system using PubMed data for the medical domain. 
\disrevisedcamclean{
In this study, we explore the feasibility of using LLM-based techniques to support human-in-the-loop research ideation through the simulation of expert personas. 
Prior systems aimed to automate research ideation or offer interaction through a single conversational assistant or agent, while few provided support to explore knowledge from multiple diverse perspectives. The \systemName\ system fills this gap by enabling users to engage with diverse perspectives represented by LLM-simulated customizable expert personas. We show that through the user study the use of such expert perspectives can improve ideation quality and also stimulate critical‑thinking behaviors without added cognitive burden.
}

\subsection{Simulating Diverse Perspectives using LLMs}
LLM-related research has widely discussed the use of persona-based methods in terms of task-oriented scenarios. Studies showed that the use of persona-based role-playing benefits the task-based performance of LLMs in a varied range of tasks, such as software development~\cite{qian2024chatdev}, mathematical problem solving~\cite{wu2023autogen}, and general instruction following~\cite{xu2023expertprompting}.
Collaboration among multiple LLM-based personas has been noted as beneficial in similar task-based environments \cite{wang2023unleashing, wang2023tpe, chan2023chateval, wu2023large}. 

The concept of using personas is not new and has been previously utilized in design and ideation processes to foster creativity and generate diverse ideas~\cite{kurtzberg2001guilford, so2017does, pruitt2010persona}.
With recent advances in LLM research, there is growing evidence that LLMs can simulate human behaviors and cognitive processes~\cite{park2023generative, dubois2024alpacafarm, park2022social}.
For instance, past works explored using the LLM-based persona simulation to curate feedback from the perspectives of different audience groups~\cite{petridis2023anglekindling,park2022social,li2024user,chiang2024enhancing}. 
From a broader perspective, studies have utilized LLM-simulated personas to gain insights into social science and economics research \cite{ziems2024can, li2024econagent}.
Research has suggested that taking perspectives from different team members during the collaborative ideation process helps to improve both ideation creativity and diversity \cite{hoever2012fostering}. Similar benefits have also been observed in the context of collaborative scientific ideation \cite{sun2022students}. 
\citet{benharrak2024writer} introduced and assessed a tool that aimed to support writers in improving their work by incorporating feedback provided by individual agents based on user-constructed personas.
However, despite these advancements, there has been limited discussion on how multi-persona simulation by LLM can enhance human-AI co-creation and collaborative ideation. In this study, we aim to explore the potential of multi-persona LLM simulations to enhance the scientific ideation processes.

\subsection{Cognitive Theories for Collaborative Ideation}
\disrevised{Scientific discovery requires both extensive knowledge and creative thinking processes \cite{langley1987scientific, aschauer2022contribution, foster2004nonlinear}. Research has sought to unveil the underlying cognitive mechanisms involved in scientific creativity, aiming to understand how scientists generate innovative ideas and make breakthroughs in their respective fields. 
Previous studies have highlighted the importance of various thinking processes in fostering scientific creativity, such as critical thinking \cite{newton2010creativity}, divergent thinking\cite{sak2013creative, hu2010creative}, and convergent thinking \cite{agnoli2016estimating}.
Moreover, critical thinking has been extensively examined within the domain of science education~\cite{santos2017role,bailin2002critical,vieira2011critical}, establishing it as a crucial skill for development especially for novice researchers and those interested in engaging with new fields.
Past research indicates that exposure to different perspectives and opinions improves critical thinking~\cite{Walton1989DialogueTF,Kuhn2019CriticalTA} during scientific research~\cite{Nussbaum2012TheTF,Hendriks2020ConstraintsAA}. Critical thinking skills can also be fostered through the encouragement of perspective-taking~\cite{Southworth2022BridgingCT} and the integration of diverse viewpoints~\cite{Lai2011CriticalTA}.}

The process of research ideation often also involves collaborative efforts, such as brainstorming sessions, peer discussions, and interdisciplinary collaboration, which can significantly enhance the potential for novel discoveries and innovative solutions \cite{siangliulue2015toward}.  
On the other hand, studies \cite{ulrich2016product, kornish2011opportunity, terwiesch2009innovation} have examined the approach of parallel search in a group-based ideation setting, both in teams and as individuals, whose findings emphasize the importance of incorporating individual contributions that are independent of each other to improve ideation quality. 
Relevant to understanding the collaboration dynamics between members during group-based ideation, Transactive Memory System (TMS) theory \cite{lewis2011transactive} has been proposed as a theoretical framework for understanding the information dynamics in team-based collaboration, where the three main factors include specialization, coordination and credibility.
Empirical studies \cite{ali2019mechanism, dastmalchi2021exploring, hanke2006team} have suggested the benefits of a well-developed transactive memory system for collaborative ideation in a team setting. Closely related, studies~\cite{mayo2017metatheoretical, harrison2007s} also highlight the importance of diversity in team members' expertise and background in promoting creative problem-solving and enhancing the overall efficacy of the group ideation process.
Past research has sought to understand users' behavior of interacting with AI-simulated experts under the context of writing \cite{benharrak2024writer} and chit-chatting \cite{ha2024clochat}. 
\disrevisedcamclean{
Building on the literature reviewed above, we conceptualize research ideation as a four‑phase cycle of divergent exploration, evidence gathering, feedback gathering, and convergent consolidation. We later designed \systemName\ to support each of the four phases. In this study, we also aim to deepen the understanding of users' expert profile customization behavior for scientific ideation through empirical observations. We examine the empirical results through the lens of TMS and critical thinking theories, thus providing implications of designs using LLM-simulated personas in the context of ideation. 
Our work hence extends prior LLM‑assisted ideation tools and explores a transferable design framework for future systems targeting different stages of ideation.
}

\section{\systemName: System Design and Implementation}

In this section, we detail the system design and implementation of \systemName. 
We first describe the formative interview study conducted with 10 researchers to identify users' key challenges and inform the design of our system. Then, we present the detailed feature design and backend implementation of \systemName, including an example user walk-through to illustrate the flow of interaction.

\subsection{Formative Study: Informing the Design of \systemName}
\label{sec:formative-study}
We conducted a formative study by interviewing 10 researchers\footnote{all participants later enrolled in the formal user studies, we use the same participant IDs for quotes} from different background domains to identify their needs when conducting interdisciplinary research. According to the interviewees' self-identification, 5 were from the Human-Computer Interaction (HCI) field, 3 were from the Bioinformatics field, and 2 had backgrounds in Psychology.

All of the interviewees mentioned the challenges in interdisciplinary learning, including difficulties in understanding methods and jargon from other domains. 
Identifying and framing meaningful research problems in unfamiliar domains was also deemed challenging to participants, which involves selecting suitable methods and understanding their relevance and application in the new context (\userquote{... we always use like different kind of new research methods ... all these kind of things were new to me, which is a big challenge for me ... ,} P19). 
Participants also noted the challenge of finding and understanding relevant past research articles. The rapidly evolving nature of some fields (e.g., HCI and Biomedical), makes it challenging to find comprehensive and up-to-date literature.
We also identified challenges in accessing domain-specific knowledge from experts. When seeking advice from peers, professors, or domain experts without already having an understanding of the domain, the guidance researchers receive is typically limited to being directed to research papers or resources, rather than receiving detailed explanations (\userquote{... it can be kind of hard to reach out and then try to explain what I want ... some of the concepts require a lot of background knowledge and previous experience that I feel like they can't really communicate to me just through an email, or like a 5 minutes talk ... ,} P12).
Participants also highlighted difficulties in finding the right person to consult, and there is often a gap in understanding between them and the experts they approach, as noted by P7 \userquote{... a gap of understanding between the two (researchers) ... professors have different kind of understanding and expectation about our capabilities, they might be thinking that I might be aware of a statistical concept ... but I do not know how to do it, and they're like, but I thought you knew it already ... .}
\disrevised{The participants were also asked about their past experience using AI tools during the ideation stage when conducting interdisciplinary research. P10 mentioned that they would use ChatGPT for explaining unfamiliar concepts and terminology, while noting its tendency of being generic \userquote{... but ChatGPT sometimes can be confused ... it sometimes would start repeating [the] same concepts ... the responses are generic and not specific ...} Other participants also shared similar remarks, including P12 \userquote{... hard to prompt it well enough to give more than generic responses, or sometimes it just gets too complicated.}}


\label{sec:design-requirements}
Considering the identified challenges and needs, along with existing literature on scientific ideation, we propose the following \textbf{design requirements} for the system:
\begin{itemize}
    \item \textbf{[D1] Facilitating rapid iteration and exploration of ideation space}:
    In addition to our formative study findings, prior studies have also identified the need for AI-assisted generation of research ideas and hypotheses~\cite{spangler2014automated,baek2024researchagent}.
    To address researchers' challenges in framing and identifying research questions in an unfamiliar domain, the system should enable users to help the user quickly iterate on their ideas and explore a broad ideation space by providing tools that allow for easy modification and visualization. 
    \item \textbf{[D2] Gathering feedback from the perspectives of domain experts}: 
    As highlighted by the formative study interviewees, in real-life research settings, getting feedback and sharing knowledge from domain experts is crucial for advancing novel ideas, but the process is often hindered by the lack of access and opportunities for communication to obtain timely and constructive input. Past theories~\cite{mayo2017metatheoretical} have also suggested the significance of having access to the knowledge of a diverse set of expertise in the context of collaborative ideation. 
    The system should have mechanisms for actively seeking and incorporating feedback from domain experts with diverse background knowledge in a timely manner. 
    \item \textbf{[D3] Supporting literature discovery in unfamiliar domains}: 
    A wide range of past research has highlighted the importance of literature discovery \cite{foster2004nonlinear} in scientific ideation. In addition to our findings from the formative study, empirical studies have also revealed researchers' need for assistance in navigating vast literature spaces, especially in unfamiliar domains~\cite{newby2011entering, palmer2009scholarly}. Given the need for comprehensive and efficient literature discovery, the system should provide support for literature discovery and review in unfamiliar domains of the user. 
\end{itemize}

\subsection{System Interaction Design}
\disrevisedcamclean{
Informed by the findings from our formative studies, we designed multiple features of the \systemName\ system to address each design requirement.  \systemName\ aims to support exploration of interdisciplinary research topics and ideas by involving diverse perspectives through LLM-simulated domain experts. The system is designed to benefit research exploration of interdisciplinary researchers, especially those planning to explore unfamiliar domains. 
}
The overall system workflow follows an interactive ideation process, as described in Figure \ref{fig:overview-diagram}.
\systemName\ offers five major features: 
\begin{itemize}
    \item \textit{Mind-Map Flow Editor}: interactive canvas where users can edit and create their ideation flow and interact with each node. 
    \item \textit{Persona Node}: node representing the profile of each simulated expert, with a customizer interface allowing users to edit the \disrevisedcamclean[traits]{characteristics} of each expert initially populated by the system.
    \item \textit{Literature Node}: node containing information about related scholarly publications, retrieved based on the expertise of a given persona. 
    \item \textit{Critique Node}: node displaying feedback and critiques given by an expert persona based on a set of relevant literature. 
    \item \textit{RQ Node and Research Outline Panel}: node facilitating the review and refinement of ideation outcomes in the form of research questions (RQs), with a structured panel for outlining the research gap behind the proposed research directions and a hypothetical plan for addressing the research question. 
\end{itemize}
We provide more detailed descriptions for each feature in \Cref{sec:system-features}.
\disrevisedcamclean{
The system does not require prior research expertise, but users need an initial idea to begin exploration. The system is primarily designed to support exploration from a rough idea and provide diverse perspectives, ultimately helping the user narrow down to a more specific concept.
}



\begin{figure*}[!ht]
    \centering
    \includegraphics[width=0.85\linewidth]{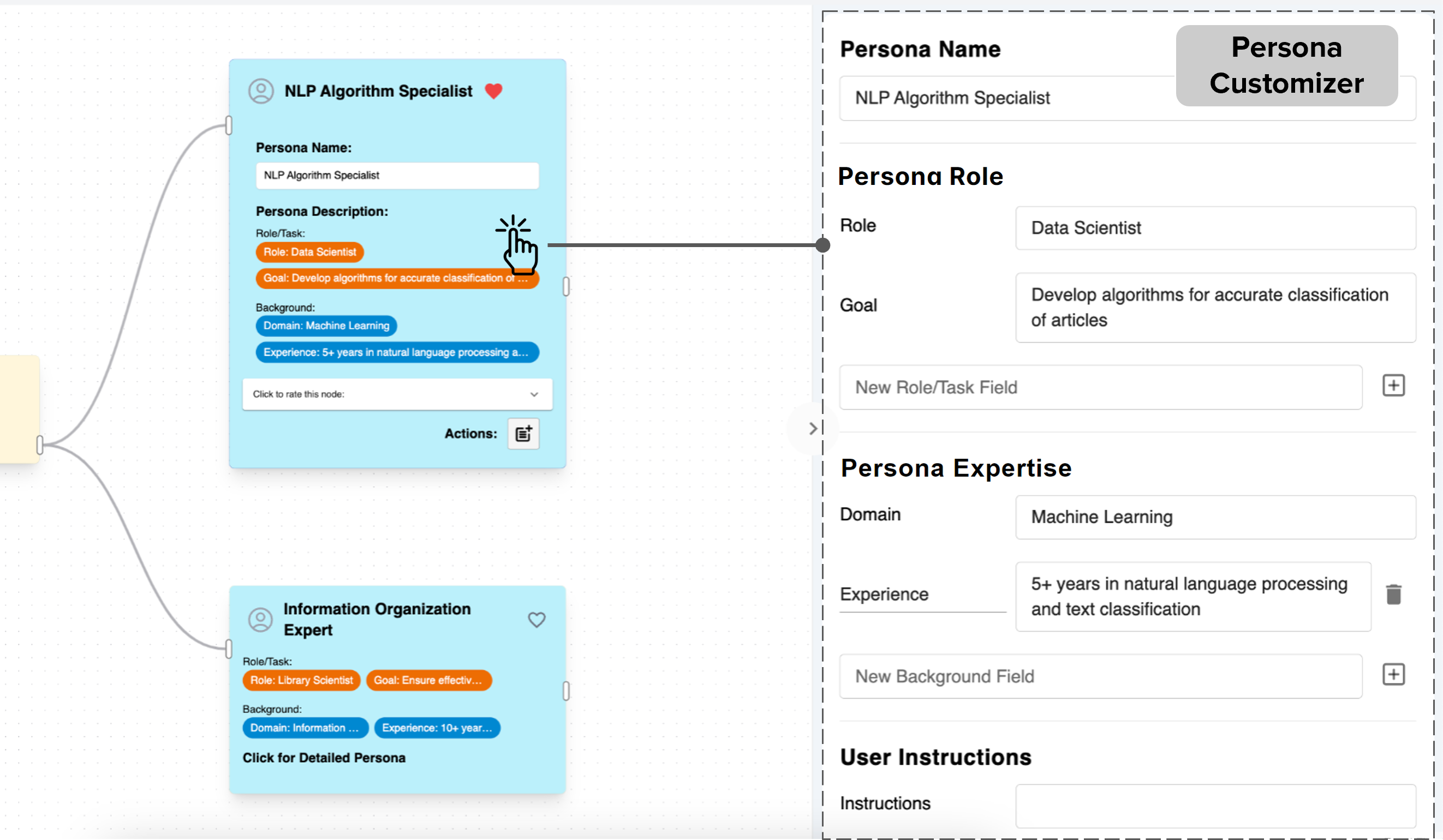}
    \caption{
    \textit{Persona Node and Customizer}: The \textit{Persona Node} interface provides an overview of each expert persona's profile information including name and \disrevisedcamclean[traits]{characteristics}. Upon clicking the node, a \textit{Persona Customizer} panel shows up on the right side of the screen displaying the detailed \disrevisedcamclean[traits]{characteristics} and allowing the user to customize them.
    }
    \Description{Three components: Persona Node ‘NLP Algorithm Specialist’, Persona Node ‘Information Organization Expert’ and Persona Customizer side panel for ‘NLP Algorithm Specialist’. Upon clicking Persona Node ‘NLP Algorithm Specialist’, the Persona Customizer side panel shows detailed Persona information which contains traits such as ‘Role’, ‘Goal’, ‘Domain’ and ‘Expertise’, allowing the user to customize based on user preference.}
    \label{fig:persona-node}
\end{figure*}

\subsubsection{System Features}
\label{sec:system-features}
The \systemName\ system provides a combination of four major features to address the three design requirements as proposed in \Cref{sec:design-requirements}.

\textbf{Mind-Map Flow Editor [D1]: knowledge organization and idea scaffolding.}
As shown in \Cref{fig:overview-diagram}, the system uses graph-based visualization to support users’ ideation process in the form of a mind map. The interactive graph canvas allows users to navigate through different research questions and personas’ feedback for ideation. 
The effect of mind maps as ideation support tools in collaborative research ideation have been suggested in prior research \cite{sun2022students}, as they facilitate easier explication of ideas and assist users in organizing ideas into structured thinking.
The system uses nodes to represent entities including personas, critiques, collections of literature, and research questions. The node-edge connections are used to reflect the logical connections between the iteration of research questions and ideas based on relevant literature insights and critiques from personas with different expertise.

\textbf{Persona Node [D2]: expert knowledge and specialization in human-AI collaborative ideation.}
Existing research~\cite{ali2019mechanism, dastmalchi2021exploring} about Transactive Memory Systems has suggested the importance of clear awareness of specialization in a team ideation setting. Thus, we introduce the design of \textit{persona nodes} to represent domain-specific expertise, allowing users to interact with LLM-simulated experts with a clear awareness of their specializations and knowledge background. By utilizing expert personas, we aim to stimulate the LLM's knowledge and assist in bridging the research efforts of researchers from different backgrounds.
This design intends to mitigate users' cognitive load, and improve the quality of feedback obtained from the perspective of each dedicated expert. The \textit{persona node} contains information about the detailed profile of each expert persona described in \disrevisedcamclean[traits]{characteristics}, as shown in Figure \ref{fig:persona-node}. 
When a user clicks on the \textit{persona node}, the system displays a \textit{persona customizer panel} on the right side where they can review and edit the details of the \disrevisedcamclean[traits]{characteristics} belonging to each expert. We took inspiration from \citet{benharrak2024writer}'s work when designing the schema of expert traits, allowing users to modify both the role and detailed expertise of each expert. 
We also provide a field for users to specify any instructions they might have for the persona during the ideation process. 
The system synthesizes each persona's initial profile based on the existing contextual information during the ideation. The detailed method we used to generate the persona profiles is presented in section \ref{sec:persona-synthesis}. 
Users can modify persona profiles at any time during exploration by adding, deleting, or editing \disrevisedcamclean[traits]{characteristics} based on their needs and preferences. We do not automatically update or overwrite cascading nodes when changes are made to personas, but users can manually regenerate additional nodes to reflect the updated persona \disrevisedcamclean[traits]{characteristics}.

\begin{figure*}
    \centering
    \includegraphics[width=0.85\linewidth]{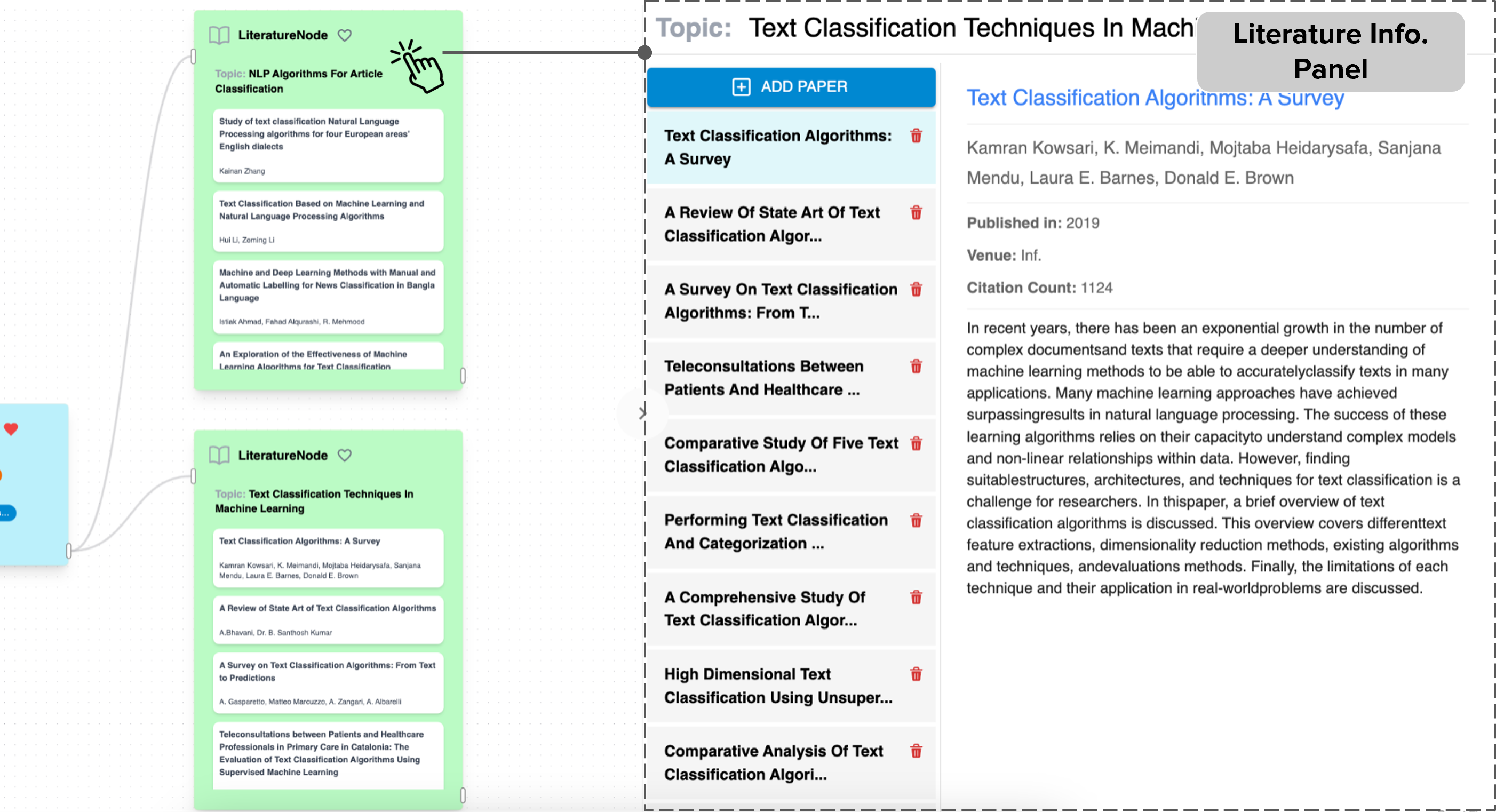}
    \caption{\textit{Literature Node and Panel}: The \textit{Literature Node} shows the preview (i.e., titles and authors) for a collection of literature relevant to a certain topic suggested by the preceding expert persona, based on the previous research question.
    Upon clicking the \textit{Literature Node}, a detailed \textit{Literature Panel} will be displayed containing information about each paper, through which users can also add and remove papers. 
    }
    \Description{Three components: LiteratureNode ‘NLP Algorithms for Article Classification’, LiteratureNode ‘Text Classification Techniques in Machine Learning’, and a Literature Information side panel for the selected paper ‘Text Classification Algorithms: A Survey’. Upon selecting a paper from the LiteratureNode ‘Text Classification Techniques in Machine Learning’, the Literature Information side panel displays detailed information about the paper, including authors, publication year, venue, citation count, and an abstract summarizing the paper's content and key findings.}
    \label{fig:literature-node}
\end{figure*}

\textbf{Literature Node [D3]: rapid literature discovery in various domains.} 
Identifying literature groundings and research gaps is an essential step during the ideation stage of the research lifecycle~\cite{foster2004nonlinear, palmer2009scholarly}. To assist users in efficiently navigating the vast amount of relevant literature, especially in unfamiliar domains, we introduced the design of \textit{Literature Node}, as shown in Figure \ref{fig:literature-node}.
Each of the \textit{Literature Nodes} presents a specific research topic, with a collection of scholarly publications proposed by the preceding expert persona. The collection of papers is retrieved from an online publication database using the pipeline described in \Cref{sec:paper-retrieval}.
When a user clicks on the node, a right-side panel is revealed with detailed information about each publication. The user will be able to delete irrelevant papers or add additional papers using a search engine-like interface.  

\textbf{Critique Node [D2]: feedback from the perspective of experts.}
Based on a \textit{Persona Node} and \textit{Literature Node}, the user will be able to generate a follow-up \textit{Critique Node}, as shown in component 4 of Figure \ref{fig:overview-diagram}. The \textit{Critique Node} contains a user-editable critique of the initial research idea based on the given \textit{Literature Node} and impersonation of the expert persona from the \textit{Persona Node}. 
The critique nodes are further used to illustrate the process of how peer feedback is given from the perspective of a persona other than the researcher using the system. 
We use the design of the three types of nodes to model the process of peer discussions and knowledge solicitation with researchers or experts from different disciplines and domains.

\begin{figure*}
    \centering
    \includegraphics[width=0.85\linewidth]{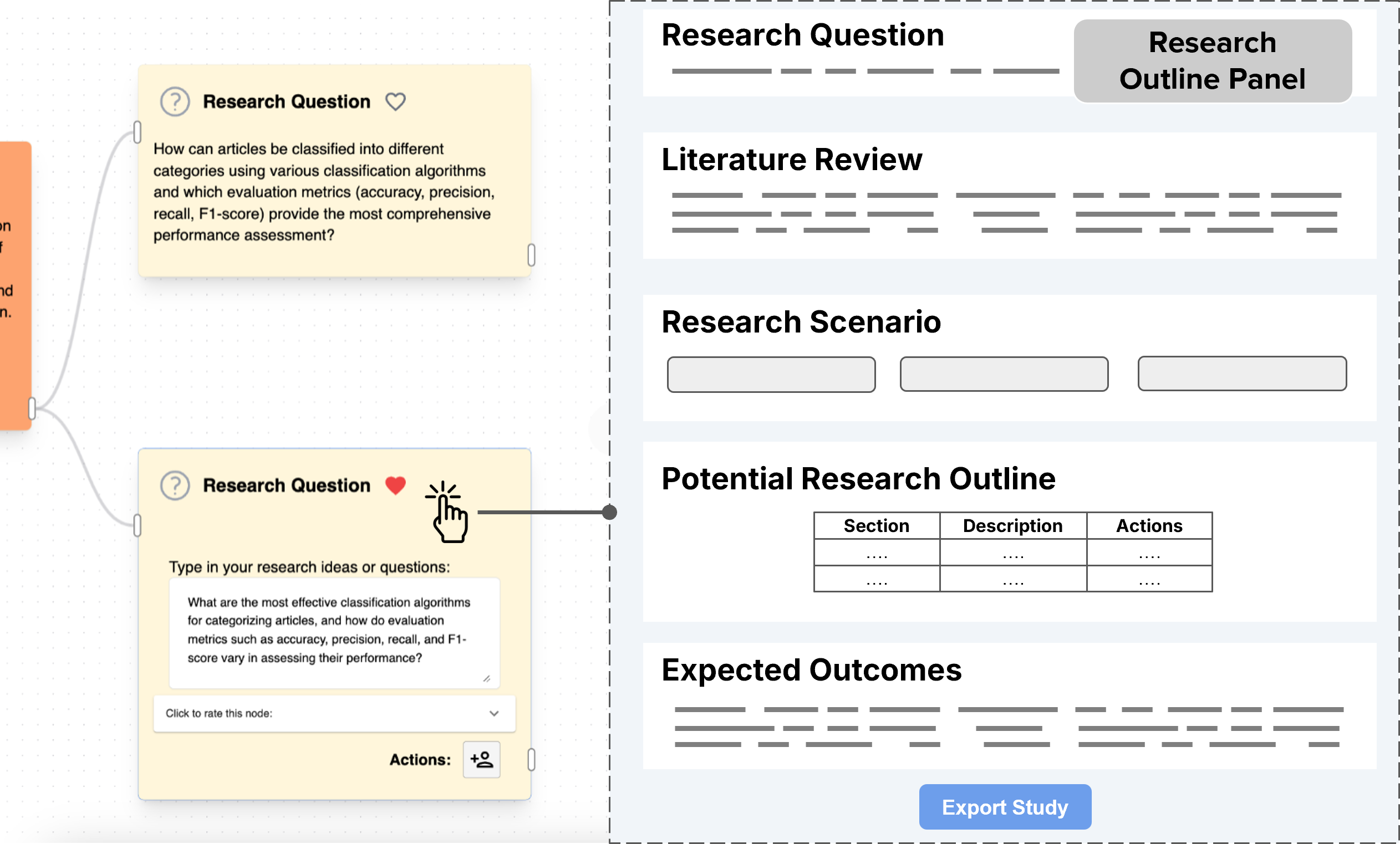}
    \caption{\textit{RQ Node} and \textit{Research Outline Panel}: The \textit{RQ Node} each displays a research question representing the topic of research. The node can be expanded upon clicking, and showcase a detailed research outline given the research question and previously identified literature. The \textit{Research Outline Panel} contains five sections: research question, literature review, research scenario, potential research outline, and expected outcomes. 
    }
    \Description{Three components: Research Question Node A, Research Question Node B and a Research Outline Panel. The Research Question Node A presents a predefined question on ‘How articles can be classified into different categories using various classification algorithms and which evaluation metrics (accuracy, precision, recall, F1-score) provide the most comprehensive performance assessment ?’. The Research Question Node B presents ‘What are the most effective classification algorithms for categorizing articles, and how evaluation metrics such as accuracy, precision, recall and F1-score vary in assessing their performance?’, and allows the user to edit the research question. Research Question Node B has a red heart clicked on top of the node. Upon selecting the Research Question Node B, the Research Outline Panel on the right displays sections including ‘Research Question’, ‘Literature Review’, ‘Research Scenario’, ‘Potential Research Outline’, and ‘Expected Outcomes’, enabling users to structure and export their study.}
    \label{fig:RQ-node}
\end{figure*}

\textbf{RQ Node and Research Outline Panel [D1]: iterative research ideation and refinement through expert feedback.}
Figure \ref{fig:RQ-node} demonstrates the \textit{RQ Node} and \textit{Research Outline Panel} feature.
This feature allows users to iteratively improve their research ideas by generating and refining research questions (RQs) and their corresponding outlines. 
The design decision of making the system RQ-driven is due to the crucial role of RQ in research development and its nature of being iteratively refined during research ideation~\cite{elio2011computing,foster2004nonlinear}.
When a user selects an \textit{RQ Node}, the system automatically generates a \textit{Research Outline panel}, which provides a structured overview of a potential research project derived from the RQ. 
The \textit{Research Outline Panel} contains five sections: research question, literature review, research scenario, potential research outline, and expected outcomes. 
The literature review section provides a summary of relevant literature, prompting the user to think critically about the existing research landscape.
The system then encourages the user to specify a detailed research scenario, with several system-generated suggestions. Based on this input, the system dynamically generates a comprehensive research outline with key sections such as motivation, objectives, methodology, expected outcomes, and potential limitations.
This interactive process allows users to iteratively refine their research outline by reviewing and incorporating system-generated suggestions and insights.

\subsubsection{Example Use Case Scenario --- a System Walkthrough}
\disrevisedcamclean{
Imagine Linh, an HCI researcher in Health Informatics, who became interested in exploring the potential application of LLMs in their familiar domain. Linh began by creating a new \textit{Research Question (RQ) Node} in the mind map canvas to guide the exploration of “fine-tuning Large Language Models to promote preventive health behavior.” After writing down the idea in the initial \textit{RQ node}, Linh clicked the action button on the node to generate follow-up nodes. The system then generated three \textit{Persona Nodes} based on the initial idea, and Linh selected “Behavior Psychology Researcher,” edited its profile, and proceeded to generate more subsequent nodes.
Three \textit{Literature Nodes} were revealed, each containing a preview of suggested papers. Interested in “Promoting Health Behavior Change With AI,” Linh expanded that node, reviewed its papers, and edited the list. Next, Linh generated three \textit{Critique Nodes} providing feedback on the initial idea; one focused on “Theoretical Framework” and referenced psychology concepts like the “Transtheoretical Model of Change (TTM).” By requesting more follow-up nodes, Linh discovered three revised versions of the initial RQ. Selecting one incorporating TTM, Linh opened a \textit{Research Outline Panel} summarizing past work and prompting a detailed research scenario. After specifying “Fine-tuning LLMs to deliver personalized stage-targeted health behavior change messages,” the system generated a research outline table with sections such as motivation, objectives, methodology, and limitations, which Linh further refined to finalize the study plan.
}

\subsection{\systemName\ System Backend and Implementation}

\label{sec:implementation_detail}
In this section, we detail the implementation details of the \systemName\  system backend\footnote{https://github.com/yiren-liu/personaflow/}. 
\disrevisedcamclean{The backend uses GPT-4o for text generation, and is designed to provide on-demand results by generating and synthesizing expert personas, identifying relevant literature based on research questions and personas, providing critiques by impersonating the personas, and generating research outlines.}
The \systemName\ system is developed as a web application in Typescript, utilizing ReactJS and TailwindCSS for the frontend. The interactive flow editor is implemented using React Flow\footnote{https://github.com/wbkd/react-flow/}. For the application backend, Python with FastAPI\footnote{https://github.com/tiangolo/fastapi/} is employed as the RESTful API server framework. We present our detailed prompts used in \Cref{apdx:system-prompts}.

\subsubsection{Synthesizing expert profiles}
\label{sec:persona-synthesis}
The system generates different personas to provide relevant literature and critique, based on the research context. The expert profiles can be generated under 3 different scenarios: 
1) Initial User RQ: When an RQ is initially provided by the user, the system uses the RQ to prompt our LLM to generate three relevant personas that simulate real-world individuals to augment ideation and brainstorming;
2) \textit{Literature Node}-based Persona: When a researcher is recommended a \textit{literature node} relevant to their RQ, they can generate personas using the \textit{literature node} to ensure the generation of relevant personas. The persona is generated by utilizing the abstracts from the previous literature, which are processed using a scientific sentence classification pipeline built with a fine-tuned BERT model~\cite{cohan2019pretrained, beltagy2019scibert}. This pipeline extracts the background, methodology, and conclusion of the research. Finally, it summarizes the literature node to provide context for generating personas;
3) Author Profiling-based Persona Generation: To leverage the expertise of scholarly authors whose papers were included in the \textit{literature node}, we developed an author profiling-based persona generation. Using the Semantic Scholar API~\cite{kinney2023semantic}, we extracted papers from authors whose work was being recommended in the \textit{literature node}. We identified the top three papers for each first author (since first authors are typically the primary contributors and experts in the research area the paper addresses), based on cosine similarity\footnote{with a threshold score of .65} using FAISS~\cite{douze2024faiss}. These papers were summarized and used as context to generate personas aiming to capture the expertise and background of the authors.
    
Researchers can also use the persona node to manually create personas from scratch by providing various characteristics, such as the persona's role, goal, background information, and specific user instructions. These \textit{persona nodes} can then be linked to previously generated \textit{RQ nodes} and \textit{literature nodes} to enhance the overall flow.
It is important to note that we do not aim to evaluate the prompting method in this study. As in prior studies~\cite{choi2024proxona,benharrak2024writer}, our focus is not on directly assessing the accuracy of generated personas. Instead, we concentrate on evaluating the ideation outcomes and how users interact with the simulated expert personas.



\subsubsection{Expert perspective-driven Paper Retrieval}
\label{sec:paper-retrieval}
\begin{figure}
    \centering
    \includegraphics[width=\linewidth]{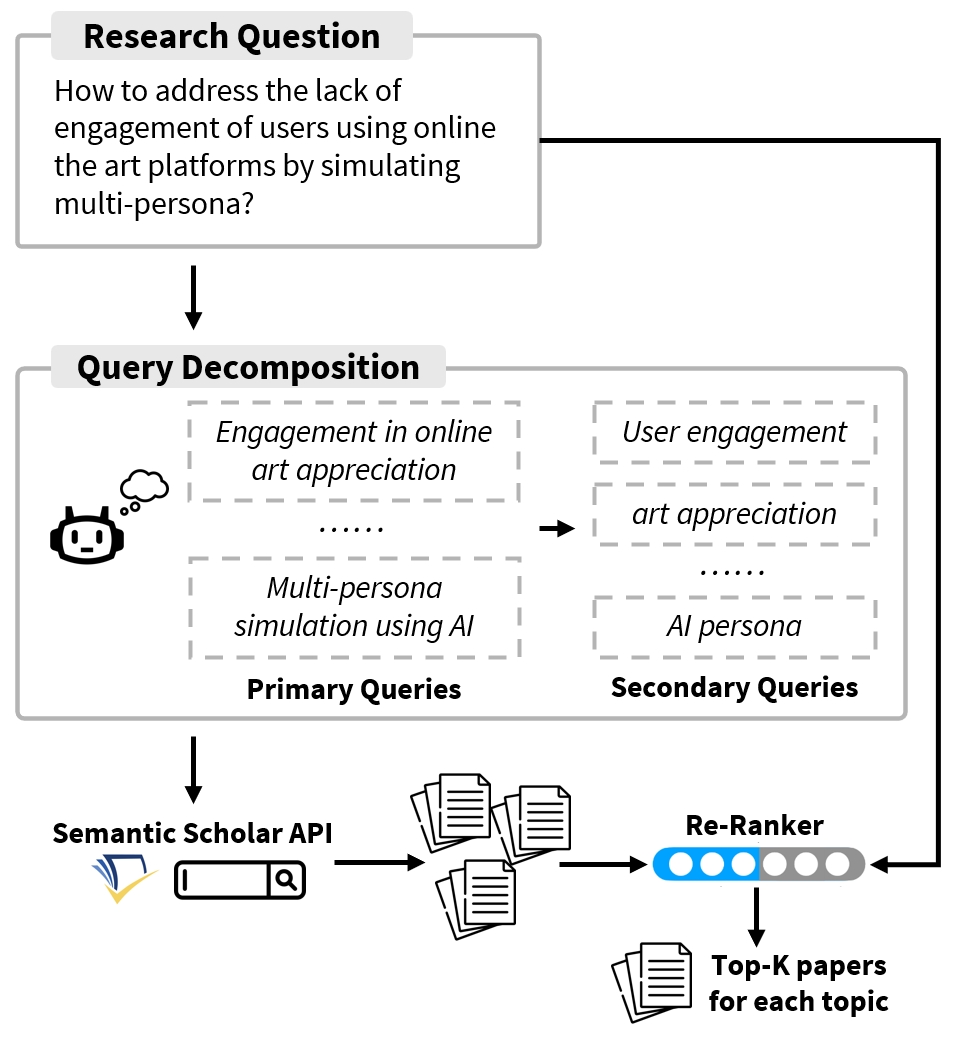}
    \caption{
    \disrevisedcamclean{
    The method used for retrieving papers using agents with expert personas.
    } 
    }
    \Description{Figure illustrates an iterative workflow in which expert‑persona agents help retrieve research papers. The process begins with a box labeled “Research Question,” exemplified by the inquiry: “How to address the lack of engagement of users using online art platforms by simulating multi‑persona?” An arrow guides this question to a “Query Decomposition” panel where a thinking robot icon generates two primary queries—one about engagement in online art appreciation and another about multi‑persona simulation using AI. Each primary query is then dissected into more focused secondary queries, such as “user engagement,” “art appreciation,” and “AI persona.” These refined queries flow downward to the Semantic Scholar API, represented by a search‑bar icon, which returns stacks of papers. The retrieved papers pass into a re‑ranking module—visualized as a bar of colored circles—that filters and selects the top‑K papers for each topic. A looping arrow from this output back to the original research‑question box signals that the entire process can repeat, continually refining results with the aid of expert personas.}
    \label{fig:literature-module}
\end{figure}

We designed and implemented an expert perspective-driven scholarly paper retrieval module to assist users in identifying the relevant papers with higher coverage. 
However, using a research question directly as a search query to retrieve papers from online databases often leads to very few or irrelevant results. 
\disrevisedcamclean{
As shown in Figure \ref{fig:literature-module}, the \systemName\ system conducts two steps when retrieving relevant papers: 1) \textbf{query decomposition} to yield high coverage of papers under different sub-topics; 2) \textbf{semantic-based re-ranking} to reduce the number of papers and only present the topic-relevant ones to the user based on user's prior research question.
}
During the query decomposition step, we prompt the LLM to first break down the user's initial research question into primary queries that capture various aspects of the main topic. Then, each of the primary queries will be decomposed into several secondary sub-queries to optimize the search results using the Semantic Scholar search API~\cite{kinney2023semantic}. A similar approach has also been found to be used in existing work for optimizing literature discovery results~\cite{zheng2024disciplink}.
For instance, the research question ``How to address the lack of engagement of users using online art platforms by simulating different AI personas?'' can be broken down into primary queries including ``engagement in online art appreciation'' and ``multi-persona simulation using AI.'' Subsequently, each primary query will be further decomposed into secondary queries that are shorter keywords. For example, the primary query ``engagement in online art appreciation'' will be decomposed into ``user engagement,'' ``art appreciation'' and ``AI persona.''
After the search results for each primary and secondary query are gathered, they are aggregated and re-ranked using a dense retrieval model~\cite{chen2024bge} that aligns the retrieved papers with the user's original research question. We take the top-k most relevant papers based on semantic similarity and present them to the user. 
During the later user studies, we present the top 10 most relevant papers to the user.



\section{User Studies Evaluating Co-Creation with \textit{\systemName}}
In this section, we describe the methodology, participant demographics, and key findings from our user studies which aimed at evaluating the co-creation experience facilitated by \systemName. We designed two studies to answer the following research questions about our system:
\begin{enumerate}
    \item[]  \textbf{RQ1:} \textit{How do multiple perspectives generated by \systemName \ affect users' perception of the research ideation process and its outcomes?}
    
    \item[]  \textbf{RQ2:} \textit{How does the design of \systemName \ support users' critical thinking activities in research ideation?}

 \item[]  \textbf{RQ3:} \textit{How does the use of different \systemName \ features impact users' perceived experience and outcomes?}
    
    
\end{enumerate}

\subsection{User Study Design}
We conducted a two-phase exploratory user study to evaluate the \systemName\ system.
All studies were completed remotely online over Zoom, where participants were asked to share their screens. Study procedures were approved by the IRB of the researchers’ institution. Participants were compensated at \$20 per hour for all user studies.

\disrevised{
\textbf{Phase 1: Task-based exploration with single and multiple expert perspectives.}
We designed a task-based exploratory user study to understand the impact of LLM-simulated experts on users' research ideation process and how they interact with these personas. Simulating the scenario where the user seeks help from domain experts when exploring a research topic, we asked participants to develop their own topics of interest and utilize our systems for ideation support. We then analyze how the ideation outcome and user perception change temporally throughout the exploration session as participants interact with different expert perspectives.
In order to examine our system's utility in supporting interdisciplinary research ideation, we recruited 21 participants from 8 different research institutions and different backgrounds (e.g., HCI, Bio-informatics, Machine Learning, etc.) with prior experience in conducting interdisciplinary research. 
Detailed information about participants' backgrounds can be found in \Cref{apdx:demographics}.
}

\disrevised{
Specifically, we focus on examining users' interactions with multiple perspectives provided by different AI experts temporally throughout the ideation process. 
At the start of each session, the user was first asked to interact with a single AI expert and complete a single iteration of research ideation. 
We instructed the user to pick one AI persona out of the first batch of LLM synthesized personas, gather literature, and receive critiques until the first set of revised research questions is reached. The user was then provided with a post-survey form for quantitative evaluation of their experience and perceptions pertaining to the first iteration.
Subsequently, the user was asked to interact with additional experts by building on the outcomes from the previous iteration and continue exploring by generating and involving multiple AI personas until they reached the second set of research questions. A post-survey was also provided inquiring about their experience and perceptions of the second iteration.
By the end of the study, we conducted an exit interview to gather qualitative feedback on their experience and rationalization of their interactions when using the system.
We chose not to run a controlled experiment with a single agent (e.g., ChatGPT) setup as baseline for two reasons: 1) Participants suggested during the formative study (\Cref{sec:formative-study}), off-the-shelf LLM chatbots like ChatGPT tend to generate generic responses and are hard to prompt; 2) during the pilot of the study design, we found that each session would take more than 1.5 hour to complete, and including an additional baseline would drastically increase the cognitive burden and cause potential fatigue for the recruited participants. More importantly, our study aims mainly to understand users' engagement and behavior changes during the ideation process, thus we decided to adopt a single session exploratory study which focuses on gather more in-depth insights into users' interaction with multiple expert perspective.
}

\disrevised{
\textbf{Phase 2: Optional open-ended free exploration.}
At the end of the exploratory study, we invited participants to an optional 30-minute free exploration session where they could freely use the system's features with minimal guidance and no procedural constraints. 
Participants were free to build on their existing flow from the previous session or create new flows using another topic. In total, 10 participants enrolled in the open-ended exploratory study.
This study session aims to provide a deeper understanding of how users would use the system in a real-world exploration setting, as well as the benefits and shortcomings of the \systemName\ system. 
}

\subsection{Data Collection and Analysis}

\subsubsection{On-the-fly rating collection during system use.} 
\label{sec:node-ratings}
The ratings of personas, critiques, and research questions were gathered through a mini-survey embedded in each node on the fly during the exploration task, as displayed in \Cref{fig:node-ratings}.
For persona, critique, and research question nodes, participants were requested to rate at least one node of their choice whenever a set of new nodes has been generated. 
The survey questionnaire for the persona nodes was formulated using insights from previous research~\cite{salminen2020effect, salminen2018persona}. Through the collection of these ratings, our goal is to capture users' immediate perceptions regarding the feedback and outputs produced by the system.

\subsubsection{Understanding user perception: post-session survey for each condition and exit interview} 
To assess participants' perceived experience with our system, we gathered their survey ratings after each task and held exit interview views prior to concluding each study. 


\textbf{Post-Session Survey}  
Participants provided their overall evaluation of the co-creation experience and outcomes through a survey. 
The survey includes a set of 5-point Likert scale questions intended to assess participants' experience in terms of ideation outcome quality, ideation experience, and system usability. 
The complete list of survey questions can be found in Appendix \ref{apdx:survey_ideation_experience}.

\textbf{Exit Interview}
After the participant completed the post-task rating survey, we also conducted a ten-minute semi-structured interview to obtain a deeper understanding of the participant's experience with the system. 
These interviews were audio-recorded and transcribed verbatim, given participants' consent, for qualitative analysis to help us capture insights into how participants interacted with the system and perceived its impact on their creative processes.

\subsubsection{Behavior analysis: think-aloud data and system log.}
To gain insights into user interactions with our system, we analyzed users' behavior through the use of think-aloud transcripts and system log analysis.
For users' perception of the system, the primary source of data was the think-aloud method. Two researchers analyzed the think-aloud data using thematic analysis~\cite{braun2012thematic} to identify patterns within the data.
We also gathered system logs to interpret user interactions while using the system, including users' edits over each node, details of all nodes generated, and snapshots of mind maps created by users. Both quantitative and qualitative analyses were performed using the collected data.

\label{sec:critical-thinking-def}
\disrevised{
We also further analyzed their behavior and think-aloud data and presented case studies demonstrating how they interacted with the system and how the system promoted their critical thinking activities.
We analyzed users' cognitive activities through the lens of critical thinking~\cite{gopinathan2017accumulation,facione1990critical,facione2011critical}, as critical thinking is crucial to the process of interdisciplinary research ideation~\cite{newton2010creativity}. 
We refer to the following critical thinking framework~\cite{facione2011critical} during analysis:
\begin{enumerate}
\item \textbf{Interpretation}: To comprehend and express the meaning or significance of a wide variety of experiences, situations, data, events, judgments, conventions, beliefs, rules, procedures, or criteria.
\item \textbf{Analysis}: To identify the intended and actual inferential relationships between different arguments and the implicit assumptions in the reasoning.
\item \textbf{Evaluation}: To assess the credibility of statements or other representations that are accounts or descriptions, and to assess the logical strength of the actual or intended inferential relationships among statements, descriptions, questions, or other forms of representation.
\item \textbf{Inference}: To identify and secure elements needed to draw reasonable conclusions, to form conjectures and hypotheses, to consider relevant information, and to reduce the consequences flowing from evidence.
\item \textbf{Explanation}: To state and justify one’s opinions and arguments based on reasons and awareness of possible counterarguments, and to present one’s reasoning in the form of cogent arguments.
\item \textbf{Self-regulation}: To self-consciously monitor one's cognitive activities, the elements used in those activities, and the results educed, particularly by applying skills in analysis, and evaluation to one’s own inferential judgments with a view toward questioning, confirming, validating, or correcting either one’s reasoning or one’s results.
\end{enumerate}  
}

\section{Findings}

\subsection{Users’ Perceptions of Experiences and Outcomes in Research Ideation (RQ1)}
We evaluated users' perception of the ideation process from two aspects: generated feedback (critiques) and users' overall perception of their ideation experience. 
\label{sec:RQ1-findings}

\subsubsection{Multi-perspectives improved users' perceived \textit{Relevance} and \textit{Helpfulness} of the generated critiques.}
\label{sec:RQ1-node-ratings}
\begin{figure*}[!h]
    \centering
    \includegraphics[width=1.\linewidth]{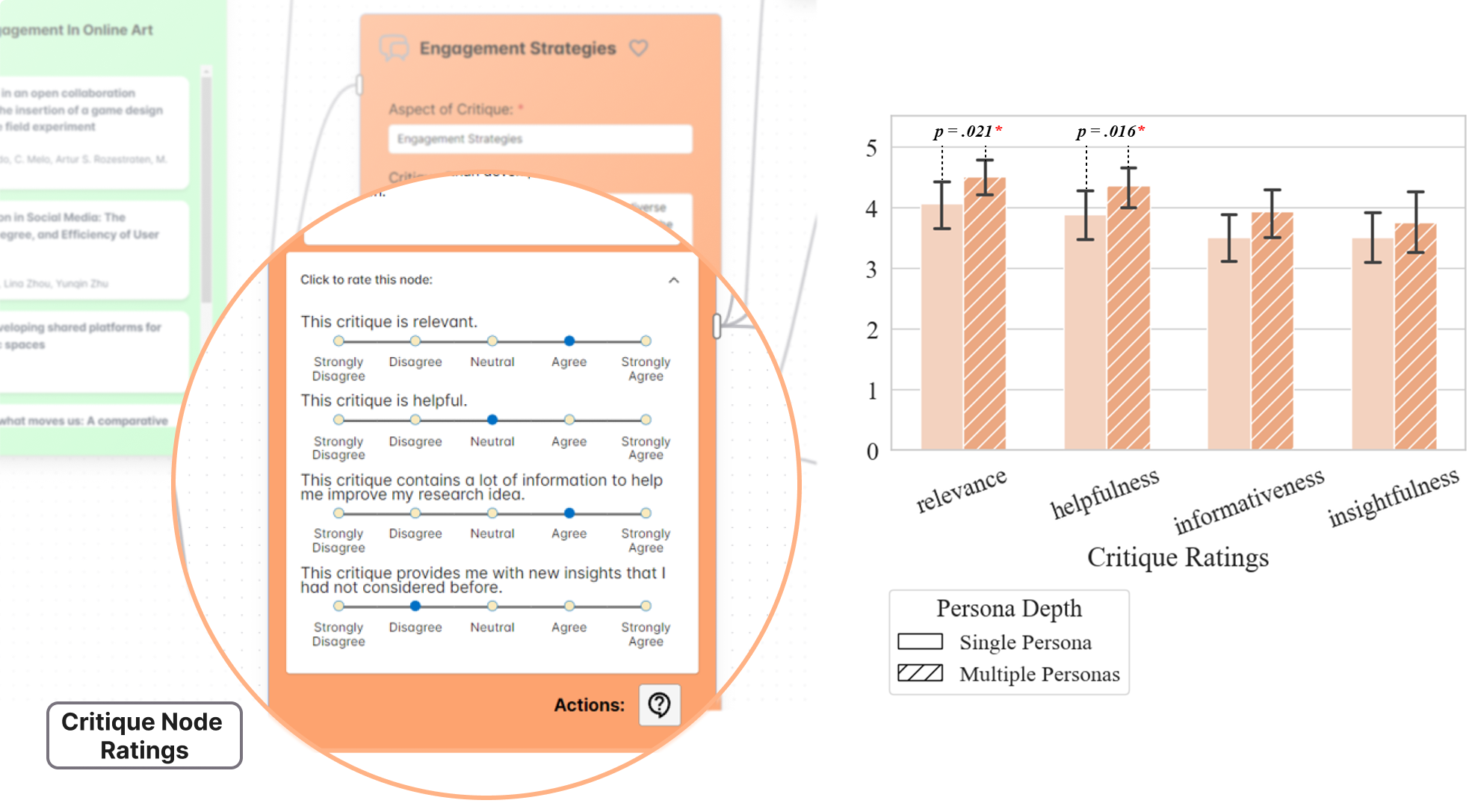}
    \caption{ 
    The left figure is an example screenshot showcasing the embedded rating surveys (the critique node) used to collect users' perceptions on the fly.
    Users were instructed to rate at least one node out of any newly generated batch of critique or RQ nodes. 
    The right plot presents the average of users' 5-point Likert scale ratings towards the generated critiques and RQs: a comparison between different dimensions of ideation outcomes between the earlier stage, where the user interacted with one single expert, and the latter stage, where multiple expert perspectives were introduced. The ratings are collected on the fly through rating forms embedded in each node during the use of the system, as detailed in section \ref{sec:node-ratings}.
    }
    \Description{Two components: Critique Node Ratings (left) and Research Question (RQ) Node Ratings (right). Each node contains sliders to rate specific statements on a scale from ‘Strongly Disagree’ to ‘Strongly Agree’. The Critique Node allows users to rate aspects such as relevance, helpfulness, information richness, and whether the critique provides new insights. The RQ Node is used to evaluate research questions based on relevance to the research topic, provision of new insights, feasibility for further research, and clarity. These ratings and sliders are magnified. charts of users' 5-point Likert scale ratings for Critique Ratings and RQ Ratings across two conditions: Single Persona and Multiple Personas. In the Critique Ratings chart, four categories are displayed: relevance, helpfulness, informativeness, and insightfulness. For relevance, the bar for the Single Persona condition is slightly below 4, while the Multiple Personas bar is closer to 4.5, with a p-value of .021 noted above. In the helpfulness category, the Single Persona condition is rated just above 3.5, while Multiple Personas reaches around 4, with a p-value of .016 indicated above the bars. Informativeness and insightfulness show similar ratings for both conditions, around 3.5, with the Multiple Personas bars being slightly higher. In the RQ Ratings chart, the four categories include relevance, creativity, feasibility, and specificity. Both conditions show nearly identical ratings for relevance, with both bars around 4.5. In creativity, feasibility, and specificity, ratings are similar between Single Persona and Multiple Personas, with both conditions consistently around 4.}
    \label{fig:node-ratings}
\end{figure*}

During participants' use of the system, we provide an embedded rating survey interface on each generated critique and RQ node to collect users' immediate evaluation of the system's outputs. 
Based on users' ratings given when using the system, we found that the involvement of multiple personas improves the ratings towards both system-generated critiques and RQs in multiple dimensions, as shown in Figure \ref{fig:node-ratings}.

There is a significant difference in the critique relevance ($t = 2.42, p = 0.021^{*}$) and helpfulness ($t = 2.54, p = 0.016^{*}$), indicating that the critiques provided by multiple expert personas enhance users' perceived relevance and helpfulness by introducing more details and perspectives. 
P17 noted that \userquote{... the system-generated critiques are good to be used for gathering new ideas and thoughts ... .} Others also mentioned that critiques generated by personas with unfamiliar domain backgrounds contained information that the user has not considered previously. 

\subsubsection{Positive correlation between the diversity of perspectives and the perceived creativity and specificity of the generated RQs.}
We also found a positive correlation between the number of expert perspectives the user engaged with during ideation and the perceived creativity ($r=0.37, p=0.041^*$) and specificity ($r=0.29, p=0.014^*$) of research questions, indicating a positive impact of LLM-simulated personas on users' perception of ideation outcomes.
Qualitative findings revealed that expert perspectives introduced new factors or concepts that users had not previously considered. One participant noted, \userquote{... Personas added important factors such as behavior psychology, into the final research questions ...} (P1). 
In other examples, participants emphasized that the personas' critiques introduced helpful new concepts for understanding a new domain, such as \userquote{generative item retrieval}, \userquote{ranking}, and \userquote{user preference modeling} (P3), or \userquote{the impact of cognitive load} during the exploration of the topic ``LLM assistant on AR smart glasses'' (P5).
Additionally, personas influenced the direction of exploration regarding topics. One participant stated, \userquote{... help drive more directions like 'AI ethics,' the exploration direction is changed and shifted ...}(P9). 
This information was drawn from the expertise of the expert, the views and perspectives of the community, and the background of the expert which is helpful in generating this type of literature.

\begin{figure*}[!h]
    \centering
    \includegraphics[width=1.\linewidth]{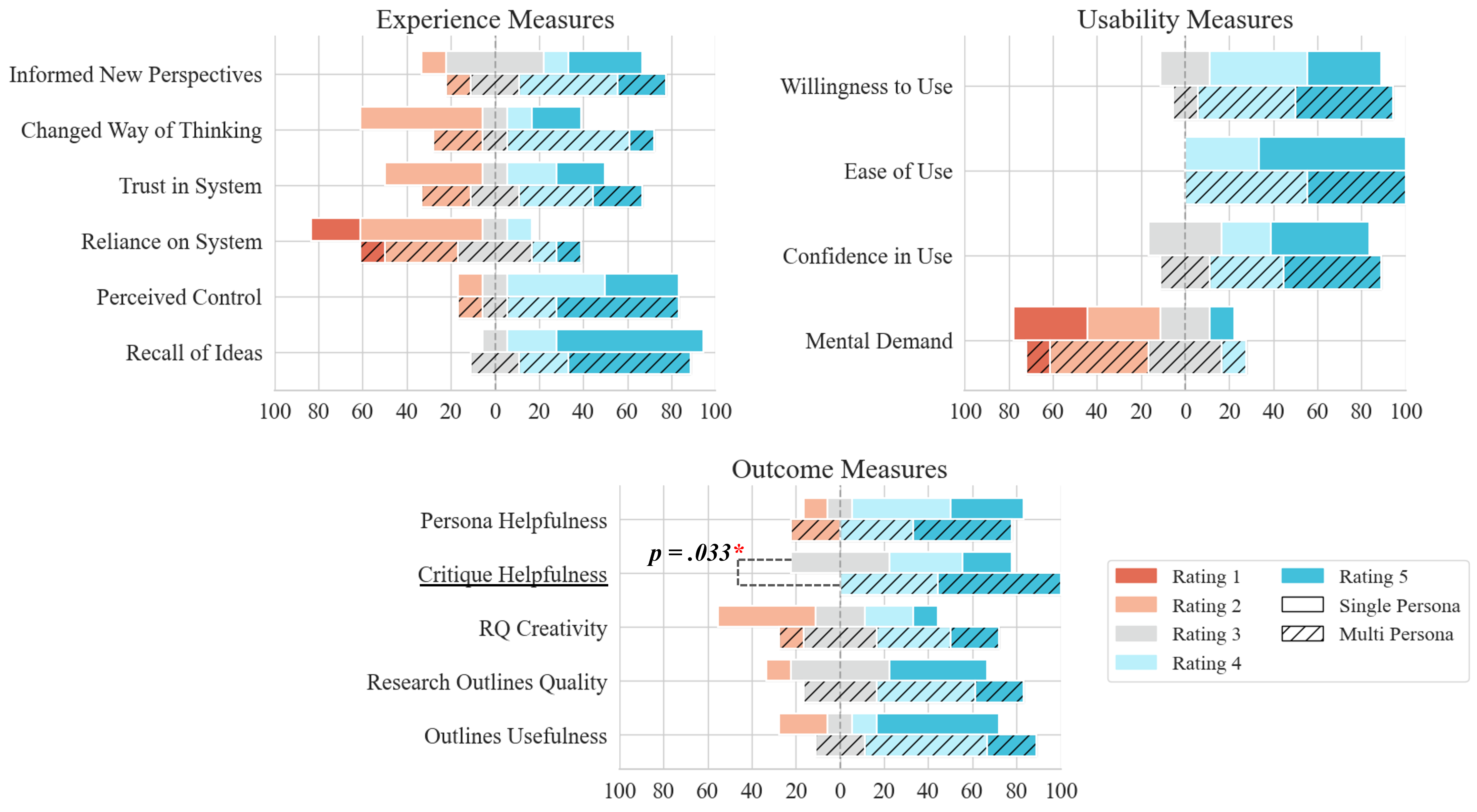}
    \caption{
    \disrevisedcamclean{
    Post-session survey results comparisons between single and multiple persona conditions. The x-axis denotes the percentage of Likert scale ratings given by participants by value. 
    The post-survey questions can be found in \cref{apdx:survey_ideation_experience}. Ratings 1-5 indicate strong disagreement to strong agreement, respectively.
    A t-test showed that users' overall perceptions of critique helpfulness significantly improved under the multi-persona condition when compared to single-persona ($t = 2.36, p = 0.033^{*}$). 
    }
    }
    \Description{Stacked bar charts showing post-session Likert scale ratings for Experience Measures, Usability Measures, and Outcome Measures across two conditions: Single Persona and Multiple Personas. The X-axis represents the percentage of ratings, while each bar is divided into five rating categories (Rating 1 through Rating 5). In the Experience Measures chart, categories include Informed New Perspectives, Changed Way of Thinking, Trust in System, Reliance on System, Perceived Control, and Recall of Ideas. The Multi Persona condition shows higher percentages of Rating 5 in categories like Informed New Perspectives and Recall of Ideas, while the Single Persona condition exhibits a wider spread across lower ratings. The Usability Measures chart includes Willingness to Use, Ease of Use, Confidence in Use, and Mental Demand. Multi Persona generally shows higher percentages of Rating 5 in Willingness to Use and Ease of Use, while Mental Demand has more weight in lower ratings (Rating 1 and Rating 2) for both conditions, indicating lower mental demand. The Outcome Measures chart shows Persona Helpfulness, Critique Helpfulness, RQ Creativity, Research Outlines Quality, and Outlines Usefulness. Critique Helpfulness shows a significant difference, with Multi Persona having a higher percentage of Rating 5, indicated by a p-value of .033, while the Single Persona condition has a wider distribution across lower ratings. Other categories show more balanced distributions between conditions.}
    \label{fig:survey-rating-stacked}
\end{figure*}

\subsubsection{Engaging with multiple expert perspectives did not increase users' perceived cognitive load.}
\label{sec:RQ1-overall}
We collected users' post-session survey ratings to understand their perception of the overall ideation process. 
We found a significant difference ($t = 2.36, p = 0.033^{*}$) between user-perceived critique helpfulness before ($M=3.78, SD=0.83$) and after ($M=4.56, SD=0.53$) engaging with multiple expert perspectives, as shown in Figure \ref{fig:survey-rating-stacked}.
Usability-related items were identified to be of no significant difference before and after users engaged with multiple perspectives, as using the system after multiple perspectives were introduced did not incur significant changes in users' perception of system functionality and usage experience. 
However, we also observed a higher level of user-perceived reliance on the system when multiple expert perspectives were involved during ideation, although not significant.

We analyzed users' behavioral and think-aloud data to understand why the use of multiple expert perspectives was perceived as beneficial by participants.
We observed that users (N=3) who were highly familiar with the chosen topic reported that the system helped them articulate their ideas, which might have previously been implicit.
P10 identified a research question that they have been considering and discussing with other researchers, but the system provided helpful information for them to elicit the method to be used and additional factors to consider during evaluation. The participant noted that \userquote{... it definitely gave new information (which) really helped me solidify which approach to use ... ,} leading to a clearer research idea and a more structured plan for their study. 
We also observed that another user (P17) manually added a new expert, ``Basic Science Educator Mentor,'' after encountering the ``Clinical Training Mentor'' persona. This decision was driven by their understanding of the medical training process, which involves two distinct stages: foundational science comprehension and practical skill application, as stated by the participant. P17 expressed the need for insights from both the ``Clinical Skill Expert'' and the ``Assessment Specialist'' to gain a holistic perspective that aligns with their exploration scope. This approach allowed the participant to consider different viewpoints, especially recognizing that users' questions often emerge from their specific domain expertise. 
Moreover, another participant (P8) emphasized the importance of incorporating multiple expert perspectives in their study, stating that \userquote{... (the expertise of) the other two personas are needed for the study to proceed ... .} This reflects how the user rationalizes the selection of personas based on the need for a combination of different skill sets in terms of the feasibility of research project.
Participants also found the breakdown of expertise into subsequent expert perspectives particularly helpful when conducting divergent exploration. P19 mentioned that \userquote{It is too much information if all expertise is within one persona ... Breakdown of personas is more helpful for dig deep, and choose the direction to go forward ...}


\subsection{ Case Studies: \systemName \ Design Supporting Diverse Critical Thinking Activities (RQ2)}
\label{sec:RQ2-findings}
\disrevised{
Based on the findings of RQ1, we identify potential shifts in users' ways of thinking before and after their engagement with multiple expert perspectives. In order to further understand how users utilized the systems and the implications of providing diverse perspectives, we first analyzed their critical thinking activities reflected through their think-aloud data during the exploration. Additionally, we examined the association between user behaviors during system use and their overall perception of the ideation process by performing a linear regression analysis using the system logs and correlating the quantitative findings with users' think-aloud data. 
}

\label{sec:RQ3-findings}
\disrevised{
After the task-based study session, we invited participants to an optional 30-minute free exploration session where they freely used the system's features without guidance or constraints. Researchers observed their usage of different nodes. This section analyzes their behavior and think-aloud data (N=10) by combining data collected from the exploratory study with the open-ended exploration session, and presents cases demonstrating how they collaborated with AI to co-create research ideas and questions.
We analyzed how the system promoted users' cognitive thinking process, more specifically critical thinking skills, as defined in \cref{sec:critical-thinking-def}. 
We found that most users engaged in critical thinking activities including interpretation, analysis and evaluation. In this section, we highlight case studies demonstrating how participants moved beyond these critical thinking processes and utilized the system to support unique critical thinking activities. 
}
\label{sec:RQ3-diverse-use-case}

\begin{figure*}[ht!]
    \centering
    \includegraphics[width=1\linewidth]{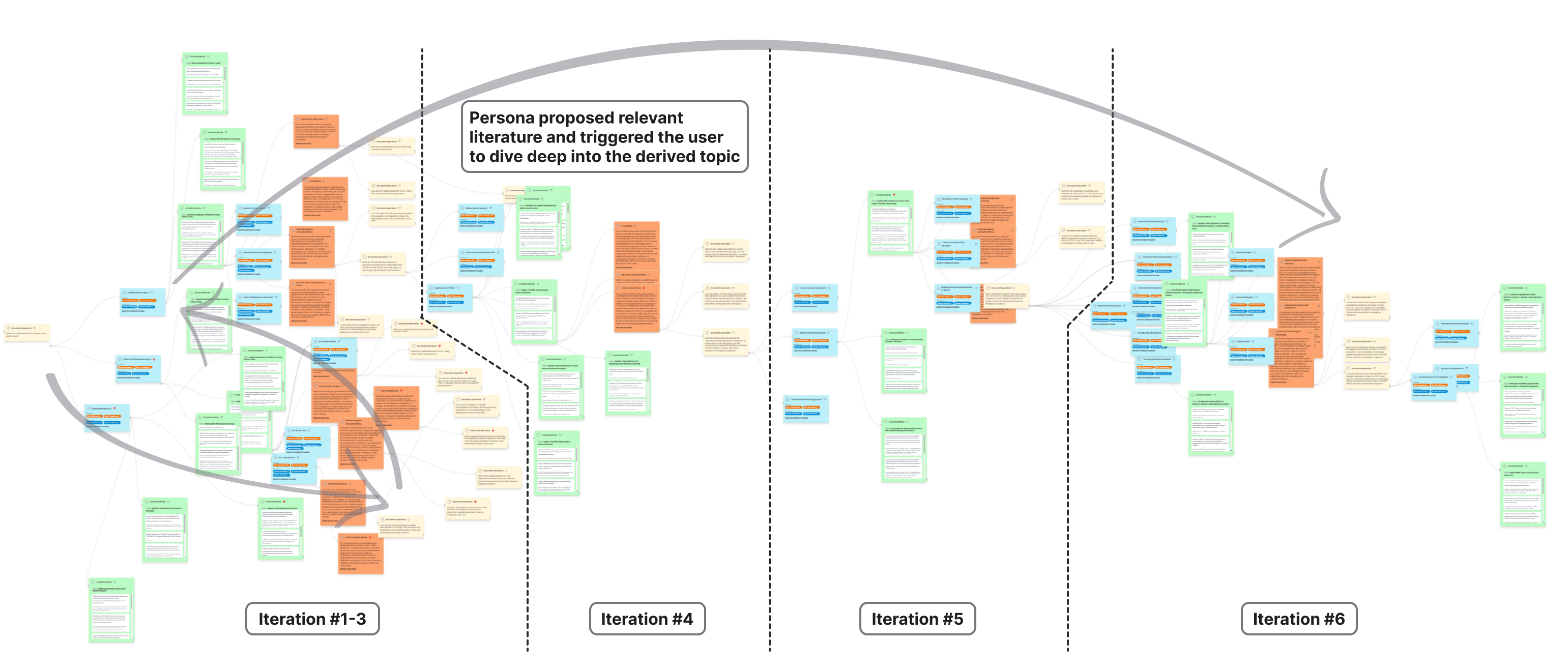}
    \caption{
    P20's thought processes: The user started by exploring different personas and RQ nodes broadly, and then decided to dive deep into interacting with one ``remote'' expert profile (``Spatial Gene Profiler'') with the topic ``single-cell RNA sequencing and spatial transcriptomics.'' Though the expert was not immediately related to the P20's research, the recommended literature was surprisingly relevant to P20's work and exactly what they read recently. This ``synergy'' quickly elicited P20's interest in a deeper exploration of this expert. \disrevisedcamclean{A larger version of the workflow screenshot can be found in \cref{appdx:p30-flow-snapshot-large}.}
    }
    \Description{
    The diagram is divided into four panels, each representing different stages of P30's interaction with persona and RQ nodes over multiple iterations. The panels are labeled as Iteration #1-3, Iteration #4, Iteration #5, and Iteration #6, showing a progressive flow of exploration across different topics and personas. In the first panel (Iteration #1-3), the user explores various persona and RQ nodes broadly without focusing on a specific topic. By Iteration #4, the user engages with the more focused ‘Spatial Gene Profiler’ persona, initially unrelated to P30's work, but offering relevant literature aligned with their interests. In Iteration #5 and #6, the user delves deeper into this topic, finding the suggested research highly relevant for further interaction. The flow of nodes, marked by curved arrows across the four panels, demonstrate the deepening focus across iterations.
    }
    \label{fig:p30-flow-snapshot}
\end{figure*}

\subsubsection{Typical critical thinking activities (Interpretation, Evaluation, and Inference) applied for ideating RQs}
\disrevised{
\Cref{fig:p30-flow-snapshot} shows an example of participant P20 generated a flow with a depth of six iterations. The participant first examined all expert perspectives generated using P20's initial research question.
At the beginning of using the system, P20 \textbf{\textit{interpreted}} their own context and needs by reviewing all expert perspectives to see which persona node aligned with P20’s domain in bioinformatics and cancer research. Among the initial expert perspectives, “Bio Data Interpreter” quickly drew the participant’s attention because its literature recommendations matched some publications P20 had already read. 
\disrevisedcamclean{
This led P20 to \textbf{\textit{evaluate}} the system’s credibility,
\userquote{Some of the papers are exactly the same paper I just read... that’s pretty interesting),} strengthening the participant’s trust in the tool.
}
In the first three rounds, P20 mostly analyzed expert perspectives that were closely aligned with their existing research focus. However, P20 then became curious about a persona labeled “Spatial Gene Profiler.” Although this expert perspective was slightly outside the participant’s typical scope, P20 recognized unexpected overlaps after examining its suggested articles.
\userquote{Some of the papers I didn’t expect to be relevant actually linked back to my research.}
Based on the new information, P20 \textbf{\textit{inferred}} a possible link to their domain, triggering a deeper dive into “single-cell RNA sequencing and spatial transcriptomic,” which is a more specific topic than the participant’s original broad question on “spatial distribution of RNA.”
}

\begin{figure*}[ht!]
    \centering
    \includegraphics[width=1\linewidth]{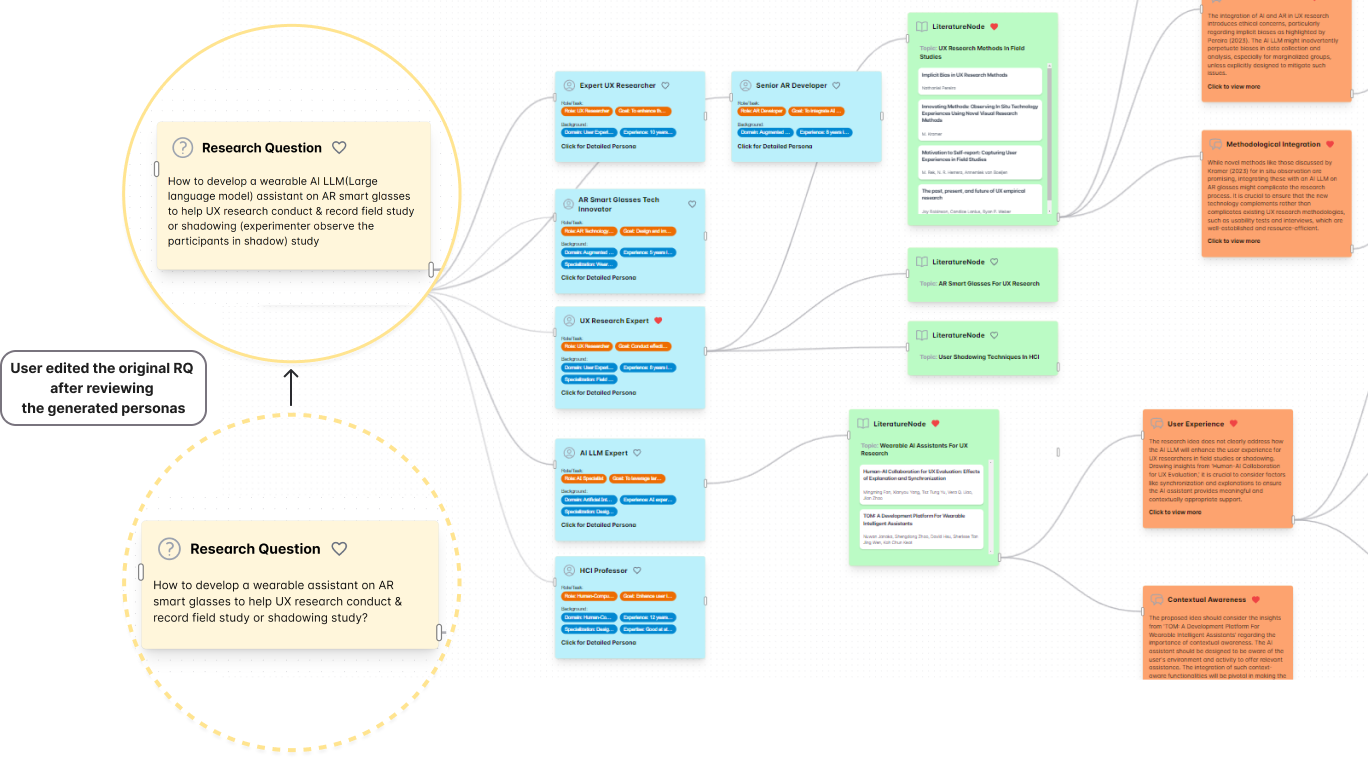}
    \caption{
    P5's thought-process (displaying 15 out of 63 nodes). The participant revised their initial research question after reviewing the generated expert profiles and literature retrieved based on the experts' perspectives.
    } 
    \Description{The diagram demonstrates Participant T5's partial block flow. It includes a Research Question node where the participant initially asked, ‘How to develop a wearable AI LLM (Large Language Model) assistant on AR smart glasses to help UX research conduct and record field study or shadowing?’. This node is circled and magnified, and a note indicates that the user edited this original research question after reviewing the generated personas. The revides research question indicates ‘How to develop a wearable AI LLM(Large language model) assistant on AR smart glasses to help UX research conduct & record field study or shadowing (experimenter observe the participants in shadow) study’. The right panel displays various personas generated from the initial research question. Multiple Literature nodes are displayed which are linked to the previous Persona Nodes. Additional Critique nodes are linked to the previous Literature Node. This flow illustrates how the participant iterated on their research question after exploring the personas and literature nodes, refining the question to focus more on the development of a wearable AI LLM for AR smart glasses in UX research.}
    \label{fig:t5-flow-snapshot}
\end{figure*}

\subsubsection{Self-regulation activities triggered to reflect on the initial RQ}
\disrevised{
Figure \ref{fig:t5-flow-snapshot} depicted an example where the participant went back and re-framed their original RQs after evaluating the expert profiles and critiques, to better align the RQs more closely with their evolving exploration interests. This was a common observation among the participants. 
Participant P5 first generated three expert profiles from their initial RQ node. After reviewing the literature nodes, the user found them-self interested in none of the suggested literature. The participant reflected \userquote{... in our language or in our lab, we typically talk about AI assistant, but here [it] doesn't show any [key]words about AI ...  I feel there's a misalignment between human and the AI agent ... So I need to make my words more specific,} indicating \textbf{\textit{self-regulation}} by re-framing and clarifying their initial idea after reading system-generated output.
As a result, the participant revised the original research question from ``How to develop a wearable assistant on AR smart glasses to help UX research conduct \& record field study or shadowing study?'' to ``How to develop a wearable AI LLM (Large language model) assistant on AR smart glasses to help UX research conduct \& record field study or shadowing (experimenter observe the participants in shadow) study?'' 
The revision led to three additional expert perspectives, two of which the participant found helpful, prompting further exploration through follow-up literature suggestions. Specifically, the participant found the newly generated persona ``UI Research Expert,'' with experience in ``field studies and user shadowing,'' particularly helpful in providing relevant literature. Additionally, the persona ``AI LLM Expert,'' specializing in ``designing user-friendly interfaces for AR devices,'' closely aligned with their exploration interests. We also observed another instance of \textbf{\textit{self-regulation}} activity, where the user commented \userquote{... it’s suggesting what I was thinking. I trust it because it aligns with what I know ...} by reflecting on the critique suggestions provided by one of the experts. 
}

\subsection{Relationships between \systemName \ Interactions and User Perceptions (RQ3)} 
As shown in \cref{sec:RQ3-findings}, we observed participants engaged in critical thinking activities when interacting with expert perspectives suggested by the system. In order to uncover deeper insights into users' usage patterns and perceptions of the system, we further analyzed their behavior data through system logs.

\subsubsection{Feature usage impacted users' perceived experience, outcome and usability.}
\label{sec:RQ2-regression}
In this section, we explored how users' behavior, such as modifications and adaptations made by users to the expert profiles, influence the outcomes of the research questions.
We analyzed the association between users' behavior factors and their perception of the ideation process measured through ratings collected through the post-session survey. 
A set of linear regression analyses was conducted with post-session survey ratings as dependent variables and users' behavioral factors (including the number of edits over RQ/critique/persona nodes and the percentage of nodes used by the user for subsequent generation) as predictors (independent variables). 
We measured the user's percentages of nodes used by computing the percentage of nodes used for generating subsequent nodes out of all nodes generated. The regression analysis results are shown in Table \ref{tab:survey-regression-results}.

\begin{table*}[!h]
    \centering
    \resizebox{\textwidth}{!}{
    \begin{tabularx}{1.20\textwidth}{llRRRRR}
        \toprule
        & & \multicolumn{5}{c}{\textbf{Dependent Variables}} \\
        \cmidrule(lr){3-7}
        \textbf{Theme} & \textbf{Survey Items} & \textbf{\# of edits on RQ nodes} & \textbf{\# of edits on Critique nodes} & \textbf{\# of edits on Persona nodes} & \textbf{\% of Nodes Used} & \textbf{Total Node Count} \\
        & & $\beta$\ (S.E.)\phantom{\textbf{***}} & $\beta$\ (S.E.)\phantom{\textbf{**}} & $\beta$\ (S.E.)\phantom{\textbf{**}} & $\beta$\ (S.E.)\phantom{\textbf{**}} & $\beta$\ (S.E.)\phantom{\textbf{***}} \\
        \midrule
        \multirow{5}{*}{\textbf{Outcome}} 
        & Persona Helpfulness & -0.226 (0.21)\phantom{\textbf{***}} & -0.285 (0.25)\phantom{\textbf{**}} & 0.069 (0.07)\phantom{\textbf{**}} & 3.402 (1.83)\phantom{\textbf{**}} & -0.001 (0.02)\phantom{\textbf{***}} \\
        & Critique Helpfulness & -0.120 (0.11)\phantom{\textbf{***}} & -0.037 (0.04)\phantom{\textbf{**}} & -0.044 (0.05)\phantom{\textbf{**}} & \textbf{2.472 (0.81)**} & \textbf{0.041 (0.01)***} \\
        & RQ Creative & \textbf{-0.376 (0.18)*}\phantom{\textbf{**}} & -0.211 (0.14)\phantom{\textbf{**}} & 0.108 (0.09)\phantom{\textbf{**}} & 2.186 (1.49)\phantom{\textbf{**}} & 0.007 (0.02)\phantom{\textbf{***}} \\
        & Research Outline Quality & -0.272 (0.14)\phantom{\textbf{***}} & -0.075 (0.07)\phantom{\textbf{**}} & 0.125 (0.08)\phantom{\textbf{**}} & \textbf{3.015 (0.99)**} & -0.008 (0.02)\phantom{\textbf{***}} \\
        & Outline Usefulness & -0.129 (0.12)\phantom{\textbf{***}} & -0.016 (0.10)\phantom{\textbf{**}} & 0.040 (0.06)\phantom{\textbf{**}} & 0.680 (1.19)\phantom{\textbf{**}} & -0.009 (0.01)\phantom{\textbf{***}} \\
        \midrule
        \multirow{6}{*}{\textbf{Experience}} 
        & Informed New Perspectives & -0.205 (0.18)\phantom{\textbf{***}} & -0.167 (0.25)\phantom{\textbf{**}} & 0.050 (0.09)\phantom{\textbf{**}} & 2.771 (1.51)\phantom{\textbf{**}} & 0.014 (0.03)\phantom{\textbf{***}} \\
        & Changed Way of Thinking & \textbf{-0.556 (0.16)***} & -0.177 (0.18)\phantom{\textbf{**}} & 0.039 (0.08)\phantom{\textbf{**}} & 2.959 (1.53)\phantom{\textbf{**}} & 0.023 (0.02)\phantom{\textbf{***}} \\
        & Trust in System & -0.179 (0.30)\phantom{\textbf{***}} & -0.250 (0.17)\phantom{\textbf{**}} & 0.092 (0.10)\phantom{\textbf{**}} & 2.090 (2.45)\phantom{\textbf{**}} & 0.000 (0.02)\phantom{\textbf{***}} \\
        & Reliance on System & \textbf{-0.396 (0.13)**}\phantom{\textbf{*}} & -0.125 (0.16)\phantom{\textbf{**}} & -0.045 (0.07)\phantom{\textbf{**}} & 2.026 (1.57)\phantom{\textbf{**}} & -0.000 (0.02)\phantom{\textbf{***}} \\
        & Perceived Control & 0.067 (0.14)\phantom{\textbf{***}} & 0.004 (0.08)\phantom{\textbf{**}} & \textbf{0.149 (0.06)**} & 1.190 (1.38)\phantom{\textbf{**}} & -0.024 (0.02)\phantom{\textbf{***}} \\
        & Recall of Ideas & -0.040 (0.15)\phantom{\textbf{***}} & 0.033 (0.10)\phantom{\textbf{**}} & \textbf{0.185 (0.07)**} & 1.586 (1.62)\phantom{\textbf{**}} & \textbf{-0.034 (0.01)**}\phantom{\textbf{*}} \\
        \midrule
        \multirow{4}{*}{\textbf{Usability}} 
        & Willingness to Use & 0.064 (0.15)\phantom{\textbf{***}} & -0.048 (0.06)\phantom{\textbf{**}} & -0.033 (0.05)\phantom{\textbf{**}} & -0.720 (1.11)\phantom{\textbf{**}} & 0.010 (0.01)\phantom{\textbf{***}} \\
        & Ease of Use & -0.080 (0.08)\phantom{\textbf{***}} & -0.039 (0.05)\phantom{\textbf{**}} & 0.045 (0.04)\phantom{\textbf{**}} & 0.651 (0.89)\phantom{\textbf{**}} & 0.014 (0.01)\phantom{\textbf{***}} \\
        & Confidence in Use & 0.163 (0.18)\phantom{\textbf{***}} & -0.155 (0.12)\phantom{\textbf{**}} & 0.018 (0.07)\phantom{\textbf{**}} & 0.873 (1.28)\phantom{\textbf{**}} & 0.015 (0.01)\phantom{\textbf{***}} \\
        & Mental Demand & 0.112 (0.17)\phantom{\textbf{***}} & 0.046 (0.14)\phantom{\textbf{**}} & \textbf{-0.217 (0.10)*}\phantom{\textbf{*}} & 1.064 (1.74)\phantom{\textbf{**}} & 0.017 (0.02)\phantom{\textbf{***}} \\
        \bottomrule
        \multicolumn{7}{l}{\footnotesize *: p$<$0.05, **: p$<$0.01, ***: p$<$0.001}
    \end{tabularx}
    }
    \caption{
    Regression results for overall survey items (dependent variables) with RQ edit count, critique edit count, persona edit count, percentage of Nodes Used, and Total Node Count as independent variables.
    }
    \Description{The Table presents regression results for overall survey items based on independent variables such as RQ node edit count, Critique node edit count, Persona node edit count, percentage of nodes used, and total node count. These findings indicate how different edits and node usage impact survey item responses. Significant findings include a positive impact of the percentage of nodes used (β = 2.472, p < 0.01) and total node count (β = 0.041, p < 0.001) on Critique Helpfulness, while RQ Creativity shows a negative association with RQ edits (β = -0.376, p < 0.05). Research Outline Quality is positively associated with the percentage of nodes used (β = 3.015, p < 0.01). Reliance on System has a negative relationship with RQ edits (β = -0.396, p < 0.01). Both Perceived Control and Recall of Ideas show positive relationships with persona edits (β = 0.149, p < 0.01) and (β = 0.185, p < 0.01), respectively, while Mental Demand is negatively associated with total node count (β = -0.034, p < 0.01) and positively with persona edits (β = -0.217, p < 0.05).}
    \label{tab:survey-regression-results}
\end{table*}

In summary, the regression results indicate positive associations between users' degree of exploration (percentage of nodes used, and total number of nodes created) and their perceived quality of generated critiques and research outlines. We also found that the extent to which users edited the personas significantly influenced their perceived control and recall of ideas.
We detail the findings in the following paragraphs.

\subsubsection{Editing the RQ Nodes lowered users' perceived reliance on the system}
Our findings showed that users' number of edits on RQ nodes was negatively associated with the overall perceived creativity of RQ nodes ($t(36)=-2.060, p=0.039^{*}$), the perceived extent to which the system changed their way of thinking ($t(36)=-3.548, p<0.001^{***}$) and reliance on the system ($t(36)=-3.164, p=0.002^{**}$). 
The number of users' edits over RQ nodes was negatively associated with the creativity of RQs, their perceived extent to the system changed their ways of thinking, as they identified the need to modify RQ nodes when they felt that the generated research questions were inadequate or misaligned with their ideation goals. Thus by allowing users to make edits over the RQ nodes, the system provided users with greater control over the content, which consequently led to a lower perceived reliance on the system.

\subsubsection{A higher \% of nodes used associated with the improved quality of critiques and outlines.} 
We found significantly positive associations between users' percentages of nodes used (\textit{\% of Nodes Used}) and their ratings given towards the helpfulness of generated critiques (\textit{Critique Helpfulness}) during the post-session survey ($t(36)=3.070, p=0.002^{**}$). A similar positive association also exists between the total number of nodes created by each user (\textit{Total Node Count}) and their critique helpfulness (\textit{Critique Helpfulness}) survey ratings ($t(36)=3.639, p<0.001^{***}$). 
The results indicate that as users explored more nodes, they received higher-quality critiques from personas.
As mentioned by the participants, new insights were discovered through reading critiques during the exploration. P13 mentioned that, \userquote{the system provided a new approach for exploration that I hadn't initially considered,} emphasizing the importance of diverse exploration in enhancing the quality of user-perceived critiques.

\subsubsection{More edits on expert profiles improve users’ perceived recall of the generated ideas and perceived control during use.}
More edits made on expert profiles (\textit{\# of edits on
Persona nodes}) led to users' enhanced recall (\textit{Recall of Ideas}) of the generated ideas  ($t(36)=2.830, p=0.005^{**}$), As shown in Table \ref{tab:survey-regression-results}.
Users' perceived control of the system (\textit{Perceived Control}) was also improved through more edits over expert profiles (\textit{\# of edits on
Persona nodes}) during the exploration ($t(36)=2.617, p=0.009^{**}$). 
The perceived relevance and helpfulness of critiques did not show a significant correlation with users' expert profile edits. While we cannot state that expert profile edits enhance ideation outcomes, our findings suggest that utilizing a higher percentage of nodes contributes to fostering a stronger sense of recall over the ideas co-created between the user and the system and allows users to have a stronger sense of agency in shaping the generated ideas. 
Some participants also mentioned the mind-map-based design specifically helped them identify and recall system-generated research ideas, \userquote{... I personally also create workflows to have a clear direction (during research) or just brainstorming ... it helped me recall and identify the ideas I had during the process ... } as noted by P9.


\subsubsection{Users’ customization of LLM-generated expert profiles.}
\label{sec:RQ2-behavior-patterns}
To provide a more in-depth understanding of participants' customization interactions over expert profiles, we qualitatively examined system logs showing users' edits over expert profiles.
We focus on understanding how users engaged with LLM-generated personas by customizing and editing the persona profiles based on what was initially suggested by the system.
An overview of all edits made by participants are presented in \cref{tab:persona-edits-count}.

\begin{table*}[h]
\resizebox{\textwidth}{!}{%
\small 
\begin{tabular}{lclc}
\hline
\textbf{Category} &
  \textbf{Trait} &
  \textbf{Example Modification} &
  \textbf{Total \# Edits} \\ \hline
\multirow{7}{*}{\textbf{EXPERTISE}} &
  \textsc{Experience} &
  \begin{tabular}[c]{p{8cm}} 
  ``\textcolor{grey}{\st{12 years in AI research and development, specializing in natural language processing and}} \textcolor{darkgreen}{[AI expert in]} large language models'' \textcolor{darkgreen}{[, with HCI background as well]}'' 
  \end{tabular} &
  14 \\ \cline{2-4} 
 &
  \textsc{Domain} &
  \begin{tabular}[c]{p{8cm}} 
  ``\textcolor{grey}{\st{Data Science,}} Recommender Systems\textcolor{darkgreen}{[, Research, Generative language models]}'' 
  \end{tabular} &
  12 \\ \cline{2-4} 
 &
  \textsc{Skills} &
  \begin{tabular}[c]{p{8cm}} 
  ``Eudaimonia, positive psychological interventions, digital well-being, \textcolor{darkgreen}{[ludology, games]}'' 
  \end{tabular} &
  6 \\ \cline{2-4} 
 &
  \textsc{Method} &
  \begin{tabular}[c]{p{8cm}} 
  ``qualitative \textcolor{grey}{\st{analyses}} and \textcolor{darkgreen}{[quantitative]}'' 
  \end{tabular} &
  2 \\ \cline{2-4} 
 &
  \textsc{Education} &
  \begin{tabular}[c]{p{8cm}} 
  ``Master's or PhD in Computer Science, AI, or related field, with \textcolor{darkgreen}{[a minor in public health]}'' 
  \end{tabular} &
  2 \\ \cline{2-4} 
 &
  \textsc{Knowledge} &
  \begin{tabular}[c]{p{8cm}} 
  ``\textcolor{darkgreen}{[Have a basic understanding of UX and UI design standards, have skills related to user data analysis]}'' (Skills)
  \end{tabular} &
  4 \\ \hline
\multirow{3}{*}{\textbf{ROLES}} &
  \textsc{Functional Role} &
  \begin{tabular}[c]{p{8cm}} 
  ``To leverage large language models for practical applications in \textcolor{darkgreen}{[facilitating HCI research]}'' 
  \end{tabular} &
  19 \\ \cline{2-4} 
 &
  \textsc{Social Identity} &
  \begin{tabular}[c]{p{8cm}} 
  ``Assistive Technology \textcolor{grey}{\st{Developer}} \textcolor{darkgreen}{[Researcher]}'' 
  \end{tabular} &
  11 \\ \cline{2-4} 
 &
  \textsc{Social Status} &
  \begin{tabular}[c]{p{8cm}} 
  ``\textcolor{darkgreen}{[CS Professor at Stanford]}'' 
  \end{tabular} &
  3 \\ \hline
\end{tabular}%
}
\caption{Number of edits user performed over expert personas and examples by different traits, using the interface showcases in \cref{fig:persona-node}.
} 
\Description{The table summarizes the number of edits made by users to expert personas, organized into two main categories: Expertise and Roles. The Expertise category includes seven traits: Experience, Domain, Skills, Method, Education, and Knowledge. Each trait shows an example of a user modification, with the original text struck through in grey and the new text added in green. For example, under Experience, “12 years in AI research and development” is replaced with “AI expert in large language models, with HCI background.” The total number of edits per trait is listed, with 14 edits for Experience, 12 for Domain, 6 for Skills, 2 for Method, 2 for Education, and 4 for Knowledge. The Roles category includes three traits: Functional Role, Social Identity, and Social Status. Examples of modifications are shown, such as changing "Assistive Technology Developer" to "Researcher" under Social Identity, with a total of 19 edits for Functional Role, 11 for Social Identity, and 3 for Social Status.}
\label{tab:persona-edits-count}
\end{table*}

\textbf{Users added traits to personas that reflect their own background.}
When choosing and editing personas, participants mostly prioritized choosing personas with similar backgrounds to theirs or from a familiar domain. 
Users found it very helpful that personas from their own field help explain terms in a familiar language (\userquote{... explains concepts in more simple and familiar terms... ,} P10). 
In total, 13 participants made edits to personas that reflect their own backgrounds and expertise.
Users are found more inclined to personalize and relate to personas that reflect their own domain background since it makes the personas more tangible and relatable, as mentioned by P16 \userquote{... it's easier for me to envision them as an actual person that I know... because it's like my field, so it's like easier for me to assign a face, or a name... easier to imagine, and to personalize a person ...} 

\textbf{Users clarified the personas' functional role with concrete research objectives.}
Participants are observed to have made various interactions and edits to these personas. 
In total, 19 edits related to personas' roles (57.6\%) were performed over each persona's functional role, as shown in Table \ref{tab:persona-edits-count}. 
Importantly, users adapt the goals of the personas more closely to certain research goals. For instance, the participant changed it from ``To leverage large language models for practical applications in various domains'' to ``To leverage large language models for practical applications in facilitating HCI research`` (P4). 
For another participant, the objective was slightly changed from ``Develop and implement data-driven recommendation algorithms for various applications'' to ``Research and propose novel recommendation algorithms, in particular, using Large Language Models as generative components of the algorithms'' (P3). 
Participants also found it helpful to include their exploration goal or expectation in the user instruction field for the persona. P13 expressed their intent of ``Help me think of what tech will involve in the system development'' directly as user instruction to the persona.

\textbf{Users specified the social statuses of expert personas.}
Interestingly, some participants (N=8) edited the personas not only in terms of expertise but also intentionally adding social traits the persona, or to \userquote{humanize} the persona, as per one of the participants (P16).
Several participants have mentioned their intentions of improving the credibility of the personas by enriching the descriptions of both the personas' experiences and roles. For instance, P16 added a new trait, ``social status'' of ``CS Professor at Stanford'', to one of the generated personas. The participant noted that \userquote{... it's easier for me to interact (with AI) if they have traits that make them more human-like ...} 
The participant also mentioned authority when rationalizing their persona edits. 
\userquote{If you're a professor from a reputable school, when you recommend literature, I guess it can be like a bit more picky ... because I expect someone like a reputable professional (to) only pick like reputable sources ...}


\textbf{Users borrowed traits from other generated personas during persona customization.}
The generated personas were also found to contain traits that inspired users, even if those traits were not part of the persona the user chose (N=5). In the case of P8, the user was intrigued by two out of the three generated personas ``Human-Computer Interaction Researcher'' and ``Audio Description Producer''. 
P8 chose the ``Audio Description Producer'' persona but would also prefer the persona to have research experience in the general HCI domain, thus adding a new trait of ``experience'' of ``being an HCI researcher. '' The participant's editing of moving traits from one persona to another was driven by their need for a persona to have the background or expertise originally presented in another persona generated by the system. 
Closely related, P15 also edited the ``role'' trait to include ``pharmacologist of precision therapies for leukemia'' for the persona ``Genetics Expert'' with the ``functional role'' of ``to offer expertise in the genetic and molecular mechanisms underlying different leukemia subtypes, '' noting that \userquote{... the two research topics seems inseparable.} The added trait was inspired by another generated persona ``Precision Therapy Pharmacologist, '' demonstrating the participant's intention to create more comprehensive personas by integrating traits from different domains suggested by the system.

\section{Discussion}
In this section, we provide further insights by discussing the implications of our findings situated in past research and cognitive theories.
Furthermore, we offer design recommendations for future systems utilizing LLM-simulated personas for research ideation.

\subsection{Bridging Interdisciplinary Research Ideation and User Agency through Customizable Multi-Persona AI}
\disrevisedcamclean{
Evaluating the \systemName\ system with researchers showed that its multi-persona design enhances research ideation without adding cognitive load (\cref{sec:RQ1-findings}) over the use of singular personas. 
}
Beyond past studies focusing on the perspective of audience~\cite{petridis2023anglekindling}, we demonstrated that incorporating expert AI personas with domain-specific expertise fosters interdisciplinary research. 
Users effectively expressed different thinking patterns and addressed diverse use cases (\cref{sec:RQ3-diverse-use-case}), with the help of the interactive mind map and easy expert profile customization, especially when refining ideas through iterative revisions.
The system also benefits users exploring less familiar domains by providing domain-specific concepts and theories (\cref{sec:RQ2-regression}), providing instant feedback and scaffolding for knowledge acquisition. Additionally, allowing users to edit AI-generated personas enhanced their perceived control and autonomy (\cref{sec:RQ1-findings}).
For researchers new to a field, challenges include knowing the right questions to ask or bridging communication gaps~\cite{august2024know, macleod2018makes,macleod2018makes}. In the context of the \systemName\ system, the interaction between a user and AI-simulated expert personas offers instant, iterative feedback and scaffolding to key concepts, which in turn, improves knowledge acquisition and facilitates future communication with human experts.
our findings (\cref{sec:RQ1-findings}) indicated the design of supporting users' edits over AI-generated expert personas enhanced their perceived control over the system, likely due to reduced cognitive load and increased user autonomy.
By further examining users' behaviors during expert profile customization (\cref{sec:RQ2-findings}), we identified that the uniqueness of an expert profile as an interface of communication with the system has provided users with a means to project their own intentional states~\cite{dennett1971intentional, dennett1987he} onto the AI, often with the intention of obtaining corresponding informational responses from the AI system.
This design promotes user agency while maintaining ideation support, drawing from the knowledge from different domains, meanwhile improving trust, reducing cognitive load~\cite{de2017people}, and addressing concerns about reduced idea ownership~\cite{guo2024exploring}, as modifying expert profiles helps users retain ownership and better understand the system’s capabilities.
This can also potentially improve users' trust in the persona and the system's subsequently generated outcomes while reducing their cognitive load~\cite{de2017people}. 
Prior work \cite{guo2024exploring} has raised concerns of reduced idea ownership during brainstorming using AI for creativity support, while our work demonstrates that allowing users to modify AI-generated personas can potentially alleviate this issue as a design that also improves the transparency and explainability of the system.


\subsection{Cultivating Critical Thinking Through Multi-Perspective Expert Simulation}
\disrevised{
Our findings indicate that the multi-perspective design of \systemName\ not only supports interdisciplinary exploration but also triggers higher-order thinking processes crucial for research ideation. In particular, we observed users repeatedly engaging in core critical thinking elements defined by past research~\cite{facione1990critical}.
\systemName's design of simulating experts tends to incur participants' activities related to \textit{self-regulation} and \textit{evaluation} by driving users to think about the alignment of background and interest between themselves and system-generated expert profiles. 
We observed participants re-stating the scope of their own research domains and clarifying terminology when browsing various expert profiles. For instance, by scanning expert profiles, some users reformulated their original research questions in more precise terms (e.g., adding the keyword “LLM” to reflect their actual research goal). 
Some instances of \textit{evaluation} (e.g., where the user identifies an expert profile to have a similar background to theirs) were also found to be associated with confirmation of users' pre-existing research ideas and consolidation of their trust and willingness to further engage with the expert perspective, contributing to trust-building but also raising the risk of over-reliance on AI.
Additionally, the system’s mind-map structure and iterative interactions promoted user-led reflection, through activities of inference from newly suggested ideas or concepts by different experts during the exploration.
Participants were observed to frequently speculate on new connections or follow-up topics based on expert suggestions (e.g., pivoting from a broad question about “spatial distribution of RNA” to a narrower focus on single-cell transcriptomic). 
Our findings suggest that the multi-perspective interaction design appears to contribute to users’ interdisciplinary research ideation process, by promoting different levels of critical thinking activities introducing both new and more specific research topics and directions. Such observations highlight the potential application of using AI-simulated experts for fostering higher-order critical thinking skills in interdisciplinary research and education~\cite{halpern1998teaching,lee2024conversational}.
On the other hand, we observed limited instances of users' \textit{explanation} activities beyond what was prompted by the think-aloud protocol. This is potentially because \systemName\ did not design features to support or stimulate explicit rationalization of their behavior and decisions. In future iterations, designers could explore using AI-simulated personas that pose reflective questions~\cite{kumar2024supporting}, prompting users to elaborate further on their thought processes and rationales.
}

\subsection{Navigating Cognitive Biases in Human-AI Persona Interaction: Confirmation and Social Biases}
\label{sec:discussion-biases}
\disrevised{
Cognitive biases have been a common concern in human-AI co-creation systems~\cite{bertrand2022cognitive}. Although the findings demonstrated the utility of the system in providing research ideation support, we noted potential biases when users interacted with the expert personas during system use. 
In \systemName, confirmation bias arose from both users’ expert choices and the LLM’s inherent biases \cite{rastogi2022deciding,Buschek2021NinePP}.
Participants tended to select or edit personas with backgrounds aligning with their own (\cref{sec:RQ2-behavior-patterns}), reinforcing existing beliefs \cite{liu2024ai}.
They also trusted personas more when they aligned with personal experiences, potentially limiting ideation diversity. This aligns with Cognitive Dissonance Theory \cite{braga1972role,harmon2019introduction}, suggesting individuals prefer assistance aligning with their beliefs. Our observations further echo this phenomenon.
Another potential cognitive bias we noticed was social bias, specifically authority bias~\cite{milgram1963behavioral,juarez2018analyzing} and social conformity~\cite{stallen2015neuroscience}. 
Users favored expert personas perceived as more authoritative (\cref{sec:RQ2-behavior-patterns}), even modifying profiles to enhance perceived expertise. This can lead to over-reliance on the system and overshadow less authoritative but equally informative expert personas.
Combined with the anchoring effect~\cite{de2018anchoring}, these factors risk collaborative fixation \cite{kohn2011collaborative}, constraining alternative perspectives.
}

\subsection{Design Implications}

\subsubsection{Mitigating cognitive biases and fixation in AI-supported ideation.}
Our study revealed the benefits of using multiple LLM-simulated personas in research ideation (\cref{sec:RQ1-overall}), while also indicating several potential concerns about inducing cognitive biases, as discussed in \Cref{sec:discussion-biases}. 
Research questions generated by the system after multiple iterations tend to become narrowly focused and even repetitive, when a certain level of specificity has been reached.
We argue that mitigating collaborative fixation~\cite{kohn2011collaborative} in AI-supported ideation is essential for promoting critical thinking and exploring diverse perspectives.
Recent studies on language model alignment also pointed out a potential misalignment between the objective of instruction-following, which is the most common paradigm in current practices of instruction tuning, and boosting models' ability to suggest out-of-box ideas and thinking against users' provided instructions ~\cite{sharma2023towards,mohammadi2024creativity}. 
Our findings suggest that human-AI co-creation systems should recognize when to stop generating ideas to avoid repetition and lack of insight due to idea fixation.
Future designs can be introduced to address fixation during collaborative ideation to help users diversify the exploration process and consider different perspectives while combating challenges including information overload~\cite{zhu2018explainable, louie2020novice} through dynamically assessing the novelty~\cite{Wang2023SciMONSI} of generated ideas.
In addition to the user-suggested feature of the mechanism to facilitate a review of past ideation outcomes to promote divergent exploration from the RQ3 findings (\cref{sec:RQ3-diverse-use-case}), a retrospective summary of historical ideas and takeaways produced during the earlier stage of ideation would serve as another potential design for future human-AI co-creation systems.

\disrevised{
\subsubsection{Beyond Research Ideation: Facilitating Diverse Scaffolding Through AI-Simulated Personas}
Leveraging the fast and adaptive responses of LLMs, prior studies have explored their effectiveness for idea iteration, scaffolding, and immediate feedback~\cite{dhillon2024shaping,heyman2024supermind}. While our study focused on supporting the research ideation process, the core design principle of using customizable AI personas can be extended to a variety of other fields such as design, consulting, content creation, product development, and decision-making in strategic planning.
During free exploration sessions (\cref{sec:RQ3-diverse-use-case}), participants used multiple personas to examine literature from various research communities. This behavior suggests that personas, whether intended for research or other domains, should highlight the shared values, norms, and conceptual frameworks typical of those communities. For instance, service designers~\cite{han2018service} or consulting teams could develop customizable personas representing different user groups to surface their diverse perspectives and values.
}

\disrevised{
Additionally, participants (P15, P3) emphasized the importance of adapting literature resources (e.g., foundational concepts vs.\ cutting-edge methods) to a researcher’s prior knowledge. Incorporating predicted user intent and background as contextual information~\cite{pfautz2015general} and supporting the establishment of a mutual theory of mind (MToM) \cite{wang2021towards} can help the system deliver content at the appropriate level of detail. Analogously in other domains, newcomers may require broader overviews, while experts benefit from deep-dives into specialized methodologies or emerging trends. Future systems could personalize system outputs through the interface of simulated personas, such as adopting the complexity of outputs, based on user’s current knowledge or project scope conveyed through customizable persona profiles. 
}

\disrevised{
Our findings also suggest that the persona customization interface offers users a stronger sense of ownership towards co-creation outcomes, helping them to more intuitively shape AI-driven feedback toward their immediate needs. Beyond research, this same principle can foster rapid ideation in general. For instance, an online content creation team might instantiate multiple personas representing different audience segments, and iterate on their profiles to better align with shifting project requirements or newly discovered audience insights. By allowing fine-grained control over persona attributes, systems can encourage diverse and adaptable exploration of multiple perspectives and critical thinking.
}

\section{Limitation and Future Research}
While our study demonstrates the potential of using LLM-simulated expert personas to facilitate the research ideation process, several limitations should be noted. 
We did not conduct systematic evaluation of the quality and accuracy of the personas that the system generated nor the literature they recommended. The personas' knowledge might inherently be limited by the dataset on which large language models were trained on when generating search queries.
Another limitation of our study is the absence of a direct comparative baseline with similar ideation tools. While there are currently no ideation instruments that closely match the functionality of \systemName, this limits our ability to objectively compare its usefulness. Future research should consider evaluating possible benchmarks or developing simplistic frameworks for comparative analysis.
Last but not least, our study does not address two important aspects: system interactions outside direct persona manipulation and social interactions other than solicitation of feedback and literature information, such as natural conversations. 
Future research should explore the extent to which these social elements genuinely impact the ideation process versus merely reflecting users' inherent biases or psychological tendencies, potentially through controlled experimental designs that isolate the effects of social traits against informational traits.

\section{Conclusion}
\disrevisedcamclean{
In this study, we introduced \systemName, an LLM-based research ideation system that aims to address the gap of a lack of access to domain experts during interdisciplinary research ideation.
The results showed that working with diverse personas significantly improved perceived relevance and creativity of RQ and critique without raising users' cognitive load. 
We also found that \systemName\ promoted participants' critical thinking activities under different use cases, showing different cognitive approaches in formulating research ideas.
Participants adapted personas across perspectives and the ability to edit persona characteristics increased participants' feeling of agency, improving the recall of concepts from unknown domains. 
Moreover, the persona editing feature allowed for deeper insights into the users' customization behaviors and what those mean for leveraging AI in research ideation. 
The results thus provide insights into both the opportunities and challenges of using LLM-simulated personas for research exploration, and hence provide design implications and point to future directions for improving human-AI collaboration in interdisciplinary research ideation and beyond.
}


\begin{acks}
\disrevisedcamclean{
This material is based upon work supported by the National Science Foundation under Grant \#2119589. Any opinions, findings, and conclusions or recommendations expressed in this material are those of the author(s) and do not necessarily reflect the views of the National Science Foundation. The authors would also like to thank the anonymous reviewers for their valuable feedback.
Additionally, results presented in this paper were obtained using CloudBank \cite{norman2021cloudbank}, which is supported by the National Science Foundation under award \#1925001.
}
\end{acks}

\bibliographystyle{ACM-Reference-Format}
\bibliography{references}


\begin{thebibliography}{117}


\ifx \showCODEN    \undefined \def \showCODEN     #1{\unskip}     \fi
\ifx \showDOI      \undefined \def \showDOI       #1{#1}\fi
\ifx \showISBNx    \undefined \def \showISBNx     #1{\unskip}     \fi
\ifx \showISBNxiii \undefined \def \showISBNxiii  #1{\unskip}     \fi
\ifx \showISSN     \undefined \def \showISSN      #1{\unskip}     \fi
\ifx \showLCCN     \undefined \def \showLCCN      #1{\unskip}     \fi
\ifx \shownote     \undefined \def \shownote      #1{#1}          \fi
\ifx \showarticletitle \undefined \def \showarticletitle #1{#1}   \fi
\ifx \showURL      \undefined \def \showURL       {\relax}        \fi
\providecommand\bibfield[2]{#2}
\providecommand\bibinfo[2]{#2}
\providecommand\natexlab[1]{#1}
\providecommand\showeprint[2][]{arXiv:#2}

\bibitem[Agnoli et~al\mbox{.}(2016)]%
        {agnoli2016estimating}
\bibfield{author}{\bibinfo{person}{Sergio Agnoli}, \bibinfo{person}{Giovanni~E Corazza}, {and} \bibinfo{person}{Mark~A Runco}.} \bibinfo{year}{2016}\natexlab{}.
\newblock \showarticletitle{Estimating creativity with a multiple-measurement approach within scientific and artistic domains}.
\newblock \bibinfo{journal}{\emph{Creativity Research Journal}} \bibinfo{volume}{28}, \bibinfo{number}{2} (\bibinfo{year}{2016}), \bibinfo{pages}{171--176}.
\newblock


\bibitem[Al~Qunayeer(2020)]%
        {al2020supporting}
\bibfield{author}{\bibinfo{person}{Huda~Suleiman Al~Qunayeer}.} \bibinfo{year}{2020}\natexlab{}.
\newblock \showarticletitle{Supporting postgraduates in research proposals through peer feedback in a Malaysian university}.
\newblock \bibinfo{journal}{\emph{Journal of Further and Higher Education}} \bibinfo{volume}{44}, \bibinfo{number}{7} (\bibinfo{year}{2020}), \bibinfo{pages}{956--970}.
\newblock


\bibitem[Ali et~al\mbox{.}(2019)]%
        {ali2019mechanism}
\bibfield{author}{\bibinfo{person}{Ahsan Ali}, \bibinfo{person}{Hongwei Wang}, {and} \bibinfo{person}{Ali~Nawaz Khan}.} \bibinfo{year}{2019}\natexlab{}.
\newblock \showarticletitle{Mechanism to enhance team creative performance through social media: a transactive memory system approach}.
\newblock \bibinfo{journal}{\emph{Computers in Human Behavior}}  \bibinfo{volume}{91} (\bibinfo{year}{2019}), \bibinfo{pages}{115--126}.
\newblock


\bibitem[Aschauer et~al\mbox{.}(2022)]%
        {aschauer2022contribution}
\bibfield{author}{\bibinfo{person}{Wolfgang Aschauer}, \bibinfo{person}{Kurt Haim}, {and} \bibinfo{person}{Christoph Weber}.} \bibinfo{year}{2022}\natexlab{}.
\newblock \showarticletitle{A contribution to scientific creativity: A validation study measuring divergent problem solving ability}.
\newblock \bibinfo{journal}{\emph{Creativity Research Journal}} \bibinfo{volume}{34}, \bibinfo{number}{2} (\bibinfo{year}{2022}), \bibinfo{pages}{195--212}.
\newblock


\bibitem[August et~al\mbox{.}(2024)]%
        {august2024know}
\bibfield{author}{\bibinfo{person}{Tal August}, \bibinfo{person}{Kyle Lo}, \bibinfo{person}{Noah~A Smith}, {and} \bibinfo{person}{Katharina Reinecke}.} \bibinfo{year}{2024}\natexlab{}.
\newblock \showarticletitle{Know Your Audience: The benefits and pitfalls of generating plain language summaries beyond the" general" audience}. In \bibinfo{booktitle}{\emph{Proceedings of the CHI Conference on Human Factors in Computing Systems}}. \bibinfo{pages}{1--26}.
\newblock


\bibitem[Baek et~al\mbox{.}(2024)]%
        {baek2024researchagent}
\bibfield{author}{\bibinfo{person}{Jinheon Baek}, \bibinfo{person}{Sujay~Kumar Jauhar}, \bibinfo{person}{Silviu Cucerzan}, {and} \bibinfo{person}{Sung~Ju Hwang}.} \bibinfo{year}{2024}\natexlab{}.
\newblock \showarticletitle{Researchagent: Iterative research idea generation over scientific literature with large language models}.
\newblock \bibinfo{journal}{\emph{arXiv preprint arXiv:2404.07738}} (\bibinfo{year}{2024}).
\newblock


\bibitem[Bailin(2002)]%
        {bailin2002critical}
\bibfield{author}{\bibinfo{person}{Sharon Bailin}.} \bibinfo{year}{2002}\natexlab{}.
\newblock \showarticletitle{Critical thinking and science education}.
\newblock \bibinfo{journal}{\emph{Science \& education}}  \bibinfo{volume}{11} (\bibinfo{year}{2002}), \bibinfo{pages}{361--375}.
\newblock


\bibitem[Beltagy et~al\mbox{.}(2019)]%
        {beltagy2019scibert}
\bibfield{author}{\bibinfo{person}{Iz Beltagy}, \bibinfo{person}{Kyle Lo}, {and} \bibinfo{person}{Arman Cohan}.} \bibinfo{year}{2019}\natexlab{}.
\newblock \showarticletitle{SciBERT: A pretrained language model for scientific text}.
\newblock \bibinfo{journal}{\emph{arXiv preprint arXiv:1903.10676}} (\bibinfo{year}{2019}).
\newblock


\bibitem[Benharrak et~al\mbox{.}(2024)]%
        {benharrak2024writer}
\bibfield{author}{\bibinfo{person}{Karim Benharrak}, \bibinfo{person}{Tim Zindulka}, \bibinfo{person}{Florian Lehmann}, \bibinfo{person}{Hendrik Heuer}, {and} \bibinfo{person}{Daniel Buschek}.} \bibinfo{year}{2024}\natexlab{}.
\newblock \showarticletitle{Writer-defined AI personas for on-demand feedback generation}. In \bibinfo{booktitle}{\emph{Proceedings of the CHI Conference on Human Factors in Computing Systems}}. \bibinfo{pages}{1--18}.
\newblock


\bibitem[Bertrand et~al\mbox{.}(2022)]%
        {bertrand2022cognitive}
\bibfield{author}{\bibinfo{person}{Astrid Bertrand}, \bibinfo{person}{Rafik Belloum}, \bibinfo{person}{James~R Eagan}, {and} \bibinfo{person}{Winston Maxwell}.} \bibinfo{year}{2022}\natexlab{}.
\newblock \showarticletitle{How cognitive biases affect XAI-assisted decision-making: A systematic review}. In \bibinfo{booktitle}{\emph{Proceedings of the 2022 AAAI/ACM Conference on AI, Ethics, and Society}}. \bibinfo{pages}{78--91}.
\newblock


\bibitem[Braga(1972)]%
        {braga1972role}
\bibfield{author}{\bibinfo{person}{Joseph~L Braga}.} \bibinfo{year}{1972}\natexlab{}.
\newblock \showarticletitle{Role theory, cognitive dissonance theory, and the interdisciplinary team}.
\newblock \bibinfo{journal}{\emph{Interchange}} \bibinfo{volume}{3}, \bibinfo{number}{4} (\bibinfo{year}{1972}), \bibinfo{pages}{69--78}.
\newblock


\bibitem[Braun and Clarke(2012)]%
        {braun2012thematic}
\bibfield{author}{\bibinfo{person}{Virginia Braun} {and} \bibinfo{person}{Victoria Clarke}.} \bibinfo{year}{2012}\natexlab{}.
\newblock \bibinfo{booktitle}{\emph{Thematic analysis.}}
\newblock \bibinfo{publisher}{American Psychological Association}.
\newblock


\bibitem[Buschek et~al\mbox{.}(2021)]%
        {Buschek2021NinePP}
\bibfield{author}{\bibinfo{person}{Daniel Buschek}, \bibinfo{person}{Lukas Mecke}, \bibinfo{person}{Florian Lehmann}, {and} \bibinfo{person}{Hai Dang}.} \bibinfo{year}{2021}\natexlab{}.
\newblock \showarticletitle{Nine Potential Pitfalls when Designing Human-AI Co-Creative Systems}.
\newblock \bibinfo{journal}{\emph{ArXiv}}  \bibinfo{volume}{abs/2104.00358} (\bibinfo{year}{2021}).
\newblock
\urldef\tempurl%
\url{https://api.semanticscholar.org/CorpusID:232478535}
\showURL{%
\tempurl}


\bibitem[Chan et~al\mbox{.}(2023)]%
        {chan2023chateval}
\bibfield{author}{\bibinfo{person}{Chi-Min Chan}, \bibinfo{person}{Weize Chen}, \bibinfo{person}{Yusheng Su}, \bibinfo{person}{Jianxuan Yu}, \bibinfo{person}{Wei Xue}, \bibinfo{person}{Shanghang Zhang}, \bibinfo{person}{Jie Fu}, {and} \bibinfo{person}{Zhiyuan Liu}.} \bibinfo{year}{2023}\natexlab{}.
\newblock \showarticletitle{Chateval: Towards better llm-based evaluators through multi-agent debate}.
\newblock \bibinfo{journal}{\emph{arXiv preprint arXiv:2308.07201}} (\bibinfo{year}{2023}).
\newblock


\bibitem[Chen et~al\mbox{.}(2024)]%
        {chen2024bge}
\bibfield{author}{\bibinfo{person}{Jianlv Chen}, \bibinfo{person}{Shitao Xiao}, \bibinfo{person}{Peitian Zhang}, \bibinfo{person}{Kun Luo}, \bibinfo{person}{Defu Lian}, {and} \bibinfo{person}{Zheng Liu}.} \bibinfo{year}{2024}\natexlab{}.
\newblock \bibinfo{title}{BGE M3-Embedding: Multi-Lingual, Multi-Functionality, Multi-Granularity Text Embeddings Through Self-Knowledge Distillation}.
\newblock
\newblock
\showeprint[arxiv]{2402.03216}~[cs.CL]


\bibitem[Chen et~al\mbox{.}(2011)]%
        {chen2011can}
\bibfield{author}{\bibinfo{person}{Xiantao Chen}, \bibinfo{person}{Ying Liu}, \bibinfo{person}{Ning Liu}, {and} \bibinfo{person}{Xiaojie Wang}.} \bibinfo{year}{2011}\natexlab{}.
\newblock \showarticletitle{Can persona facilitate ideation? A comparative study on effects of personas in brainstorming}. In \bibinfo{booktitle}{\emph{Human-Computer Interaction--INTERACT 2011: 13th IFIP TC 13 International Conference, Lisbon, Portugal, September 5-9, 2011, Proceedings, Part IV 13}}. Springer, \bibinfo{pages}{491--494}.
\newblock


\bibitem[Chiang et~al\mbox{.}(2024)]%
        {chiang2024enhancing}
\bibfield{author}{\bibinfo{person}{Chun-Wei Chiang}, \bibinfo{person}{Zhuoran Lu}, \bibinfo{person}{Zhuoyan Li}, {and} \bibinfo{person}{Ming Yin}.} \bibinfo{year}{2024}\natexlab{}.
\newblock \showarticletitle{Enhancing AI-Assisted Group Decision Making through LLM-Powered Devil's Advocate}. In \bibinfo{booktitle}{\emph{Proceedings of the 29th International Conference on Intelligent User Interfaces}}. \bibinfo{pages}{103--119}.
\newblock


\bibitem[Choi et~al\mbox{.}(2024)]%
        {choi2024proxona}
\bibfield{author}{\bibinfo{person}{Yoonseo Choi}, \bibinfo{person}{Eun~Jeong Kang}, \bibinfo{person}{Seulgi Choi}, \bibinfo{person}{Min~Kyung Lee}, {and} \bibinfo{person}{Juho Kim}.} \bibinfo{year}{2024}\natexlab{}.
\newblock \showarticletitle{Proxona: Leveraging LLM-Driven Personas to Enhance Creators' Understanding of Their Audience}.
\newblock \bibinfo{journal}{\emph{arXiv preprint arXiv:2408.10937}} (\bibinfo{year}{2024}).
\newblock


\bibitem[Chugh et~al\mbox{.}(2022)]%
        {chugh2022supervisory}
\bibfield{author}{\bibinfo{person}{Ritesh Chugh}, \bibinfo{person}{Stephanie Macht}, {and} \bibinfo{person}{Bobby Harreveld}.} \bibinfo{year}{2022}\natexlab{}.
\newblock \showarticletitle{Supervisory feedback to postgraduate research students: a literature review}.
\newblock \bibinfo{journal}{\emph{Assessment \& Evaluation in Higher Education}} \bibinfo{volume}{47}, \bibinfo{number}{5} (\bibinfo{year}{2022}), \bibinfo{pages}{683--697}.
\newblock


\bibitem[Cohan et~al\mbox{.}(2019)]%
        {cohan2019pretrained}
\bibfield{author}{\bibinfo{person}{Arman Cohan}, \bibinfo{person}{Iz Beltagy}, \bibinfo{person}{Daniel King}, \bibinfo{person}{Bhavana Dalvi}, {and} \bibinfo{person}{Daniel~S Weld}.} \bibinfo{year}{2019}\natexlab{}.
\newblock \showarticletitle{Pretrained language models for sequential sentence classification}.
\newblock \bibinfo{journal}{\emph{arXiv preprint arXiv:1909.04054}} (\bibinfo{year}{2019}).
\newblock


\bibitem[Dastmalchi et~al\mbox{.}(2021)]%
        {dastmalchi2021exploring}
\bibfield{author}{\bibinfo{person}{Mohammad~Reza Dastmalchi}, \bibinfo{person}{Bimal Balakrishnan}, {and} \bibinfo{person}{Danielle Oprean}.} \bibinfo{year}{2021}\natexlab{}.
\newblock \showarticletitle{Exploring the role of transactive memory systems in team decision-making during ideation phase}.
\newblock \bibinfo{journal}{\emph{Proceedings of the Design Society}}  \bibinfo{volume}{1} (\bibinfo{year}{2021}), \bibinfo{pages}{1529--1536}.
\newblock


\bibitem[De~Graaf and Malle(2017)]%
        {de2017people}
\bibfield{author}{\bibinfo{person}{Maartje~MA De~Graaf} {and} \bibinfo{person}{Bertram~F Malle}.} \bibinfo{year}{2017}\natexlab{}.
\newblock \showarticletitle{How people explain action (and autonomous intelligent systems should too)}. In \bibinfo{booktitle}{\emph{2017 AAAI Fall Symposium Series}}.
\newblock


\bibitem[de~Wilde et~al\mbox{.}(2018)]%
        {de2018anchoring}
\bibfield{author}{\bibinfo{person}{Tim~RW de Wilde}, \bibinfo{person}{Femke~S Ten~Velden}, {and} \bibinfo{person}{Carsten~KW De~Dreu}.} \bibinfo{year}{2018}\natexlab{}.
\newblock \showarticletitle{The anchoring-bias in groups}.
\newblock \bibinfo{journal}{\emph{Journal of Experimental Social Psychology}}  \bibinfo{volume}{76} (\bibinfo{year}{2018}), \bibinfo{pages}{116--126}.
\newblock


\bibitem[Dennett(1971)]%
        {dennett1971intentional}
\bibfield{author}{\bibinfo{person}{Daniel~C Dennett}.} \bibinfo{year}{1971}\natexlab{}.
\newblock \showarticletitle{Intentional systems}.
\newblock \bibinfo{journal}{\emph{The journal of philosophy}} \bibinfo{volume}{68}, \bibinfo{number}{4} (\bibinfo{year}{1971}), \bibinfo{pages}{87--106}.
\newblock


\bibitem[Dennett(1987)]%
        {dennett1987he}
\bibfield{author}{\bibinfo{person}{Daniel~C Dennett}.} \bibinfo{year}{1987}\natexlab{}.
\newblock \bibinfo{title}{he Intentional Stance}.
\newblock
\newblock


\bibitem[Dhillon et~al\mbox{.}(2024)]%
        {dhillon2024shaping}
\bibfield{author}{\bibinfo{person}{Paramveer~S Dhillon}, \bibinfo{person}{Somayeh Molaei}, \bibinfo{person}{Jiaqi Li}, \bibinfo{person}{Maximilian Golub}, \bibinfo{person}{Shaochun Zheng}, {and} \bibinfo{person}{Lionel~Peter Robert}.} \bibinfo{year}{2024}\natexlab{}.
\newblock \showarticletitle{Shaping Human-AI Collaboration: Varied Scaffolding Levels in Co-writing with Language Models}. In \bibinfo{booktitle}{\emph{Proceedings of the CHI Conference on Human Factors in Computing Systems}}. \bibinfo{pages}{1--18}.
\newblock


\bibitem[Douze et~al\mbox{.}(2024)]%
        {douze2024faiss}
\bibfield{author}{\bibinfo{person}{Matthijs Douze}, \bibinfo{person}{Alexandr Guzhva}, \bibinfo{person}{Chengqi Deng}, \bibinfo{person}{Jeff Johnson}, \bibinfo{person}{Gergely Szilvasy}, \bibinfo{person}{Pierre-Emmanuel Mazar{\'e}}, \bibinfo{person}{Maria Lomeli}, \bibinfo{person}{Lucas Hosseini}, {and} \bibinfo{person}{Herv{\'e} J{\'e}gou}.} \bibinfo{year}{2024}\natexlab{}.
\newblock \showarticletitle{The faiss library}.
\newblock \bibinfo{journal}{\emph{arXiv preprint arXiv:2401.08281}} (\bibinfo{year}{2024}).
\newblock


\bibitem[Dubois et~al\mbox{.}(2024)]%
        {dubois2024alpacafarm}
\bibfield{author}{\bibinfo{person}{Yann Dubois}, \bibinfo{person}{Chen~Xuechen Li}, \bibinfo{person}{Rohan Taori}, \bibinfo{person}{Tianyi Zhang}, \bibinfo{person}{Ishaan Gulrajani}, \bibinfo{person}{Jimmy Ba}, \bibinfo{person}{Carlos Guestrin}, \bibinfo{person}{Percy~S Liang}, {and} \bibinfo{person}{Tatsunori~B Hashimoto}.} \bibinfo{year}{2024}\natexlab{}.
\newblock \showarticletitle{Alpacafarm: A simulation framework for methods that learn from human feedback}.
\newblock \bibinfo{journal}{\emph{Advances in Neural Information Processing Systems}}  \bibinfo{volume}{36} (\bibinfo{year}{2024}).
\newblock


\bibitem[Elio et~al\mbox{.}(2011)]%
        {elio2011computing}
\bibfield{author}{\bibinfo{person}{Renee Elio}, \bibinfo{person}{Jim Hoover}, \bibinfo{person}{Ioanis Nikolaidis}, \bibinfo{person}{Mohammad Salavatipour}, \bibinfo{person}{Lorna Stewart}, {and} \bibinfo{person}{Ken Wong}.} \bibinfo{year}{2011}\natexlab{}.
\newblock \bibinfo{title}{About computing science research methodology}.
\newblock
\newblock


\bibitem[Facione(1990)]%
        {facione1990critical}
\bibfield{author}{\bibinfo{person}{Peter Facione}.} \bibinfo{year}{1990}\natexlab{}.
\newblock \showarticletitle{Critical thinking: A statement of expert consensus for purposes of educational assessment and instruction (The Delphi Report)}.
\newblock  (\bibinfo{year}{1990}).
\newblock


\bibitem[Facione et~al\mbox{.}(2011)]%
        {facione2011critical}
\bibfield{author}{\bibinfo{person}{Peter~A Facione} {et~al\mbox{.}}} \bibinfo{year}{2011}\natexlab{}.
\newblock \showarticletitle{Critical thinking: What it is and why it counts}.
\newblock \bibinfo{journal}{\emph{Insight assessment}} \bibinfo{volume}{1}, \bibinfo{number}{1} (\bibinfo{year}{2011}), \bibinfo{pages}{1--23}.
\newblock


\bibitem[Foster(2004)]%
        {foster2004nonlinear}
\bibfield{author}{\bibinfo{person}{Allen Foster}.} \bibinfo{year}{2004}\natexlab{}.
\newblock \showarticletitle{A nonlinear model of information-seeking behavior}.
\newblock \bibinfo{journal}{\emph{Journal of the American society for information science and technology}} \bibinfo{volume}{55}, \bibinfo{number}{3} (\bibinfo{year}{2004}), \bibinfo{pages}{228--237}.
\newblock


\bibitem[Gonzalez et~al\mbox{.}(2024)]%
        {gonzalez2024collaborative}
\bibfield{author}{\bibinfo{person}{Gabriel~Enrique Gonzalez}, \bibinfo{person}{Dario Andres~Silva Moran}, \bibinfo{person}{Stephanie Houde}, \bibinfo{person}{Jessica He}, \bibinfo{person}{Steven~I Ross}, \bibinfo{person}{Michael~J Muller}, \bibinfo{person}{Siya Kunde}, {and} \bibinfo{person}{Justin~D Weisz}.} \bibinfo{year}{2024}\natexlab{}.
\newblock \showarticletitle{Collaborative Canvas: A Tool for Exploring LLM Use in Group Ideation Tasks.}. In \bibinfo{booktitle}{\emph{IUI Workshops}}.
\newblock


\bibitem[Gopinathan et~al\mbox{.}(2017)]%
        {gopinathan2017accumulation}
\bibfield{author}{\bibinfo{person}{Girija Gopinathan}, \bibinfo{person}{Laurie-Ann~M Hellsten}, {and} \bibinfo{person}{Lynnette Leeseberg~Stamler}.} \bibinfo{year}{2017}\natexlab{}.
\newblock \showarticletitle{Accumulation of Content Validation Evidence for the Critical Thinking Self-Assessment Scale.}
\newblock \bibinfo{journal}{\emph{Journal of nursing measurement}} \bibinfo{volume}{25}, \bibinfo{number}{1} (\bibinfo{year}{2017}).
\newblock


\bibitem[Gratch et~al\mbox{.}(2014)]%
        {gratch2014s}
\bibfield{author}{\bibinfo{person}{Jonathan Gratch}, \bibinfo{person}{Gale~M Lucas}, \bibinfo{person}{Aisha~Aisha King}, {and} \bibinfo{person}{Louis-Philippe Morency}.} \bibinfo{year}{2014}\natexlab{}.
\newblock \showarticletitle{It's only a computer: the impact of human-agent interaction in clinical interviews}. In \bibinfo{booktitle}{\emph{Proceedings of the 2014 international conference on Autonomous agents and multi-agent systems}}. \bibinfo{pages}{85--92}.
\newblock


\bibitem[Gu and Krenn(2024)]%
        {gu2024generation}
\bibfield{author}{\bibinfo{person}{Xuemei Gu} {and} \bibinfo{person}{Mario Krenn}.} \bibinfo{year}{2024}\natexlab{}.
\newblock \showarticletitle{Generation and human-expert evaluation of interesting research ideas using knowledge graphs and large language models}.
\newblock \bibinfo{journal}{\emph{arXiv preprint arXiv:2405.17044}} (\bibinfo{year}{2024}).
\newblock


\bibitem[Guo et~al\mbox{.}(2024)]%
        {guo2024exploring}
\bibfield{author}{\bibinfo{person}{Alicia Guo}, \bibinfo{person}{Pat Pataranutaporn}, {and} \bibinfo{person}{Pattie Maes}.} \bibinfo{year}{2024}\natexlab{}.
\newblock \showarticletitle{Exploring the Impact of AI Value Alignment in Collaborative Ideation: Effects on Perception, Ownership, and Output}. In \bibinfo{booktitle}{\emph{Extended Abstracts of the CHI Conference on Human Factors in Computing Systems}}. \bibinfo{pages}{1--11}.
\newblock


\bibitem[Ha et~al\mbox{.}(2024)]%
        {ha2024clochat}
\bibfield{author}{\bibinfo{person}{Juhye Ha}, \bibinfo{person}{Hyeon Jeon}, \bibinfo{person}{Daeun Han}, \bibinfo{person}{Jinwook Seo}, {and} \bibinfo{person}{Changhoon Oh}.} \bibinfo{year}{2024}\natexlab{}.
\newblock \showarticletitle{CloChat: Understanding How People Customize, Interact, and Experience Personas in Large Language Models}. In \bibinfo{booktitle}{\emph{Proceedings of the CHI Conference on Human Factors in Computing Systems}}. \bibinfo{pages}{1--24}.
\newblock


\bibitem[Halpern(1998)]%
        {halpern1998teaching}
\bibfield{author}{\bibinfo{person}{Diane~F Halpern}.} \bibinfo{year}{1998}\natexlab{}.
\newblock \showarticletitle{Teaching critical thinking for transfer across domains: Disposition, skills, structure training, and metacognitive monitoring.}
\newblock \bibinfo{journal}{\emph{American psychologist}} \bibinfo{volume}{53}, \bibinfo{number}{4} (\bibinfo{year}{1998}), \bibinfo{pages}{449}.
\newblock


\bibitem[Han et~al\mbox{.}(2018)]%
        {han2018service}
\bibfield{author}{\bibinfo{person}{Nayoung Han}, \bibinfo{person}{Seung~Hee Han}, \bibinfo{person}{Hyuneun Chu}, \bibinfo{person}{Jaehyun Kim}, \bibinfo{person}{Ki~Yon Rhew}, \bibinfo{person}{Jeong-Hyun Yoon}, \bibinfo{person}{Nam~Kyung Je}, \bibinfo{person}{Sandy~Jeong Rhie}, \bibinfo{person}{Eunhee Ji}, \bibinfo{person}{Euni Lee}, {et~al\mbox{.}}} \bibinfo{year}{2018}\natexlab{}.
\newblock \showarticletitle{Service design oriented multidisciplinary collaborative team care service model development for resolving drug related problems}.
\newblock \bibinfo{journal}{\emph{PloS one}} \bibinfo{volume}{13}, \bibinfo{number}{9} (\bibinfo{year}{2018}), \bibinfo{pages}{e0201705}.
\newblock


\bibitem[Hanke(2006)]%
        {hanke2006team}
\bibfield{author}{\bibinfo{person}{Ralph~CM Hanke}.} \bibinfo{year}{2006}\natexlab{}.
\newblock \bibinfo{booktitle}{\emph{Team creativity: A process model}}.
\newblock \bibinfo{publisher}{The Pennsylvania State University}.
\newblock


\bibitem[Harmon-Jones and Mills(2019)]%
        {harmon2019introduction}
\bibfield{author}{\bibinfo{person}{Eddie Harmon-Jones} {and} \bibinfo{person}{Judson Mills}.} \bibinfo{year}{2019}\natexlab{}.
\newblock \showarticletitle{An introduction to cognitive dissonance theory and an overview of current perspectives on the theory.}
\newblock  (\bibinfo{year}{2019}).
\newblock


\bibitem[Harrison and Klein(2007)]%
        {harrison2007s}
\bibfield{author}{\bibinfo{person}{David~A Harrison} {and} \bibinfo{person}{Katherine~J Klein}.} \bibinfo{year}{2007}\natexlab{}.
\newblock \showarticletitle{What's the difference? Diversity constructs as separation, variety, or disparity in organizations}.
\newblock \bibinfo{journal}{\emph{Academy of management review}} \bibinfo{volume}{32}, \bibinfo{number}{4} (\bibinfo{year}{2007}), \bibinfo{pages}{1199--1228}.
\newblock


\bibitem[Hendriks et~al\mbox{.}(2020)]%
        {Hendriks2020ConstraintsAA}
\bibfield{author}{\bibinfo{person}{Friederike Hendriks}, \bibinfo{person}{Elisabeth Mayweg-Paus}, \bibinfo{person}{Mark Felton}, \bibinfo{person}{Kalypso Iordanou}, \bibinfo{person}{Regina Jucks}, {and} \bibinfo{person}{Maria Zimmermann}.} \bibinfo{year}{2020}\natexlab{}.
\newblock \showarticletitle{Constraints and Affordances of Online Engagement With Scientific Information—A Literature Review}.
\newblock \bibinfo{journal}{\emph{Frontiers in Psychology}}  \bibinfo{volume}{11} (\bibinfo{year}{2020}).
\newblock
\urldef\tempurl%
\url{https://api.semanticscholar.org/CorpusID:227496913}
\showURL{%
\tempurl}


\bibitem[Heyman et~al\mbox{.}(2024)]%
        {heyman2024supermind}
\bibfield{author}{\bibinfo{person}{Jennifer~L Heyman}, \bibinfo{person}{Steven~R Rick}, \bibinfo{person}{Gianni Giacomelli}, \bibinfo{person}{Haoran Wen}, \bibinfo{person}{Robert Laubacher}, \bibinfo{person}{Nancy Taubenslag}, \bibinfo{person}{Max Knicker}, \bibinfo{person}{Younes Jeddi}, \bibinfo{person}{Pranav Ragupathy}, \bibinfo{person}{Jared Curhan}, {et~al\mbox{.}}} \bibinfo{year}{2024}\natexlab{}.
\newblock \showarticletitle{Supermind Ideator: How Scaffolding Human-AI Collaboration Can Increase Creativity}. In \bibinfo{booktitle}{\emph{Proceedings of the ACM Collective Intelligence Conference}}. \bibinfo{pages}{18--28}.
\newblock


\bibitem[Hoever et~al\mbox{.}(2012)]%
        {hoever2012fostering}
\bibfield{author}{\bibinfo{person}{Inga~J Hoever}, \bibinfo{person}{Daan Van~Knippenberg}, \bibinfo{person}{Wendy~P Van~Ginkel}, {and} \bibinfo{person}{Harry~G Barkema}.} \bibinfo{year}{2012}\natexlab{}.
\newblock \showarticletitle{Fostering team creativity: perspective taking as key to unlocking diversity's potential.}
\newblock \bibinfo{journal}{\emph{Journal of applied psychology}} \bibinfo{volume}{97}, \bibinfo{number}{5} (\bibinfo{year}{2012}), \bibinfo{pages}{982}.
\newblock


\bibitem[Hu et~al\mbox{.}(2010)]%
        {hu2010creative}
\bibfield{author}{\bibinfo{person}{Weiping Hu}, \bibinfo{person}{Quan~Zhen Shi}, \bibinfo{person}{Qin Han}, \bibinfo{person}{Xingqi Wang}, {and} \bibinfo{person}{Philip Adey}.} \bibinfo{year}{2010}\natexlab{}.
\newblock \showarticletitle{Creative scientific problem finding and its developmental trend}.
\newblock \bibinfo{journal}{\emph{Creativity Research Journal}} \bibinfo{volume}{22}, \bibinfo{number}{1} (\bibinfo{year}{2010}), \bibinfo{pages}{46--52}.
\newblock


\bibitem[Ju{\'a}rez~Ramos(2018)]%
        {juarez2018analyzing}
\bibfield{author}{\bibinfo{person}{Ver{\'o}nica Ju{\'a}rez~Ramos}.} \bibinfo{year}{2018}\natexlab{}.
\newblock \bibinfo{booktitle}{\emph{Analyzing the role of cognitive biases in the decision-making process}}.
\newblock \bibinfo{publisher}{IGI Global}.
\newblock


\bibitem[Kinney et~al\mbox{.}(2023)]%
        {kinney2023semantic}
\bibfield{author}{\bibinfo{person}{Rodney Kinney}, \bibinfo{person}{Chloe Anastasiades}, \bibinfo{person}{Russell Authur}, \bibinfo{person}{Iz Beltagy}, \bibinfo{person}{Jonathan Bragg}, \bibinfo{person}{Alexandra Buraczynski}, \bibinfo{person}{Isabel Cachola}, \bibinfo{person}{Stefan Candra}, \bibinfo{person}{Yoganand Chandrasekhar}, \bibinfo{person}{Arman Cohan}, {et~al\mbox{.}}} \bibinfo{year}{2023}\natexlab{}.
\newblock \showarticletitle{The semantic scholar open data platform}.
\newblock \bibinfo{journal}{\emph{arXiv preprint arXiv:2301.10140}} (\bibinfo{year}{2023}).
\newblock


\bibitem[Kohn and Smith(2011)]%
        {kohn2011collaborative}
\bibfield{author}{\bibinfo{person}{Nicholas~W Kohn} {and} \bibinfo{person}{Steven~M Smith}.} \bibinfo{year}{2011}\natexlab{}.
\newblock \showarticletitle{Collaborative fixation: Effects of others' ideas on brainstorming}.
\newblock \bibinfo{journal}{\emph{Applied Cognitive Psychology}} \bibinfo{volume}{25}, \bibinfo{number}{3} (\bibinfo{year}{2011}), \bibinfo{pages}{359--371}.
\newblock


\bibitem[Korinek(2023)]%
        {korinek2023language}
\bibfield{author}{\bibinfo{person}{Anton Korinek}.} \bibinfo{year}{2023}\natexlab{}.
\newblock \bibinfo{booktitle}{\emph{Language models and cognitive automation for economic research}}.
\newblock \bibinfo{type}{{T}echnical {R}eport}. \bibinfo{institution}{National Bureau of Economic Research}.
\newblock


\bibitem[Kornish and Ulrich(2011)]%
        {kornish2011opportunity}
\bibfield{author}{\bibinfo{person}{Laura~J Kornish} {and} \bibinfo{person}{Karl~T Ulrich}.} \bibinfo{year}{2011}\natexlab{}.
\newblock \showarticletitle{Opportunity spaces in innovation: Empirical analysis of large samples of ideas}.
\newblock \bibinfo{journal}{\emph{Management science}} \bibinfo{volume}{57}, \bibinfo{number}{1} (\bibinfo{year}{2011}), \bibinfo{pages}{107--128}.
\newblock


\bibitem[Kuhn(2019)]%
        {Kuhn2019CriticalTA}
\bibfield{author}{\bibinfo{person}{Deanna Kuhn}.} \bibinfo{year}{2019}\natexlab{}.
\newblock \showarticletitle{Critical Thinking as Discourse}.
\newblock \bibinfo{journal}{\emph{Human Development}}  \bibinfo{volume}{62} (\bibinfo{year}{2019}), \bibinfo{pages}{146 -- 164}.
\newblock
\urldef\tempurl%
\url{https://api.semanticscholar.org/CorpusID:197704266}
\showURL{%
\tempurl}


\bibitem[Kumar et~al\mbox{.}(2024)]%
        {kumar2024supporting}
\bibfield{author}{\bibinfo{person}{Harsh Kumar}, \bibinfo{person}{Ruiwei Xiao}, \bibinfo{person}{Benjamin Lawson}, \bibinfo{person}{Ilya Musabirov}, \bibinfo{person}{Jiakai Shi}, \bibinfo{person}{Xinyuan Wang}, \bibinfo{person}{Huayin Luo}, \bibinfo{person}{Joseph~Jay Williams}, \bibinfo{person}{Anna~N Rafferty}, \bibinfo{person}{John Stamper}, {et~al\mbox{.}}} \bibinfo{year}{2024}\natexlab{}.
\newblock \showarticletitle{Supporting self-reflection at scale with large language models: Insights from randomized field experiments in classrooms}. In \bibinfo{booktitle}{\emph{Proceedings of the eleventh ACM conference on learning@ scale}}. \bibinfo{pages}{86--97}.
\newblock


\bibitem[Kurtzberg and Amabile(2001)]%
        {kurtzberg2001guilford}
\bibfield{author}{\bibinfo{person}{Terri~R Kurtzberg} {and} \bibinfo{person}{Teresa~M Amabile}.} \bibinfo{year}{2001}\natexlab{}.
\newblock \showarticletitle{From Guilford to creative synergy: Opening the black box of team-level creativity}.
\newblock \bibinfo{journal}{\emph{Creativity Research Journal}} \bibinfo{volume}{13}, \bibinfo{number}{3-4} (\bibinfo{year}{2001}), \bibinfo{pages}{285--294}.
\newblock


\bibitem[Lai et~al\mbox{.}(2011)]%
        {Lai2011CriticalTA}
\bibfield{author}{\bibinfo{person}{Emily~R Lai}, \bibinfo{person}{Michael Bay-Borelli}, \bibinfo{person}{Robert Kirkpatrick}, \bibinfo{person}{Anli Lin}, {and} \bibinfo{person}{Changjiang Wang}.} \bibinfo{year}{2011}\natexlab{}.
\newblock \showarticletitle{Critical Thinking: A Literature Review Research Report}.
\newblock
\urldef\tempurl%
\url{https://api.semanticscholar.org/CorpusID:55328336}
\showURL{%
\tempurl}


\bibitem[Langley(1987)]%
        {langley1987scientific}
\bibfield{author}{\bibinfo{person}{Pat Langley}.} \bibinfo{year}{1987}\natexlab{}.
\newblock \bibinfo{booktitle}{\emph{Scientific discovery: Computational explorations of the creative processes}}.
\newblock \bibinfo{publisher}{MIT press}.
\newblock


\bibitem[Lee et~al\mbox{.}(2024)]%
        {lee2024conversational}
\bibfield{author}{\bibinfo{person}{Soohwan Lee}, \bibinfo{person}{Seoyeong Hwang}, {and} \bibinfo{person}{Kyungho Lee}.} \bibinfo{year}{2024}\natexlab{}.
\newblock \showarticletitle{Conversational Agents as Catalysts for Critical Thinking: Challenging Design Fixation in Group Design}.
\newblock \bibinfo{journal}{\emph{arXiv preprint arXiv:2406.11125}} (\bibinfo{year}{2024}).
\newblock


\bibitem[Lee et~al\mbox{.}(2020)]%
        {lee2020hear}
\bibfield{author}{\bibinfo{person}{Yi-Chieh Lee}, \bibinfo{person}{Naomi Yamashita}, \bibinfo{person}{Yun Huang}, {and} \bibinfo{person}{Wai Fu}.} \bibinfo{year}{2020}\natexlab{}.
\newblock \showarticletitle{" I hear you, I feel you": encouraging deep self-disclosure through a chatbot}. In \bibinfo{booktitle}{\emph{Proceedings of the 2020 CHI conference on human factors in computing systems}}. \bibinfo{pages}{1--12}.
\newblock


\bibitem[Lewis and Herndon(2011)]%
        {lewis2011transactive}
\bibfield{author}{\bibinfo{person}{Kyle Lewis} {and} \bibinfo{person}{Benjamin Herndon}.} \bibinfo{year}{2011}\natexlab{}.
\newblock \showarticletitle{Transactive memory systems: Current issues and future research directions}.
\newblock \bibinfo{journal}{\emph{Organization science}} \bibinfo{volume}{22}, \bibinfo{number}{5} (\bibinfo{year}{2011}), \bibinfo{pages}{1254--1265}.
\newblock


\bibitem[Li et~al\mbox{.}(2024a)]%
        {li2024user}
\bibfield{author}{\bibinfo{person}{Jie Li}, \bibinfo{person}{Hancheng Cao}, \bibinfo{person}{Laura Lin}, \bibinfo{person}{Youyang Hou}, \bibinfo{person}{Ruihao Zhu}, {and} \bibinfo{person}{Abdallah El~Ali}.} \bibinfo{year}{2024}\natexlab{a}.
\newblock \showarticletitle{User experience design professionals’ perceptions of generative artificial intelligence}. In \bibinfo{booktitle}{\emph{Proceedings of the CHI Conference on Human Factors in Computing Systems}}. \bibinfo{pages}{1--18}.
\newblock


\bibitem[Li et~al\mbox{.}(2024b)]%
        {li2024econagent}
\bibfield{author}{\bibinfo{person}{Nian Li}, \bibinfo{person}{Chen Gao}, \bibinfo{person}{Mingyu Li}, \bibinfo{person}{Yong Li}, {and} \bibinfo{person}{Qingmin Liao}.} \bibinfo{year}{2024}\natexlab{b}.
\newblock \showarticletitle{EconAgent: Large Language Model-Empowered Agents for Simulating Macroeconomic Activities}. In \bibinfo{booktitle}{\emph{Proceedings of the 62nd Annual Meeting of the Association for Computational Linguistics (Volume 1: Long Papers)}}. \bibinfo{pages}{15523--15536}.
\newblock


\bibitem[Li et~al\mbox{.}(2023)]%
        {li2023metaagents}
\bibfield{author}{\bibinfo{person}{Yuan Li}, \bibinfo{person}{Yixuan Zhang}, {and} \bibinfo{person}{Lichao Sun}.} \bibinfo{year}{2023}\natexlab{}.
\newblock \showarticletitle{Metaagents: Simulating interactions of human behaviors for llm-based task-oriented coordination via collaborative generative agents}.
\newblock \bibinfo{journal}{\emph{arXiv preprint arXiv:2310.06500}} (\bibinfo{year}{2023}).
\newblock


\bibitem[Liu et~al\mbox{.}(2024)]%
        {liu2024ai}
\bibfield{author}{\bibinfo{person}{Yiren Liu}, \bibinfo{person}{Si Chen}, \bibinfo{person}{Haocong Cheng}, \bibinfo{person}{Mengxia Yu}, \bibinfo{person}{Xiao Ran}, \bibinfo{person}{Andrew Mo}, \bibinfo{person}{Yiliu Tang}, {and} \bibinfo{person}{Yun Huang}.} \bibinfo{year}{2024}\natexlab{}.
\newblock \showarticletitle{How ai processing delays foster creativity: Exploring research question co-creation with an llm-based agent}. In \bibinfo{booktitle}{\emph{Proceedings of the CHI Conference on Human Factors in Computing Systems}}. \bibinfo{pages}{1--25}.
\newblock


\bibitem[Louie et~al\mbox{.}(2020)]%
        {louie2020novice}
\bibfield{author}{\bibinfo{person}{Ryan Louie}, \bibinfo{person}{Andy Coenen}, \bibinfo{person}{Cheng~Zhi Huang}, \bibinfo{person}{Michael Terry}, {and} \bibinfo{person}{Carrie~J Cai}.} \bibinfo{year}{2020}\natexlab{}.
\newblock \showarticletitle{Novice-AI music co-creation via AI-steering tools for deep generative models}. In \bibinfo{booktitle}{\emph{Proceedings of the 2020 CHI conference on human factors in computing systems}}. \bibinfo{pages}{1--13}.
\newblock


\bibitem[Lozano et~al\mbox{.}(2023)]%
        {lozano2023clinfo}
\bibfield{author}{\bibinfo{person}{Alejandro Lozano}, \bibinfo{person}{Scott~L Fleming}, \bibinfo{person}{Chia-Chun Chiang}, {and} \bibinfo{person}{Nigam Shah}.} \bibinfo{year}{2023}\natexlab{}.
\newblock \showarticletitle{Clinfo. ai: An open-source retrieval-augmented large language model system for answering medical questions using scientific literature}. In \bibinfo{booktitle}{\emph{PACIFIC SYMPOSIUM ON BIOCOMPUTING 2024}}. World Scientific, \bibinfo{pages}{8--23}.
\newblock


\bibitem[MacLeod(2018)]%
        {macleod2018makes}
\bibfield{author}{\bibinfo{person}{Miles MacLeod}.} \bibinfo{year}{2018}\natexlab{}.
\newblock \showarticletitle{What makes interdisciplinarity difficult? Some consequences of domain specificity in interdisciplinary practice}.
\newblock \bibinfo{journal}{\emph{Synthese}} \bibinfo{volume}{195}, \bibinfo{number}{2} (\bibinfo{year}{2018}), \bibinfo{pages}{697--720}.
\newblock


\bibitem[Man et~al\mbox{.}(2018)]%
        {man2018understanding}
\bibfield{author}{\bibinfo{person}{Deliang Man}, \bibinfo{person}{Yiqin Xu}, {and} \bibinfo{person}{John~Mitchell O’Toole}.} \bibinfo{year}{2018}\natexlab{}.
\newblock \showarticletitle{Understanding autonomous peer feedback practices among postgraduate students: A case study in a Chinese university}.
\newblock \bibinfo{journal}{\emph{Assessment \& Evaluation in Higher Education}} \bibinfo{volume}{43}, \bibinfo{number}{4} (\bibinfo{year}{2018}), \bibinfo{pages}{527--536}.
\newblock


\bibitem[Mayo et~al\mbox{.}(2017)]%
        {mayo2017metatheoretical}
\bibfield{author}{\bibinfo{person}{Margarita Mayo}, \bibinfo{person}{Maria Kakarika}, \bibinfo{person}{Charalampos Mainemelis}, {and} \bibinfo{person}{Nicolas~Till Deuschel}.} \bibinfo{year}{2017}\natexlab{}.
\newblock \showarticletitle{A metatheoretical framework of diversity in teams}.
\newblock \bibinfo{journal}{\emph{Human Relations}} \bibinfo{volume}{70}, \bibinfo{number}{8} (\bibinfo{year}{2017}), \bibinfo{pages}{911--939}.
\newblock


\bibitem[Memmert and Tavanapour(2023)]%
        {memmert2023towards}
\bibfield{author}{\bibinfo{person}{Lucas Memmert} {and} \bibinfo{person}{Navid Tavanapour}.} \bibinfo{year}{2023}\natexlab{}.
\newblock \showarticletitle{Towards human-AI-collaboration in brainstorming: Empirical insights into the perception of working with a generative AI}.
\newblock  (\bibinfo{year}{2023}).
\newblock


\bibitem[Milgram(1963)]%
        {milgram1963behavioral}
\bibfield{author}{\bibinfo{person}{Stanley Milgram}.} \bibinfo{year}{1963}\natexlab{}.
\newblock \showarticletitle{Behavioral study of obedience.}
\newblock \bibinfo{journal}{\emph{The Journal of abnormal and social psychology}} \bibinfo{volume}{67}, \bibinfo{number}{4} (\bibinfo{year}{1963}), \bibinfo{pages}{371}.
\newblock


\bibitem[Mohammadi(2024)]%
        {mohammadi2024creativity}
\bibfield{author}{\bibinfo{person}{Behnam Mohammadi}.} \bibinfo{year}{2024}\natexlab{}.
\newblock \showarticletitle{Creativity Has Left the Chat: The Price of Debiasing Language Models}.
\newblock \bibinfo{journal}{\emph{arXiv preprint arXiv:2406.05587}} (\bibinfo{year}{2024}).
\newblock


\bibitem[Newby(2011)]%
        {newby2011entering}
\bibfield{author}{\bibinfo{person}{Jill Newby}.} \bibinfo{year}{2011}\natexlab{}.
\newblock \showarticletitle{Entering unfamiliar territory: Building an information literacy course for graduate students in interdisciplinary areas}.
\newblock \bibinfo{journal}{\emph{Reference \& User Services Quarterly}} \bibinfo{volume}{50}, \bibinfo{number}{3} (\bibinfo{year}{2011}), \bibinfo{pages}{224--229}.
\newblock


\bibitem[Newton(2010)]%
        {newton2010creativity}
\bibfield{author}{\bibinfo{person}{Lynn~D Newton}.} \bibinfo{year}{2010}\natexlab{}.
\newblock \showarticletitle{Creativity in science and science education: A response to Ghassib}.
\newblock \bibinfo{journal}{\emph{Gifted and Talented International}} \bibinfo{volume}{25}, \bibinfo{number}{1} (\bibinfo{year}{2010}), \bibinfo{pages}{105--108}.
\newblock


\bibitem[Nigam et~al\mbox{.}(2024)]%
        {nigam2024acceleron}
\bibfield{author}{\bibinfo{person}{Harshit Nigam}, \bibinfo{person}{Manasi Patwardhan}, \bibinfo{person}{Lovekesh Vig}, {and} \bibinfo{person}{Gautam Shroff}.} \bibinfo{year}{2024}\natexlab{}.
\newblock \showarticletitle{Acceleron: A Tool to Accelerate Research Ideation}.
\newblock \bibinfo{journal}{\emph{arXiv preprint arXiv:2403.04382}} (\bibinfo{year}{2024}).
\newblock


\bibitem[Norman et~al\mbox{.}(2021)]%
        {norman2021cloudbank}
\bibfield{author}{\bibinfo{person}{Michael Norman}, \bibinfo{person}{Vince Kellen}, \bibinfo{person}{Shava Smallen}, \bibinfo{person}{Brian DeMeulle}, \bibinfo{person}{Shawn Strande}, \bibinfo{person}{Ed Lazowska}, \bibinfo{person}{Naomi Alterman}, \bibinfo{person}{Rob Fatland}, \bibinfo{person}{Sarah Stone}, \bibinfo{person}{Amanda Tan}, {et~al\mbox{.}}} \bibinfo{year}{2021}\natexlab{}.
\newblock \showarticletitle{CloudBank: Managed Services to Simplify Cloud Access for Computer Science Research and Education}.
\newblock In \bibinfo{booktitle}{\emph{Practice and Experience in Advanced Research Computing}}. \bibinfo{pages}{1--4}.
\newblock


\bibitem[Nussbaum et~al\mbox{.}(2012)]%
        {Nussbaum2012TheTF}
\bibfield{author}{\bibinfo{person}{E.~Michael Nussbaum}, \bibinfo{person}{Gale~M. Sinatra}, {and} \bibinfo{person}{Marissa~C. Owens}.} \bibinfo{year}{2012}\natexlab{}.
\newblock \showarticletitle{The Two Faces of Scientific Argumentation: Applications to Global Climate Change}.
\newblock
\urldef\tempurl%
\url{https://api.semanticscholar.org/CorpusID:150852554}
\showURL{%
\tempurl}


\bibitem[Palmer et~al\mbox{.}(2009)]%
        {palmer2009scholarly}
\bibfield{author}{\bibinfo{person}{Carole~L Palmer}, \bibinfo{person}{Lauren~C Teffeau}, {and} \bibinfo{person}{Carrie~M Pirmann}.} \bibinfo{year}{2009}\natexlab{}.
\newblock \showarticletitle{Scholarly information practices in the online environment}.
\newblock \bibinfo{journal}{\emph{Report commissioned by OCLC Research. Published online at: www. oclc. org/programs/publications/reports/2009-02. pdf}} (\bibinfo{year}{2009}).
\newblock


\bibitem[Park et~al\mbox{.}(2023)]%
        {park2023generative}
\bibfield{author}{\bibinfo{person}{Joon~Sung Park}, \bibinfo{person}{Joseph O'Brien}, \bibinfo{person}{Carrie~Jun Cai}, \bibinfo{person}{Meredith~Ringel Morris}, \bibinfo{person}{Percy Liang}, {and} \bibinfo{person}{Michael~S Bernstein}.} \bibinfo{year}{2023}\natexlab{}.
\newblock \showarticletitle{Generative agents: Interactive simulacra of human behavior}. In \bibinfo{booktitle}{\emph{Proceedings of the 36th annual acm symposium on user interface software and technology}}. \bibinfo{pages}{1--22}.
\newblock


\bibitem[Park et~al\mbox{.}(2022)]%
        {park2022social}
\bibfield{author}{\bibinfo{person}{Joon~Sung Park}, \bibinfo{person}{Lindsay Popowski}, \bibinfo{person}{Carrie Cai}, \bibinfo{person}{Meredith~Ringel Morris}, \bibinfo{person}{Percy Liang}, {and} \bibinfo{person}{Michael~S Bernstein}.} \bibinfo{year}{2022}\natexlab{}.
\newblock \showarticletitle{Social simulacra: Creating populated prototypes for social computing systems}. In \bibinfo{booktitle}{\emph{Proceedings of the 35th Annual ACM Symposium on User Interface Software and Technology}}. \bibinfo{pages}{1--18}.
\newblock


\bibitem[Petridis et~al\mbox{.}(2023)]%
        {petridis2023anglekindling}
\bibfield{author}{\bibinfo{person}{Savvas Petridis}, \bibinfo{person}{Nicholas Diakopoulos}, \bibinfo{person}{Kevin Crowston}, \bibinfo{person}{Mark Hansen}, \bibinfo{person}{Keren Henderson}, \bibinfo{person}{Stan Jastrzebski}, \bibinfo{person}{Jeffrey~V Nickerson}, {and} \bibinfo{person}{Lydia~B Chilton}.} \bibinfo{year}{2023}\natexlab{}.
\newblock \showarticletitle{Anglekindling: Supporting journalistic angle ideation with large language models}. In \bibinfo{booktitle}{\emph{Proceedings of the 2023 CHI conference on human factors in computing systems}}. \bibinfo{pages}{1--16}.
\newblock


\bibitem[Pfautz et~al\mbox{.}(2015)]%
        {pfautz2015general}
\bibfield{author}{\bibinfo{person}{Stacy~Lovell Pfautz}, \bibinfo{person}{Gabriel Ganberg}, \bibinfo{person}{Adam Fouse}, {and} \bibinfo{person}{Nathan Schurr}.} \bibinfo{year}{2015}\natexlab{}.
\newblock \showarticletitle{A general context-aware framework for improved human-system interactions}.
\newblock \bibinfo{journal}{\emph{Ai Magazine}} \bibinfo{volume}{36}, \bibinfo{number}{2} (\bibinfo{year}{2015}), \bibinfo{pages}{42--49}.
\newblock


\bibitem[Pruitt and Adlin(2010)]%
        {pruitt2010persona}
\bibfield{author}{\bibinfo{person}{John Pruitt} {and} \bibinfo{person}{Tamara Adlin}.} \bibinfo{year}{2010}\natexlab{}.
\newblock \bibinfo{booktitle}{\emph{The persona lifecycle: keeping people in mind throughout product design}}.
\newblock \bibinfo{publisher}{Elsevier}.
\newblock


\bibitem[Qian et~al\mbox{.}(2024)]%
        {qian2024chatdev}
\bibfield{author}{\bibinfo{person}{Chen Qian}, \bibinfo{person}{Wei Liu}, \bibinfo{person}{Hongzhang Liu}, \bibinfo{person}{Nuo Chen}, \bibinfo{person}{Yufan Dang}, \bibinfo{person}{Jiahao Li}, \bibinfo{person}{Cheng Yang}, \bibinfo{person}{Weize Chen}, \bibinfo{person}{Yusheng Su}, \bibinfo{person}{Xin Cong}, {et~al\mbox{.}}} \bibinfo{year}{2024}\natexlab{}.
\newblock \showarticletitle{Chatdev: Communicative agents for software development}. In \bibinfo{booktitle}{\emph{Proceedings of the 62nd Annual Meeting of the Association for Computational Linguistics (Volume 1: Long Papers)}}. \bibinfo{pages}{15174--15186}.
\newblock


\bibitem[Rastogi et~al\mbox{.}(2022)]%
        {rastogi2022deciding}
\bibfield{author}{\bibinfo{person}{Charvi Rastogi}, \bibinfo{person}{Yunfeng Zhang}, \bibinfo{person}{Dennis Wei}, \bibinfo{person}{Kush~R Varshney}, \bibinfo{person}{Amit Dhurandhar}, {and} \bibinfo{person}{Richard Tomsett}.} \bibinfo{year}{2022}\natexlab{}.
\newblock \showarticletitle{Deciding fast and slow: The role of cognitive biases in ai-assisted decision-making}.
\newblock \bibinfo{journal}{\emph{Proceedings of the ACM on Human-Computer Interaction}} \bibinfo{volume}{6}, \bibinfo{number}{CSCW1} (\bibinfo{year}{2022}), \bibinfo{pages}{1--22}.
\newblock


\bibitem[Saeed et~al\mbox{.}(2021)]%
        {saeed2021integrating}
\bibfield{author}{\bibinfo{person}{Murad~Abdu Saeed}, \bibinfo{person}{Arif~Ahmed Mohammed H. Al-Ahdal}, {and} \bibinfo{person}{Huda~Suleiman Al~Qunayeer}.} \bibinfo{year}{2021}\natexlab{}.
\newblock \showarticletitle{Integrating research proposal writing into a postgraduate research method course: what does it tell us?}
\newblock \bibinfo{journal}{\emph{International Journal of Research \& Method in Education}} \bibinfo{volume}{44}, \bibinfo{number}{3} (\bibinfo{year}{2021}), \bibinfo{pages}{303--318}.
\newblock


\bibitem[Sak and Ayas(2013)]%
        {sak2013creative}
\bibfield{author}{\bibinfo{person}{Ugur Sak} {and} \bibinfo{person}{M~Bahadir Ayas}.} \bibinfo{year}{2013}\natexlab{}.
\newblock \showarticletitle{Creative Scientific Ability Test (C-SAT): A new measure of scientific creativity}.
\newblock \bibinfo{journal}{\emph{Psychological Test and Assessment Modeling}} \bibinfo{volume}{55}, \bibinfo{number}{3} (\bibinfo{year}{2013}), \bibinfo{pages}{316--329}.
\newblock


\bibitem[Salminen et~al\mbox{.}(2020)]%
        {salminen2020effect}
\bibfield{author}{\bibinfo{person}{Joni Salminen}, \bibinfo{person}{Soon-gyo Jung}, \bibinfo{person}{Jo{\~a}o~M Santos}, \bibinfo{person}{Shammur Chowdhury}, {and} \bibinfo{person}{Bernard~J Jansen}.} \bibinfo{year}{2020}\natexlab{}.
\newblock \showarticletitle{The effect of experience on persona perceptions}. In \bibinfo{booktitle}{\emph{Extended Abstracts of the 2020 CHI Conference on Human Factors in Computing Systems}}. \bibinfo{pages}{1--9}.
\newblock


\bibitem[Salminen et~al\mbox{.}(2018)]%
        {salminen2018persona}
\bibfield{author}{\bibinfo{person}{Joni Salminen}, \bibinfo{person}{Haewoon Kwak}, \bibinfo{person}{Jo{\~a}o~M Santos}, \bibinfo{person}{Soon-Gyo Jung}, \bibinfo{person}{Jisun An}, {and} \bibinfo{person}{Bernard~J Jansen}.} \bibinfo{year}{2018}\natexlab{}.
\newblock \showarticletitle{Persona perception scale: developing and validating an instrument for human-like representations of data}. In \bibinfo{booktitle}{\emph{Extended Abstracts of the 2018 CHI Conference on Human Factors in Computing Systems}}. \bibinfo{pages}{1--6}.
\newblock


\bibitem[Sandholm et~al\mbox{.}(2024)]%
        {sandholm2024randomness}
\bibfield{author}{\bibinfo{person}{Thomas Sandholm}, \bibinfo{person}{Sayandev Mukherjee}, {and} \bibinfo{person}{Bernardo~A Huberman}.} \bibinfo{year}{2024}\natexlab{}.
\newblock \showarticletitle{Randomness Is All You Need: Semantic Traversal of Problem-Solution Spaces with Large Language Models}.
\newblock \bibinfo{journal}{\emph{arXiv preprint arXiv:2402.06053}} (\bibinfo{year}{2024}).
\newblock


\bibitem[Santos(2017)]%
        {santos2017role}
\bibfield{author}{\bibinfo{person}{Luis~Fernando Santos}.} \bibinfo{year}{2017}\natexlab{}.
\newblock \showarticletitle{The role of critical thinking in science education.}
\newblock \bibinfo{journal}{\emph{Online Submission}} \bibinfo{volume}{8}, \bibinfo{number}{20} (\bibinfo{year}{2017}), \bibinfo{pages}{160--173}.
\newblock


\bibitem[Shaer et~al\mbox{.}(2024)]%
        {shaer2024ai}
\bibfield{author}{\bibinfo{person}{Orit Shaer}, \bibinfo{person}{Angelora Cooper}, \bibinfo{person}{Osnat Mokryn}, \bibinfo{person}{Andrew~L Kun}, {and} \bibinfo{person}{Hagit Ben~Shoshan}.} \bibinfo{year}{2024}\natexlab{}.
\newblock \showarticletitle{AI-Augmented Brainwriting: Investigating the use of LLMs in group ideation}. In \bibinfo{booktitle}{\emph{Proceedings of the CHI Conference on Human Factors in Computing Systems}}. \bibinfo{pages}{1--17}.
\newblock


\bibitem[Sharma et~al\mbox{.}(2023)]%
        {sharma2023towards}
\bibfield{author}{\bibinfo{person}{Mrinank Sharma}, \bibinfo{person}{Meg Tong}, \bibinfo{person}{Tomasz Korbak}, \bibinfo{person}{David Duvenaud}, \bibinfo{person}{Amanda Askell}, \bibinfo{person}{Samuel~R Bowman}, \bibinfo{person}{Newton Cheng}, \bibinfo{person}{Esin Durmus}, \bibinfo{person}{Zac Hatfield-Dodds}, \bibinfo{person}{Scott~R Johnston}, {et~al\mbox{.}}} \bibinfo{year}{2023}\natexlab{}.
\newblock \showarticletitle{Towards understanding sycophancy in language models}.
\newblock \bibinfo{journal}{\emph{arXiv preprint arXiv:2310.13548}} (\bibinfo{year}{2023}).
\newblock


\bibitem[Shen et~al\mbox{.}(2023)]%
        {shen2023convxai}
\bibfield{author}{\bibinfo{person}{Hua Shen}, \bibinfo{person}{Chieh-Yang Huang}, \bibinfo{person}{Tongshuang Wu}, {and} \bibinfo{person}{Ting-Hao~Kenneth Huang}.} \bibinfo{year}{2023}\natexlab{}.
\newblock \showarticletitle{ConvXAI: Delivering heterogeneous AI explanations via conversations to support human-AI scientific writing}. In \bibinfo{booktitle}{\emph{Companion Publication of the 2023 Conference on Computer Supported Cooperative Work and Social Computing}}. \bibinfo{pages}{384--387}.
\newblock


\bibitem[Siangliulue et~al\mbox{.}(2015)]%
        {siangliulue2015toward}
\bibfield{author}{\bibinfo{person}{Pao Siangliulue}, \bibinfo{person}{Kenneth~C Arnold}, \bibinfo{person}{Krzysztof~Z Gajos}, {and} \bibinfo{person}{Steven~P Dow}.} \bibinfo{year}{2015}\natexlab{}.
\newblock \showarticletitle{Toward collaborative ideation at scale: Leveraging ideas from others to generate more creative and diverse ideas}. In \bibinfo{booktitle}{\emph{Proceedings of the 18th ACM Conference on Computer Supported Cooperative Work \& Social Computing}}. \bibinfo{pages}{937--945}.
\newblock


\bibitem[So and Joo(2017)]%
        {so2017does}
\bibfield{author}{\bibinfo{person}{Chaehan So} {and} \bibinfo{person}{Jaewoo Joo}.} \bibinfo{year}{2017}\natexlab{}.
\newblock \showarticletitle{Does a persona improve creativity?}
\newblock \bibinfo{journal}{\emph{The Design Journal}} \bibinfo{volume}{20}, \bibinfo{number}{4} (\bibinfo{year}{2017}), \bibinfo{pages}{459--475}.
\newblock


\bibitem[Southworth(2022)]%
        {Southworth2022BridgingCT}
\bibfield{author}{\bibinfo{person}{James Southworth}.} \bibinfo{year}{2022}\natexlab{}.
\newblock \showarticletitle{Bridging critical thinking and transformative learning: The role of perspective-taking}.
\newblock \bibinfo{journal}{\emph{Theory and Research in Education}}  \bibinfo{volume}{20} (\bibinfo{year}{2022}), \bibinfo{pages}{44 -- 63}.
\newblock
\urldef\tempurl%
\url{https://api.semanticscholar.org/CorpusID:249241081}
\showURL{%
\tempurl}


\bibitem[Spangler et~al\mbox{.}(2014)]%
        {spangler2014automated}
\bibfield{author}{\bibinfo{person}{Scott Spangler}, \bibinfo{person}{Angela~D Wilkins}, \bibinfo{person}{Benjamin~J Bachman}, \bibinfo{person}{Meena Nagarajan}, \bibinfo{person}{Tajhal Dayaram}, \bibinfo{person}{Peter Haas}, \bibinfo{person}{Sam Regenbogen}, \bibinfo{person}{Curtis~R Pickering}, \bibinfo{person}{Austin Comer}, \bibinfo{person}{Jeffrey~N Myers}, {et~al\mbox{.}}} \bibinfo{year}{2014}\natexlab{}.
\newblock \showarticletitle{Automated hypothesis generation based on mining scientific literature}. In \bibinfo{booktitle}{\emph{Proceedings of the 20th ACM SIGKDD international conference on Knowledge discovery and data mining}}. \bibinfo{pages}{1877--1886}.
\newblock


\bibitem[Stallen and Sanfey(2015)]%
        {stallen2015neuroscience}
\bibfield{author}{\bibinfo{person}{Mirre Stallen} {and} \bibinfo{person}{Alan~G Sanfey}.} \bibinfo{year}{2015}\natexlab{}.
\newblock \showarticletitle{The neuroscience of social conformity: Implications for fundamental and applied research}.
\newblock \bibinfo{journal}{\emph{Frontiers in neuroscience}}  \bibinfo{volume}{9} (\bibinfo{year}{2015}), \bibinfo{pages}{337}.
\newblock


\bibitem[Sun et~al\mbox{.}(2022)]%
        {sun2022students}
\bibfield{author}{\bibinfo{person}{Meng Sun}, \bibinfo{person}{Minhong Wang}, \bibinfo{person}{Rupert Wegerif}, {and} \bibinfo{person}{Jun Peng}.} \bibinfo{year}{2022}\natexlab{}.
\newblock \showarticletitle{How do students generate ideas together in scientific creativity tasks through computer-based mind mapping?}
\newblock \bibinfo{journal}{\emph{Computers \& Education}}  \bibinfo{volume}{176} (\bibinfo{year}{2022}), \bibinfo{pages}{104359}.
\newblock


\bibitem[Ta et~al\mbox{.}(2020)]%
        {ta2020user}
\bibfield{author}{\bibinfo{person}{Vivian Ta}, \bibinfo{person}{Caroline Griffith}, \bibinfo{person}{Carolynn Boatfield}, \bibinfo{person}{Xinyu Wang}, \bibinfo{person}{Maria Civitello}, \bibinfo{person}{Haley Bader}, \bibinfo{person}{Esther DeCero}, \bibinfo{person}{Alexia Loggarakis}, {et~al\mbox{.}}} \bibinfo{year}{2020}\natexlab{}.
\newblock \showarticletitle{User experiences of social support from companion chatbots in everyday contexts: thematic analysis}.
\newblock \bibinfo{journal}{\emph{Journal of medical Internet research}} \bibinfo{volume}{22}, \bibinfo{number}{3} (\bibinfo{year}{2020}), \bibinfo{pages}{e16235}.
\newblock


\bibitem[Terwiesch and Ulrich(2009)]%
        {terwiesch2009innovation}
\bibfield{author}{\bibinfo{person}{Christian Terwiesch} {and} \bibinfo{person}{Karl~T Ulrich}.} \bibinfo{year}{2009}\natexlab{}.
\newblock \bibinfo{booktitle}{\emph{Innovation tournaments: Creating and selecting exceptional opportunities}}.
\newblock \bibinfo{publisher}{Harvard Business Press}.
\newblock


\bibitem[Ulrich and Eppinger(2016)]%
        {ulrich2016product}
\bibfield{author}{\bibinfo{person}{Karl~T Ulrich} {and} \bibinfo{person}{Steven~D Eppinger}.} \bibinfo{year}{2016}\natexlab{}.
\newblock \bibinfo{booktitle}{\emph{Product design and development}}.
\newblock \bibinfo{publisher}{McGraw-hill}.
\newblock


\bibitem[Vieira et~al\mbox{.}(2011)]%
        {vieira2011critical}
\bibfield{author}{\bibinfo{person}{Rui~Marques Vieira}, \bibinfo{person}{Celina Tenreiro-Vieira}, {and} \bibinfo{person}{Isabel~P Martins}.} \bibinfo{year}{2011}\natexlab{}.
\newblock \showarticletitle{Critical thinking: Conceptual clarification and its importance in science education.}
\newblock \bibinfo{journal}{\emph{Science education international}} \bibinfo{volume}{22}, \bibinfo{number}{1} (\bibinfo{year}{2011}), \bibinfo{pages}{43--54}.
\newblock


\bibitem[Walton(1989)]%
        {Walton1989DialogueTF}
\bibfield{author}{\bibinfo{person}{Douglas Walton}.} \bibinfo{year}{1989}\natexlab{}.
\newblock \showarticletitle{Dialogue theory for critical thinking}.
\newblock \bibinfo{journal}{\emph{Argumentation}}  \bibinfo{volume}{3} (\bibinfo{year}{1989}), \bibinfo{pages}{169--184}.
\newblock
\urldef\tempurl%
\url{https://api.semanticscholar.org/CorpusID:143100862}
\showURL{%
\tempurl}


\bibitem[Wang et~al\mbox{.}(2023d)]%
        {wang2023tpe}
\bibfield{author}{\bibinfo{person}{Hongru Wang}, \bibinfo{person}{Huimin Wang}, \bibinfo{person}{Lingzhi Wang}, \bibinfo{person}{Minda Hu}, \bibinfo{person}{Rui Wang}, \bibinfo{person}{Boyang Xue}, \bibinfo{person}{Hongyuan Lu}, \bibinfo{person}{Fei Mi}, {and} \bibinfo{person}{Kam-Fai Wong}.} \bibinfo{year}{2023}\natexlab{d}.
\newblock \showarticletitle{Tpe: Towards better compositional reasoning over conceptual tools with multi-persona collaboration}.
\newblock \bibinfo{journal}{\emph{arXiv preprint arXiv:2309.16090}} (\bibinfo{year}{2023}).
\newblock


\bibitem[Wang et~al\mbox{.}(2023a)]%
        {wang2023scimon}
\bibfield{author}{\bibinfo{person}{Qingyun Wang}, \bibinfo{person}{Doug Downey}, \bibinfo{person}{Heng Ji}, {and} \bibinfo{person}{Tom Hope}.} \bibinfo{year}{2023}\natexlab{a}.
\newblock \showarticletitle{Scimon: Scientific inspiration machines optimized for novelty}.
\newblock \bibinfo{journal}{\emph{arXiv preprint arXiv:2305.14259}} (\bibinfo{year}{2023}).
\newblock


\bibitem[Wang et~al\mbox{.}(2023b)]%
        {Wang2023SciMONSI}
\bibfield{author}{\bibinfo{person}{Qingyun Wang}, \bibinfo{person}{Doug Downey}, \bibinfo{person}{Heng Ji}, {and} \bibinfo{person}{Tom Hope}.} \bibinfo{year}{2023}\natexlab{b}.
\newblock \showarticletitle{SciMON: Scientific Inspiration Machines Optimized for Novelty}. In \bibinfo{booktitle}{\emph{Annual Meeting of the Association for Computational Linguistics}}.
\newblock
\urldef\tempurl%
\url{https://api.semanticscholar.org/CorpusID:258841365}
\showURL{%
\tempurl}


\bibitem[Wang et~al\mbox{.}(2021)]%
        {wang2021towards}
\bibfield{author}{\bibinfo{person}{Qiaosi Wang}, \bibinfo{person}{Koustuv Saha}, \bibinfo{person}{Eric Gregori}, \bibinfo{person}{David Joyner}, {and} \bibinfo{person}{Ashok Goel}.} \bibinfo{year}{2021}\natexlab{}.
\newblock \showarticletitle{Towards mutual theory of mind in human-ai interaction: How language reflects what students perceive about a virtual teaching assistant}. In \bibinfo{booktitle}{\emph{Proceedings of the 2021 CHI conference on human factors in computing systems}}. \bibinfo{pages}{1--14}.
\newblock


\bibitem[Wang and Li(2011)]%
        {wang2011tell}
\bibfield{author}{\bibinfo{person}{Ting Wang} {and} \bibinfo{person}{Linda~Y Li}.} \bibinfo{year}{2011}\natexlab{}.
\newblock \showarticletitle{‘Tell me what to do’vs.‘guide me through it’: Feedback experiences of international doctoral students}.
\newblock \bibinfo{journal}{\emph{Active learning in higher education}} \bibinfo{volume}{12}, \bibinfo{number}{2} (\bibinfo{year}{2011}), \bibinfo{pages}{101--112}.
\newblock


\bibitem[Wang et~al\mbox{.}(2023c)]%
        {wang2023unleashing}
\bibfield{author}{\bibinfo{person}{Zhenhailong Wang}, \bibinfo{person}{Shaoguang Mao}, \bibinfo{person}{Wenshan Wu}, \bibinfo{person}{Tao Ge}, \bibinfo{person}{Furu Wei}, {and} \bibinfo{person}{Heng Ji}.} \bibinfo{year}{2023}\natexlab{c}.
\newblock \showarticletitle{Unleashing the emergent cognitive synergy in large language models: A task-solving agent through multi-persona self-collaboration}.
\newblock \bibinfo{journal}{\emph{arXiv preprint arXiv:2307.05300}} (\bibinfo{year}{2023}).
\newblock


\bibitem[Wu et~al\mbox{.}(2023b)]%
        {wu2023large}
\bibfield{author}{\bibinfo{person}{Ning Wu}, \bibinfo{person}{Ming Gong}, \bibinfo{person}{Linjun Shou}, \bibinfo{person}{Shining Liang}, {and} \bibinfo{person}{Daxin Jiang}.} \bibinfo{year}{2023}\natexlab{b}.
\newblock \showarticletitle{Large language models are diverse role-players for summarization evaluation}. In \bibinfo{booktitle}{\emph{CCF International Conference on Natural Language Processing and Chinese Computing}}. Springer, \bibinfo{pages}{695--707}.
\newblock


\bibitem[Wu et~al\mbox{.}(2023a)]%
        {wu2023autogen}
\bibfield{author}{\bibinfo{person}{Qingyun Wu}, \bibinfo{person}{Gagan Bansal}, \bibinfo{person}{Jieyu Zhang}, \bibinfo{person}{Yiran Wu}, \bibinfo{person}{Shaokun Zhang}, \bibinfo{person}{Erkang Zhu}, \bibinfo{person}{Beibin Li}, \bibinfo{person}{Li Jiang}, \bibinfo{person}{Xiaoyun Zhang}, {and} \bibinfo{person}{Chi Wang}.} \bibinfo{year}{2023}\natexlab{a}.
\newblock \showarticletitle{Autogen: Enabling next-gen llm applications via multi-agent conversation framework}.
\newblock \bibinfo{journal}{\emph{arXiv preprint arXiv:2308.08155}} (\bibinfo{year}{2023}).
\newblock


\bibitem[Xu et~al\mbox{.}(2023)]%
        {xu2023expertprompting}
\bibfield{author}{\bibinfo{person}{Benfeng Xu}, \bibinfo{person}{An Yang}, \bibinfo{person}{Junyang Lin}, \bibinfo{person}{Quan Wang}, \bibinfo{person}{Chang Zhou}, \bibinfo{person}{Yongdong Zhang}, {and} \bibinfo{person}{Zhendong Mao}.} \bibinfo{year}{2023}\natexlab{}.
\newblock \showarticletitle{Expertprompting: Instructing large language models to be distinguished experts}.
\newblock \bibinfo{journal}{\emph{arXiv preprint arXiv:2305.14688}} (\bibinfo{year}{2023}).
\newblock


\bibitem[Zheng et~al\mbox{.}(2024)]%
        {zheng2024disciplink}
\bibfield{author}{\bibinfo{person}{Chengbo Zheng}, \bibinfo{person}{Yuanhao Zhang}, \bibinfo{person}{Zeyu Huang}, \bibinfo{person}{Chuhan Shi}, \bibinfo{person}{Minrui Xu}, {and} \bibinfo{person}{Xiaojuan Ma}.} \bibinfo{year}{2024}\natexlab{}.
\newblock \showarticletitle{DiscipLink: Unfolding Interdisciplinary Information Seeking Process via Human-AI Co-Exploration}.
\newblock \bibinfo{journal}{\emph{arXiv preprint arXiv:2408.00447}} (\bibinfo{year}{2024}).
\newblock


\bibitem[Zhu et~al\mbox{.}(2018)]%
        {zhu2018explainable}
\bibfield{author}{\bibinfo{person}{Jichen Zhu}, \bibinfo{person}{Antonios Liapis}, \bibinfo{person}{Sebastian Risi}, \bibinfo{person}{Rafael Bidarra}, {and} \bibinfo{person}{G~Michael Youngblood}.} \bibinfo{year}{2018}\natexlab{}.
\newblock \showarticletitle{Explainable AI for designers: A human-centered perspective on mixed-initiative co-creation}. In \bibinfo{booktitle}{\emph{2018 IEEE conference on computational intelligence and games (CIG)}}. IEEE, \bibinfo{pages}{1--8}.
\newblock


\bibitem[Ziems et~al\mbox{.}(2024)]%
        {ziems2024can}
\bibfield{author}{\bibinfo{person}{Caleb Ziems}, \bibinfo{person}{William Held}, \bibinfo{person}{Omar Shaikh}, \bibinfo{person}{Jiaao Chen}, \bibinfo{person}{Zhehao Zhang}, {and} \bibinfo{person}{Diyi Yang}.} \bibinfo{year}{2024}\natexlab{}.
\newblock \showarticletitle{Can large language models transform computational social science?}
\newblock \bibinfo{journal}{\emph{Computational Linguistics}} \bibinfo{volume}{50}, \bibinfo{number}{1} (\bibinfo{year}{2024}), \bibinfo{pages}{237--291}.
\newblock


\end{thebibliography}

\clearpage
\appendix
\onecolumn
\section{Appendices}
\label{appendix}

\subsection{Participants Demographics}
\label{apdx:demographics}
\begin{table}[!h]
\centering
\caption{Participants' Research Background and Experience}
\label{tab:participant_demographics}
\begin{tabularx}{\textwidth}{lXlX}
\toprule
\textbf{PID} & \textbf{Background} & \textbf{Level of Education} & \textbf{Years of Research Experience} \\
\midrule
T1  & HCI                   & Doctoral            & 3-4 years          \\
T2  & HCI                   & Doctoral            & 3-4 years          \\
T3  & ML/NLP                & Doctoral            & 3-4 years          \\
T4  & HCI                   & Doctoral            & 3-4 years          \\
T5  & HCI                   & Doctoral            & 3-4 years          \\
T6  & HCI                   & Doctoral (Post)     & 5+ years           \\
P7  & ML/NLP                & Doctoral            & 3-4 years          \\
P8  & HCI                   & Doctoral            & 3-4 years          \\
P9  & HCI                   & Doctoral (Post)     & 5+ years           \\
P10 & HCI                   & Doctoral            & 3-4 years          \\
P11 & Biochemistry          & Undergraduate       & Less than 1 year   \\
P12 & Microbiology          & Undergraduate       & Less than 1 year   \\
P13 & Bioinformatics        & Doctoral            & 1-2 years          \\
P14 & Critical Data Studies & Doctoral            & 3-4 years          \\
P25 & Biomedical Research   & Doctoral (Post)     & 5+ years           \\
P22 & HCI                   & Doctoral            & 3-4 years          \\
P17 & Education             & Doctoral            & 1-2 years          \\
P20 & Economics             & Master              & 1-2 years          \\
P29 & HCI                   & Doctoral            & 3-4 years          \\
P30 & Bioinformatics        & Doctoral            & 3-4 years          \\
P31 & HCI                   & Master              & 1-2 years          \\
\bottomrule
\end{tabularx}
\end{table}

\subsection{Node Rating Questions (5-point Likert Scale)}
\label{apdx:node_rating_questions}
Here, we present the node rating questions embedded on each node used to collect users' immediate feedback toward generated critiques and RQs.
\begin{table}[h!]
\centering
\caption{Node Rating Questions for Critique Nodes}
\begin{tabularx}{\textwidth}{llX}
\toprule
\textbf{Node Type} & \textbf{Aspect} & \textbf{Question} \\
\midrule
Critique Node & Relevance & This critique is relevant to the research topic. \\
& helpfulness & This critique is helpful for my research exploration. \\
& Informativeness & This critique contains a lot of information to help improve my research idea. \\
& Insightfulness & This critique provides me with new insights that I had not considered before. \\
\bottomrule
\end{tabularx}
\end{table}

\begin{table}[h!]
\centering
\caption{Node Rating Questions for RQ Nodes}
\begin{tabularx}{\textwidth}{llX}
\toprule
\textbf{Node Type} & \textbf{Aspect} & \textbf{Question} \\
\midrule
RQ Node & Relevance & This RQ is relevant to the research topic. \\
& Creativity & This RQ provides me with new insights that I had not considered before. \\
& Feasibility & This RQ is feasible to conduct further research on. \\
& Specificity & This RQ describes itself with very specific information. \\
\bottomrule
\end{tabularx}
\end{table}

\newpage
\subsection{Post-Session Survey Questions (5-point Likert Scale)}
Here, we present the post-session survey questions utilized to gather participants' feedback on their user experience and overall satisfaction with the system. This form was filled out by participants two times after finishing the single-persona and multi-persona conditions, respectively.
\label{apdx:survey_ideation_outcome}
\begin{table}[!h]
\centering
\caption{Survey Questions for Overall Ideation Outcome Evaluation}
\label{tab:survey_ideation_outcome}
\begin{tabularx}{\textwidth}{lX}
\toprule
\textbf{Aspect} & \textbf{Question} \\
\midrule
Overall Quality & The generated personas were very helpful for ideation. \\
& The generated critiques were very helpful for ideation. \\
& The generated literatures by personas are domain relevant. \\
& The generated RQs are creative. \\
& The generated RQs are feasible. \\
& The generated Research Outlines are in general of high quality. \\
& I feel like I can build on the generated Research Outlines to further establish my own research planning. \\
\bottomrule
\end{tabularx}
\end{table}

\label{apdx:survey_ideation_experience}
\begin{table}[h]
\centering
\caption{Survey Questions for Overall Ideation Experience Evaluation}
\label{tab:survey_ideation_experience}
\begin{tabularx}{\textwidth}{lX}
\toprule
\textbf{Aspect} & \textbf{Question} \\
\midrule
Persona & The generated personas provided new perspectives during my ideation process. \\
& The generated personas changed the way I think during the ideation. \\
\addlinespace
\hdashline
\addlinespace
Human-AI Relationship & I feel like I trust the generated outcomes of the system. \\
& I think I rely on the system too much. \\
& I feel like I’m in control when using the system. \\
& I can easily recall and identify the ideas I created using the system. \\
\bottomrule
\end{tabularx}
\end{table}

\label{apdx:survey_usability_cognitive_load}
\begin{table}[h]
\centering
\caption{Survey Questions for System Usability and Cognitive Load Evaluation}
\label{tab:survey_usability_cognitive_load}
\begin{tabularx}{\textwidth}{lX}
\toprule
\textbf{Aspect} & \textbf{Question} \\
\midrule
Usability & I think that I would like to use this system frequently. \\
& I thought the system was easy to use. \\
& I felt very confident using the system. \\
\addlinespace
\hdashline
\addlinespace
Cognitive Load & The task is very mentally demanding. \\
\bottomrule
\end{tabularx}
\end{table}

\newpage
\subsection{Sample Workflow Analyzed During Case Studies}
  \label{appdx:p30-flow-snapshot-large}
\begin{figure}[htbp]
  \centering
  \includegraphics[angle=90,width=.52\linewidth]{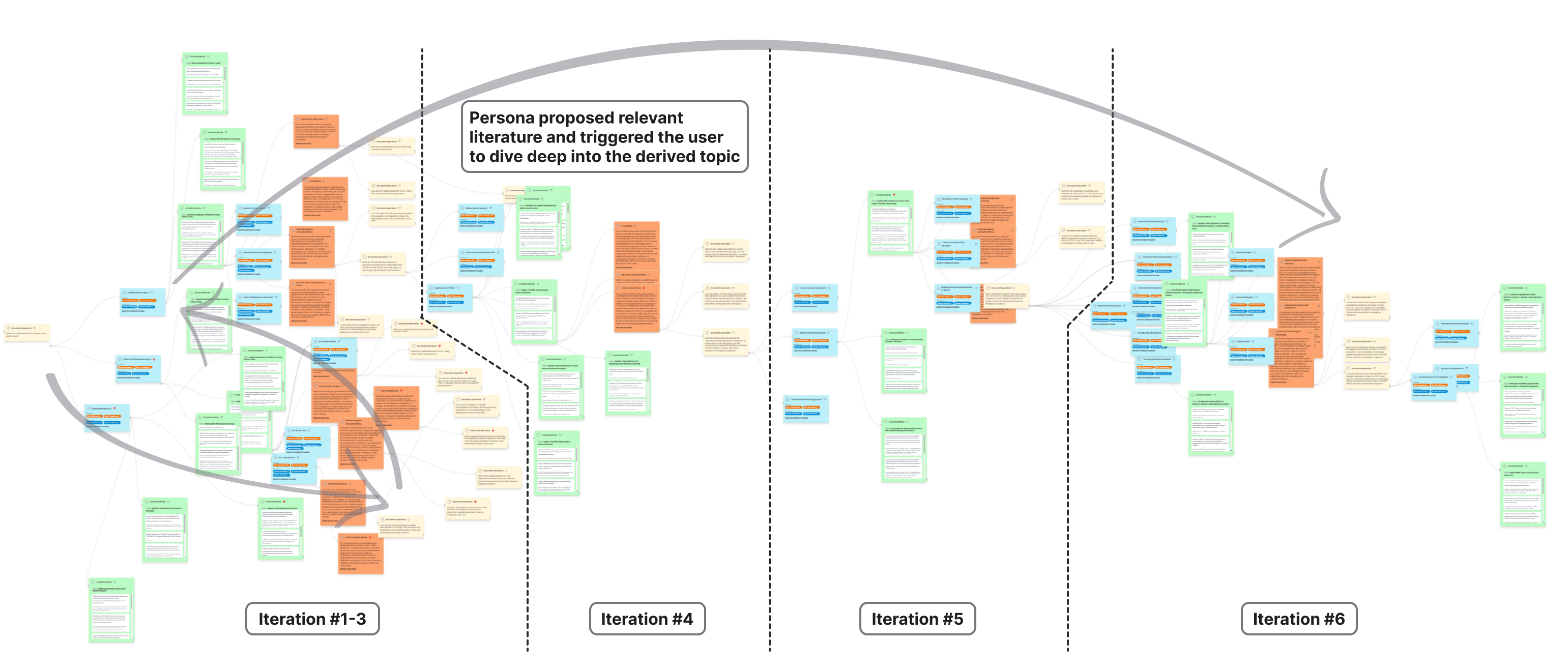}
  \caption{%
    A larger version of the workflow visualization of P20's thought processes.
  }
\end{figure}


\newpage
\subsection{Prompts Used in the System}
\label{apdx:system-prompts}
We list all prompts we designed and used in the system below. 

\vspace{.5cm}
\textbf{Generate Critiques from Persona and Literature:}
\begin{lstlisting}
You are a research assistant agent with the following characteristics:
{persona}
Your goal is to assist users by providing critiques on their research ideas and questions.

You will be provided with an original research idea/question provided by the user and a persona based on which you would need to play the role and provide critiques as if you are the persona.
You will also be provided with a list of literature, based on which you would need to provide critiques as if you are the person.

You should provide critiques from multiple different aspects.
The format of the critiques should strictly follow the following format:
[{{"critique_aspect": "...", "critique_detail": "..."}}, ...]
Keep the number of critiques to be 3. Do not generate content other than the json itself.

Now you are provided with the following research idea/question:
{rq}
And you are provided with the following literature (if any):
{lits}

Your critiques:
\end{lstlisting}

\vspace{.5cm}
\textbf{Generate Literature Queries from Persona:}
\begin{lstlisting}
You are a research assistant agent with the following characteristics:
{persona}
Your goal is to assist the user by proposing literature search queries on their research idea and questions.
You will be provided with an original research idea/question provided by the user and a persona based on which you would need to play the role and propose literature search queries as if you are the persona.
You may also be provided with a list of literature, in which case you would need to propose literature search queries by considering the scope of the provided literature.
Your role is to help users craft optimized search queries and terms for scholarly search engines like Semantic Scholar or Google Scholar, based on unstructured information they provide. 
You should guide users in refining their ideas or descriptions into precise, relevant search terms. 

You should provide search queries from multiple different aspects.
The format of the queries should strictly follow the following format:
[{{"search_query": "..."}}, ...]
Keep the number of queries to be 3. Do not generate other content besides the json itself.

Some examples of outputs could be: 
[{{"search_query": "AI for healthcare"}}, {{"search_query": "Machine learning for medical imaging"}}, {{"search_query": "Deep learning for cancer detection"}}]
Queries should be short and concise, if a query is too long, consider breaking it down into multiple queries.

Now you are provided with the following research idea/question:
{rq}
And you are provided with the following literature (if any):
{lits}

Your search queries:
\end{lstlisting}

\vspace{.5cm}
\textbf{Breaking Down Primary Literature Queries to Secondary Queries:}
\begin{lstlisting}
You are a research assistant agent that assists users by breaking down their literature search queries into more specific terms and sub-queries.
You will play the following persona:
{persona}

You will be provided with a literature search query that represents the user's interest and a persona based on which you would need to play the role. Break down the search query into more specific terms and phrases as if you are the persona.
You will also be provided with the user's initial research question/idea they would like to explore through the literature search query.
Your role is to help users craft optimized search queries and terms for scholarly search engines like Semantic Scholar, based on unstructured information they provide.
You should provide both the breakdown of the original query and a short rationale for the information you are looking for. The rationale will be later used to rerank the search results.

You should provide search queries from multiple different aspects to provide a comprehensive breakdown of the original query.
The format of the queries should strictly follow the following format:
[{{"search_query": "SEARCH_QUERY_1", "rationale": "RATIONALE_1"}}, {{"search_query": "SEARCH_QUERY_2", "rationale": "RATIONALE_2"}}, ...]
Do not generate other content besides the json itself.

Some examples of outputs could be: 
    User's original research question/idea:
        "How can we address the lack of engagement of users using online art platforms by simulating multi-persona?"
    Original query: 
        "Multi-Persona Simulation For Enhancing User Interaction"
    
    Breakdown queries:
        [{{"search_query": "Multi-Persona Simulation", "rationale": "What are the existing methods for simulating multiple personas?"}}, {{"search_query": "Simulated personas for user interaction", "rationale": "How are simulated personas used to enhance user interaction?"}}, ...]
Queries should be short and concise, if a query is too long, consider breaking it down into multiple queries or keywords.

Now you are provided with the following research idea/question:
{rq}

The user provided the following original search query:
{search_query}

Now generate a list of more specific search queries based on the provided search query.
\end{lstlisting}

\vspace{.5cm}
\textbf{Generating Personas from Initial Research Question/Idea:}
\begin{lstlisting}
You are a research assistant agent with the following characteristics:
{persona}
Your goal is to assist the user by hypothesizing personas that are best fitted to provide feedback based on their research idea and questions.
You will be provided with an original research idea/question provided by the user, based on which you would need to hypothesize personas that are best fitted to provide feedback.

You should hypothesize personas that are best fitted to provide feedback based on the research idea/question provided.
Do NOT use real names for the personas. Use high-level roles instead. Persona names should be informative and capture the essence of the persona's expertise and focus. Both the persona names and descriptions should be presented in a human-readable format.
The format of the personas should strictly follow the following format:
[{{"persona_description": {{"role_fields": {{"Role": "...", "Goal": "", ... }}, "background_fields":{{"Domain": "...", ... }}}}, "persona_name": "..."}}, ...]
Keep the number of personas to be 3. Do not generate other content besides the json itself.

Note that the user has already been provided with the following past personas:
{history_personas}
Try to generate personas that are different from the past personas.

Now you are provided with the following research idea/question:
{rq}

Generate your hypothesized personas:
\end{lstlisting}

\vspace{.5cm}
\textbf{Generating Research Questions from Persona:}
\begin{lstlisting}
You are a research assistant agent with the following characteristics:
{persona}
Your goal is to assist the user by hypothesizing personas that are best fitted to provide feedback based on their research idea and questions.
You will be provided with an original research idea/question provided by the user, based on which you would need to hypothesize personas that are best fitted to provide feedback.

You should hypothesize personas that are best fitted to provide feedback based on the research idea/question provided.
Do NOT use real names for the personas. Use high-level roles instead. Persona names should be informative and capture the essence of the persona's expertise and focus. Both the persona names and descriptions should be presented in a human-readable format.
The format of the personas should strictly follow the following format:
[{{"persona_description": {{"role_fields": {{"Role": "...", "Goal": "", ... }}, "background_fields":{{"Domain": "...", ... }}}}, "persona_name": "..."}}, ...]
Keep the number of personas to be 3. Do not generate other content besides the json itself.

Note that the user has already been provided with the following past personas:
{history_personas}
Try to generate personas that are different from the past personas.

Now you are provided with the following research idea/question:
{rq}

Generate your hypothesized personas:
\end{lstlisting}

\vspace{.5cm}
\textbf{Generating Personas from a Summary of Literature:}
\begin{lstlisting}
You are a research assistant agent with the following characteristics:
{persona}
Your goal is to assist the user by hypothesizing researchers' personas based on the literature summary provided.
You will be provided with a literature summary provided by the user, based on which you would need to hypothesize researchers' personas as the author of the literature included in the summary.

You should hypothesize researchers' personas as the author of the literature included in the summary, think about the methods they used, the expertise they have, and the research domains they are interested in.
Do NOT use real names for the personas. Use high-level roles instead. Persona names should be informative and capture the essence of the persona's expertise and focus. Both the persona names and descriptions should be presented in a human-readable format.
The format of the personas should strictly follow the following format:
[{{"persona_description": {{"role_fields": {{"Role": "...", "Goal": "", ... }}, "background_fields":{{"Domain": "...", ... }}}}, "persona_name": "..."}}, ...]
Keep the number of personas to be 3. Do not generate other content besides the json itself.

Now you are provided with the following literature summary:
{summary}
\end{lstlisting}

\vspace{.5cm}
\textbf{Generating Research Outline Table:}
\begin{lstlisting}
You are an experienced researcher tasked with assuming a specific persona based on the provided background information. 
Imagine you are playing the role of the persona described below:
{persona}

You will be provided with a research question, a specific research scenario, a critique node, and a literature node. The research question evolves from the critique node, which in turn, is informed by the literature node.
Your objective is to propose a structured outline for a research project based on the provided information. 
Focus on the feasibility and present a clear hypothesis and actionable steps for each section in a narrative format. 
Be concise and brief in your responses, ensuring that each section is well-defined and contributes to the overall research project.
Remember to keep a suggestive tone throughout and not directly state the information. 

The user has previously engaged in a discussion with the following historical context:
{context}

Additional details for your persona and research components are as follows:
Research Question:
{rq}
Research Scenario:
{scenario}
Critique Node:
{critiqueNode}
Literature Node:
{literature}


Now, based on the provided information, generate a structured outline for a research project in the following format:
    The outline is a table where each row includes a section title and a brief description of the content to be included in that section.
    Pick a suitable outline structure based on the expertise/domain of the persona or literature provided. 
    Focus on sections that are practical and actionable, always start with "Motivation and Research Gap", do not inlcude sections such as "Introduction", "Literature Review", "Conclusion", and "References".
    The description should be concise and informative in bullet points of short sentences each within 20-30 words.
    Please adhere strictly to the following output format for the outline:
    [{{"title": "TITLE OF THE SECTION", "description": "1. DESCRIPTION OF THE SECTION BULLET 1\n\n2. DESCRIPTION OF THE SECTION BULLET 2\n\n..."}}, ...]
    Only output the outline in the json format. Do not include any other content besides the json itself.
\end{lstlisting}

\vspace{.5cm}
\textbf{Generating Hypothetical Abstract:}
\begin{lstlisting}
You are a research assistant agent tasked with generating a hypothetical research abstract for a research paper based on the provided information. 
Imagine you are playing the role of the persona described below:
{persona}

You will be provided with a research question, a critique, a literature review, and a table of research outline.
Your objective is to craft a concise and informative hypothetical abstract that based on the table of research outline.
The abstract should be concise and informative, reflecting the research scenario and aligning with the persona's expertise and focus.
Your abstract should be structured and simulate a real-world research abstract, avoid using direct quotes and citations from the provided information.

The user has previously engaged in a discussion with the following historical context:
{context}

Additional details for the research components are as follows:
Research Question:
{rq}
Research Scenario:
{scenario}
Literature Review:
{literature}
Table of research outline:
{tableData}


You have already come up with the following critique for the research question:
{critiqueNode}

Generate a hypothetical abstract for the research paper based on the provided information in the following format:
{{"hypothetical_abstract": "YOUR HYPOTHETICAL ABSTRACT"}}
\end{lstlisting}

\vspace{.5cm}
\textbf{Generating Literature Review:}
\begin{lstlisting}
You are a research assistant agent tasked with generating a literature review based on the provided information. 
You will be provided with abstracts of several research papers, and a research question the literature review should be conducted to address.
Your objective is to craft a concise and informative literature review that summarizes the key findings, methodologies, and contributions of the provided papers. 
The literature review should be well-organized in bullet points, highlighting the current state of research on the topic, identifying trends, gaps, and suggesting future research directions.
The literature review should contain the following key points:
- Relevant Past Findings: Summarize the main findings from each paper.
- Existing Methods: Describe the research methodologies used in the studies.
- Contributions from Prior Works: Highlight the unique contributions of each paper.
- Research Gap and Motivation: Identify the research gap and the motivation for the study.
Each point should be concise and short, focusing on the key takeaways from the papers.


The user has previously engaged in a discussion with the following historical context:
{context}


Now with the literature abstracts are as follows:
-----------
{abstracts}
-----------
And given the following research question:
{rq}
Generate a single-paragraph literature review based on the provided papers in the format as follows:
{{"literature_review": {{"Relevant Past Findings": "...", "Existing Methods": "...", "Contributions from Prior Works": "...", "Research Gap and Motivation": "..."}}}}
Output only json format. Do not output markdown. For citations, use the following format: "Author et al. (Year)(URL) found that..."
\end{lstlisting}

\vspace{.5cm}
\textbf{Generating Research Scenario Suggestions:}
\begin{lstlisting}
You are a research assistant agent tasked with generating a few research scenario suggestions based on the provided information.
Some examples of research scenarios could be new research directions, potential studies, or innovative approaches based on the research question and literature abstracts provided.
Each research scenario should be short and concise, and within 20 words.

The user has previously engaged in a discussion with the following historical context:
{context}

Now with the literature abstracts are as follows:
-----------
{abstracts}
-----------
And given the following research question:
{rq}

Generate 3 research scenarios based on the provided information in the format as follows:
{{"research_scenarios": ["RESEARCH SCENARIO 1", "RESEARCH SCENARIO 2", ...]}}
Output only json format. Do not output markdown. Be creative and think of different research directions based on the research question and literature abstracts provided.
\end{lstlisting}

\newpage

\end{document}